\documentclass{elsart}

\usepackage[square,comma]{natbib}
\usepackage{graphicx}
\usepackage{pxfonts}

\usepackage{amssymb}

\begin{document}

\thispagestyle{empty}
\begin{Large}
\textbf{DEUTSCHES ELEKTRONEN-SYNCHROTRON}

\textbf{\large{in der HELMHOLTZ-GEMEINSCHAFT}\\}
\end{Large}

DESY 09-069

May 2009

\begin{eqnarray}
\nonumber &&\cr \nonumber && \cr \nonumber &&\cr
\end{eqnarray}
\begin{eqnarray}
\nonumber
\end{eqnarray}
\begin{center}
\begin{Large}
\textbf{Method for the determination of the three-dimensional
structure of ultrashort relativistic electron bunches}
\end{Large}
\begin{eqnarray}
\nonumber &&\cr \nonumber && \cr
\end{eqnarray}

\begin{large}
Gianluca Geloni, Petr Ilinski, Evgeni Saldin, Evgeni Schneidmiller
and Mikhail Yurkov
\end{large}
\textsl{\\Deutsches Elektronen-Synchrotron DESY, Hamburg}
\begin{eqnarray}
\nonumber
\end{eqnarray}
\begin{eqnarray}
\nonumber
\end{eqnarray}
\begin{eqnarray}
\nonumber
\end{eqnarray}
ISSN 0418-9833
\begin{eqnarray}
\nonumber
\end{eqnarray}
\begin{large}
\textbf{NOTKESTRASSE 85 - 22607 HAMBURG}
\end{large}
\end{center}
\clearpage
\newpage

\begin{frontmatter}


\title{Method for the determination of the three-dimensional
structure of ultrashort relativistic electron bunches}


\author[DESY]{Gianluca Geloni}
\author[DESY]{Petr Ilinski}
\author[DESY]{Evgeni Saldin}
\author[DESY]{Evgeni Schneidmiller}
\author[DESY]{Mikhail Yurkov}

\address[DESY]{Deutsches Elektronen-Synchrotron (DESY), Hamburg,
Germany}

\begin{abstract}
We describe a novel technique to characterize ultrashort electron
bunches in X-ray Free-Electron Lasers. Namely, we propose to use
coherent Optical Transition Radiation to measure three-dimensional
(3D) electron density distributions. Our method relies on the
combination of two known diagnostics setups, an Optical Replica
Synthesizer (ORS) and an Optical Transition Radiation (OTR)
imager. Electron bunches are modulated at optical wavelengths in
the ORS setup. When these electron bunches pass through a metal
foil target, coherent radiation pulses of tens MW power are
generated. It is thereafter possible to exploit advantages of
coherent imaging techniques, such as direct imaging, diffractive
imaging, Fourier holography and their combinations. The proposed
method opens up the possibility of real-time, wavelength-limited,
single-shot 3D imaging of an ultrashort electron bunch.
\end{abstract}


\end{frontmatter}


\clearpage

\tableofcontents

\newpage

\section{\label{sec:intro}Introduction}

Three X-ray Free-Electron Lasers (XFELs),  LCLS \cite{LCLS},  SCSS
\cite{SCSS}, and the European XFEL \cite{XFEL} are currently under
commissioning or under construction. These machines are based on
the Self-Amplified Spontaneous Emission (SASE) process
\cite{KOND}-\cite{PELL} and will be operating with electron bunch
durations of less than $100$ fs.

Operational success of XFELs will be related to the ability of
monitoring the spatio-temporal structure of these sub-$100$ fs
electron bunches as they travel along the XFEL structure. However,
the femtosecond time-scale is beyond the scale of standard
electronic display instrumentation. Therefore, the development of
methods for characterizing such short electron bunches both in the
longitudinal and in the transverse directions is a high-priority
task, which is very challenging.

A method for peak-current shape measurements of ultrashort
electron bunches was proposed in \cite{ORSS}. It uses the
undulator-based Optical Replica Synthesizer (ORS), together with
the ultrashort laser pulse shape measurement technique called
Frequency-Resolved Optical Gating (FROG) \cite{FROG}. It was
demonstrated in \cite{ORSS} that the peak-current profile for a
single, ultrashort electron bunch could be determined with a
resolution of a few femtoseconds. The ORS method is currently
being tested at the Free-electron laser in Hamburg (FLASH)
\cite{ANG2}.

Recently, feasibility studies for integrating the ORS diagnostics
setup with a timing scheme for pump-probe experiments and with a
scheme for output-power stabilization of an X-ray SASE FEL were
presented in \cite{TIME} and \cite{STAB}. Both schemes rely on an
external optical laser to modulate the energy of an electron bunch
in a short undulator. Such energy-modulation is subsequently
converted into density modulation by means of a dispersive
section. Since the ORS setup already includes seed laser, energy
modulator undulator and dispersion section, it is the most natural
option to be considered in the implementation of both timing and
stabilization schemes.

In this paper we present a feasibility study for integrating the
ORS setup with a high-resolution electron bunch imager based on
coherent Optical Transition Radiation (OTR).

Electron bunch imagers based on incoherent OTR constitute the main
device presently available for the characterization of an
ultrashort electron bunch in the transverse direction. They work
by measuring the transverse intensity distribution. However, since
no fast enough detector is presently available, the image is
actually integrated over the duration of the electron bunch.
Therefore, incoherent OTR imagers fail to measure the temporal
dependence of the charge density distribution within the bunch.
For these reasons, the use of standard incoherent OTR imagers is
limited to transverse electron-beam diagnostics, to measure e.g.
the projected transverse emittance of electrons.

However, it is primarily the emittance of electrons in short axial
slices\footnote{These slices are only a fraction of the full
($100$ fs) bunch length.}, which determines the performance of an
XFEL. There is, therefore, a compelling need for the development
of electron diagnostics capable of measuring three-dimensional
(3D) ultrashort electron bunch structures with micron-level
resolution.

The main advantages of coherent OTR imaging with respect to the
usual incoherent OTR imaging is in the coherence of the radiation
pulse, and in the high photon flux. Exploitation of these
advantages lead to applications of coherent OTR imaging that are
not confined to diagnostics of the transverse distribution of
electrons. The novel diagnostic techniques described here can
indeed be used to determine the 3D distribution of electrons in a
ultrashort single bunch. In combination with multi-shot
measurements and quadrupole scans, they can also be used to
determine the electron bunch slice emittance.

The possibility of single-shot, 3D imaging of electron bunches
with microscale resolution makes coherent OTR imaging an ideal
on-line tool for aligning the bunch formation system at XFELs.

Future XFEL operation will set tight tolerances on electron bunch
trajectories. They need to be carefully monitored along the
full-length of the machine. In order to ensure SASE lasing at
X-ray wavelengths, a very high orbit accuracy of a few microns has
to be ensured in the $200$ m long undulator. The resolution of
incoherent OTR imagers is not adequate to characterize the
position of the center of gravity of an electron bunch with such
accuracy. Our studies show that coherent OTR imaging can be
utilized as an effective tool for measuring the absolute position
of the electron bunch with the required micron accuracy.

Finally, the improvement of bunch-imaging techniques up to the
microscale level does not only yield a powerful diagnostic tool,
but opens up new possibilities in XFEL technology as well.

A pioneering experiment for integrating the ORS setup with a
coherent OTR imager was performed at FLASH. The energy of the
coherent OTR pulse was measured as a function of the position of a
relatively short seed laser pulse along the (long) bunch, i.e. the
slice peak-current was measured\footnote{The electron bunch was
not compressed at the time of those measurements.}. First attempts
to extract information about the correlation between longitudinal
and transverse distribution also took place. These results, which
should be considered as first steps into a novel direction of
electron-beam diagnostics, are reported in \cite{ANGE}.

In this work, we illustrate the potential of the proposed coherent
OTR imager scheme for the case of the European XFEL \cite{XFEL}.
We show that it naturally fits into the project. Technical
realization will be straightforward and cost-effective, since it
is essentially based on technical components (ORS, OTR diagnostics
stations), which are already included in the design of the
European XFEL. In our analysis we considered the baseline
parameters, so that our scheme can be implemented at the very
first stage of operation of the European XFEL facility. However,
the applicability of our method is not restricted to the European
XFEL setup. Other projects, e.g. LCLS or SCSS \cite{LCLS,SCSS} may
benefit from this work as well.

Our paper is organized as follows. In the next Sections
\ref{sec:setup}, \ref{sec:OTRsetup} and \ref{sec:imager} we
introduce basic concepts  and important details pertaining the
Optical Replica method, the coherent OTR generation and the image
formation of a modulated electron bunch.

In Section \ref{sec:ft} we analyze a coherent OTR imaging setup,
the so called $4f$ filtering architecture, which is relatively
simple to implement experimentally. We demonstrate that such setup
can be used to characterize electron density profiles on the
microscale level. Such resolution level can be reached by taking
advantage of a number of options, including e.g. spatial filtering
in the Fourier plane and radial-to-linear polarization conversion,
which make coherent OTR imaging more accurate.

Diffractive imaging is one of the most promising techniques for
microscale imaging of electron bunches, when a detector records
the Fraunhofer diffraction pattern radiated by the electron bunch.
Subsequently, an image can be reconstructed with the help of a
phase retrieval algorithm. This method reduces the requirements on
the optical hardware by increasing the sophistication in the
post-processing of the data collected by the system. Besides, a
diffractive imaging setup has the same ultimate resolution of the
$4f$ coherent imaging setup. This extremely simple method is
discussed for two versions, with and without the use of lenses, in
Section \ref{sec:diffite}.

Fourier-Transform Holography (FTH) is analyzed as another
promising imaging method in Section \ref{sec:holog}. FTH is a
non-iterative imaging technique, so the image can be reconstructed
in a single step deterministic computation. This is achieved by
placing a coherent point source at an appropriate distance from
the object and having the object field interfering with the
reference wave produced by this point source, detecting the
interference pattern in the Fourier plane. For optical
applications, the resolution of holographic techniques is not
limited by size and quality of the point-like source. With the
help of modern lithographic methods it is not difficult to produce
a pinhole, unresolved at optical wavelengths, and let sufficiently
bright radiation through it. The fast, unambiguous and direct
reconstruction achieved in FTH is attractive for coherent OTR
imaging of electron bunches. Moreover, FTH may also be used to
generate a low-resolution image of the bunch to support
diffractive imaging techniques. In this case, multiple references
can be added to the FTH setup in order to increase the a-priori
information available.

One of the main unsolved problems in XFEL electron bunch
diagnostics is the characterization of bunches that have
significant distortions in transverse phase space, e.g. bunches
whose transverse phase-space ellipse varies along the beam itself.
In an RF photoinjector with perfectly working emittance
compensation technique, the electron beam transverse profile is
axis-symmetric, and the Twiss parameters are equal in all slices
(excluding the emittance, which varies). For a real beam, the
variation in the space charge forces can be significant and cannot
be properly compensated with a solenoid emittance compensation
scheme. In addition, Coherent Synchrotron Radiation (CSR) related
effects in bunch compressors can lead to further deviations from
the axis-symmetric model. The knowledge of the variation of the
phase-space ellipse along the bunch at the output of the bunch
formation system could provide significant information about the
physical mechanisms responsible for the generation of ultrashort
bunches in XFELs. If the 3D structure of electron bunches could be
provided (even as the result of a multi-shot measurement), a
quadrupole-scan (which is a multi-shot measurement method too)
could be used to provide phase-space density distribution
measurements. Twiss parameters in each slice could be
reconstructed in this way.

As mentioned above, the applications of coherent OTR imaging are
not confined to diagnostics of the transverse distribution of
electrons projected along the longitudinal axis.

Simple extensions of our proposed diagnostic techniques allow for
the characterization of the 3D structure of electron bunches with
a multi-shot measurement. Our approach, described in Section
\ref{sec:diffite}, involves a combination of real and reciprocal
space imaging spectrometers. Both imaging setups use frequency
filters to obtain the spectral data of the image. When the filter
bandpass is changed, successive images are recorded at different
wavelengths. This process is repeated, wavelength by wavelength,
until the entire spatial spectral data is built up slice by slice.
The result is the simultaneous knowledge of two "3D cubes" of
spectral data, one in the real space $(\Delta \lambda, \Delta x,
\Delta y)$ and the other in the reciprocal space $(\Delta \lambda,
\Delta \omega_x, \Delta \omega_y)$, having indicated with
$\omega_{x,y}$ the spatial frequencies relative to the $x$ and $y$
axis. Application of e.g. the Gerchberg-Saxton algorithm
\cite{GERC}, which is the first practical iterative algorithm to
have been developed for solving the Fourier-phase-retrieval
problem, allows one to retrieve the spatio-temporal electron-bunch
structure.


In Section \ref{sec:diffite} we also show how the determination of
the projections of the cube of data in reciprocal space onto
specific planes of interest is sufficient to reconstruct the
electron-bunch structure, even without knowledge about the cube of
spectral data in real space. The advantage of this method is that
it requires no reference pulse shorter than the electron bunch
(i.e. no synchronization). In other words, the optical replica
pulse can be measured in the 3D Fourier domain. We name this novel
method Frequency-Resolved Optical Diffractive Imaging (FRODI).

FRODI is further developed in Section \ref{sec:diffite} from a
multi-shot to a single-shot technique to measure the 3D structure
of a single electron bunch. This is accomplished by splitting the
beam and simultaneously measuring orthogonal $(x,t)$, $(y,t)$ and
$(x,y)$ projections. The entire traces can be recorded by three
detectors, and used to reconstruct the desired 3D electron-bunch
structure.

While FROG requires the use of nonlinear-optical process, FRODI is
a linear-optical method, and linear-optical processes do not
require intense pulses. Here we discuss about measurements of
complicated pulses in three dimensions, and it is interesting to
compare FRODI with the well-known FROG technique, which can
measure complicated pulses in one dimension (1D).  As was reported
in \cite{FROG} concerning measurements of temporal structures in
optical pulses: "It can be shown that linear-optical methods
cannot completely measure ultrashort pulses". Quite
counter-intuitively, the measurement of 3D structures of optical
pulses in time and space is simpler than the measurement of
temporal 1D structures alone. However, in analogy with this fact,
it is well-known that the two-dimensional (2D) phase retrieval
problem is solvable, unlike the 1D one. To detect the temporal
structure of optical pulses, FROG requires the use of a nonlinear
optical process, which allows to extend the 1D reconstruction
problem to a 2D reconstruction problem. This involves an
artificial 2D Fourier domain. Unlike it, FRODI transforms the
problem of measurement of 3D structures in time and space into
measurements of the more natural 3D spatial-frequency and
temporal-frequency domains. No prior information about the
electron bunch structure is necessary to reconstruct the electron
bunch density distribution from the experimental traces.

Our 3D imaging technique FRODI turns out to be a relatively simple
solution to a very complicated problem. For the 3D
(spatio-temporal) electric field of optical replica pulses
produced by optically-modulated electron bunches with
spatio-temporal distortions, different spatial frequencies are
related to different temporal spectra (i.e. spatial frequency and
temporal frequency are coupled).

Multi-shot and single-shot techniques for the characterization of
the electron bunch can also be based on FTH setups. A new
multi-shot technique for 3D imaging of the electron bunch based on
frequency gated FTH is discussed in Section \ref{sec:holog}. In
the same Section \ref{sec:holog} we also consider spatio-temporal
FTH techniques.  An extension of the method opens up the
possibility for single-shot 3D imaging of ultrashort electron
bunches.

Time-gated FTH is the next class of techniques discussed in
Section \ref{sec:holog}. The principle of this method is
straightforward. A hologram records information about the object
only when it is illuminated simultaneously by a coherent reference
wave. Then, when a short reference is used, the hologram is
equivalent to a time-gated viewing system \cite{ABRA}. We propose
a method based on time-gated FTH with multiple reference sources
capable of characterizing the spatio-temporal structure of
individual electron bunches. Multiple, ultrashort (about $10$ fs)
reference pulses are generated with a varying time-delay, so that
several two-dimensional images (frames) of the electron bunch at
different position inside the bunch can be reconstructed from a
single holographic pattern. We call this technique Holography
Optical Time Resolved Imaging (HOTRI).

We conclude our work in Section \ref{sec:conc}.



\section{\label{sec:setup}Optical replica setup}

\subsection{\label{sub:mod}Optical replica pulse generation}

We propose to create a coherent pulse of optical radiation by
first modulating the electron bunch at a given optical wavelength
and, second, by letting it pass through a metal foil target, thus
producing coherent Optical Transition Radiation (OTR) at the
modulation wavelength. The radiation pulse should be produced in
such a way as to constitute an exact replica of the electron
bunch. Such optical replica can be used for the determination of
the 3D structure of electron bunches. Although other projects may
benefit from our study too, throughout this paper we will mainly
refer to parameters and design of the European XFEL.

\begin{figure}
\begin{center}
\includegraphics*[width=115mm]{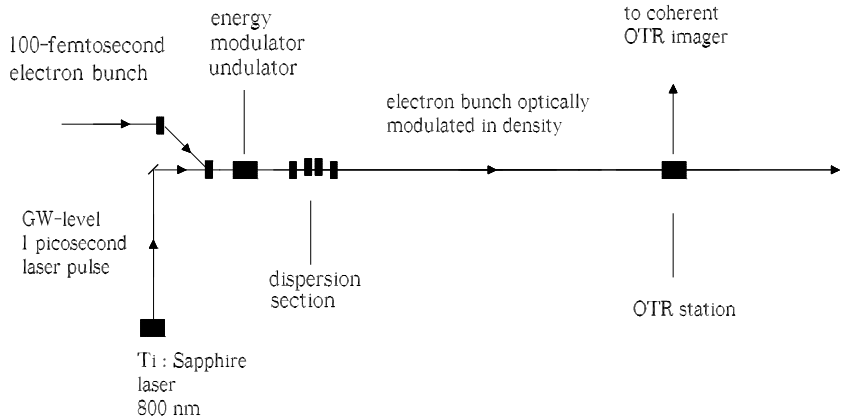}
\caption{\label{setupim}  Schematic diagram of the coherent
imager. The working principle is based on the optical modulation
of the electron bunch and on emission of coherent OTR radiation
from the metallic mirror.}
\end{center}
\end{figure}

In order to produce the optical replica we need to modulate the
electron bunch at a fixed optical wavelength. One may take
advantage of an Optical Replica Synthesizer (ORS) modulator
\cite{ORSS}, which we suppose to be installed after the BC2 bunch
compressor chicane. A basic scheme to generate coherent OTR is
shown in Fig. \ref{setupim}.

A relatively long laser pulse serves as a seed for the modulator,
consisting of a short undulator and a dispersion section. The
central area of the laser pulse should overlap with the electron
pulse. In order to ensure simple synchronization, the duration of
the laser pulse should be much longer than the electron pulse time
jitter, which is estimated to be of the order of $100$ fs.
Foreseen parameters of the seed laser are: wavelength $\lambda_m =
800$ nm, energy in the laser pulse $1$ mJ and pulse duration
(FWHM) $1$ ps. The laser beam is focused onto the electron bunch
in a short (the number of periods is $N_w = 5$) modulator
undulator resonant at the optical wavelength of $800$ nm. Optimal
conditions of focusing are met by positioning the laser beam waist
into the center of the modulator undulator, with a Rayleigh length
of the laser beam equal to the undulator length. Since the
electron betatron function $\beta$, the undulator length $L_w$ and
the Rayleigh length of the laser beam are of the same magnitude,
the size of the laser beam waist turns out to be about $20$ times
larger than the electron beam size. As a consequence, we can
approximate the laser beam with a plane wave when discussing about
the modulation of the electron bunch.

The seed laser pulse interacts with the electron beam in the
modulator undulator and produces an amplitude of the energy
modulation in the electron bunch of about $500$ keV. Subsequently,
the electron bunch passes through the dispersion section (with
momentum compaction factor $R_{56} \simeq 50 \mu$m), where the
energy modulation is converted into density modulation at the
laser wavelength. The electron bunch density modulation reaches an
amplitude of about $10\%$.

Finally, the modulated electron bunch travels through the OTR
screen. A powerful burst of OTR is emitted, which contains
coherent and incoherent parts. The coherent OTR has much greater
number of photons (up to $10^{13}$ i.e. $1 \mu$J per pulse, as we
will see), and can be used for diagnostic purposes.  A
quantitative treatment for coherent OTR is presented in Section
\ref{sec:OTRsetup}. The way we can take advantage of coherent OTR
properties is discussed in the following Sections.

It should be mentioned that OTR screens can be positioned at
various locations down the electron beam line where electrons have
substantially different energies. In the case of the European XFEL
\cite{XFEL}, the electron energy varies from $2$ GeV (second bunch
compression chicane) up to $17.5$ GeV at the undulator entrance.
For other machines, these parameters differ. In the case of LCLS
\cite{LCLS}, energies will range from about $4.5$ GeV to $13.6$
GeV.

Reference \cite{ORSS} includes a discussion about how to avoid the
influence of self-interaction effects in the ORS setup, when the
radiator is placed just behind the modulator. This case is
practically realized, for example, when we want to use a 3D OTR
imager to align the bunch formation system, and the OTR station is
placed just after the ORS setup behind the second bunch compressor
at $2$ GeV. The situation changes when the OTR imager is placed
behind the main XFEL accelerator at $17.5$ GeV, and the distance
between ORS modular and OTR imager is in the kilometer scale. In
Section \ref{sub:probl} we will present ideas how to avoid
self-interaction effects in the high-energy case. When, instead,
the ORS setup and the OTR imager are placed just behind the bunch
compression system, it sufficient to use the ideas introduced in
\cite{ORSS}.

\subsection{\label{sub:moduo}Optical modulator}

In order to use the coherent OTR burst for diagnostic purposes,
one has to ensure that an optical replica of the electron bunch is
actually produced. In fact, the electron bunch density modulation
can be perturbed by collective fields. It is therefore important
to consider collective interactions (radiation and space-charge
fields) influencing the operation of the Optical Replica
modulator, to ensure that longitudinal dynamics in the Optical
Replica modulator is governed by single-particle effects,
independently of the presence of other particles. In particular,
our method for electron-beam structure measurements is based on
the assumption that the electron bunch density modulation does not
appreciably change due to longitudinal space-charge (LSC)
interactions, i.e. plasma oscillations, as the beam propagates
through the setup behind the Optical Replica modulator, up to the
OTR station. Thus, the passage of the modulated electron bunch
through the setup must be studied.

Even when self-interactions are negligible, distortions can be
introduced due to nonuniform local energy spread within the
electron bunch.

Let us first consider under which conditions on the Optical
Replica modulator, the energy-spread effects are negligible. The
way the electron bunch is modulated in the modulator is
quantitatively described in e.g. \cite{CZON}. The current $I$ at
the exit of the dispersion section is found to be a composition of
harmonics of the modulation frequency $\omega_m$:

\begin{eqnarray}
I &=& I_0 + 2 I_0 \sum_{n=1}^{\infty} \exp\left[-\frac{n^2}{2}
\langle(\Delta \mathcal{E})^2\rangle\left(\frac{\omega_m R_{56}}{c
\mathcal{E}_0}\right)^2\right]  J_n\left(n P_0 \frac{\omega_m
R_{56}}{c \mathcal{E}_0}\right)\cr && \times \cos\left[n \omega_m
\left(\frac{z}{v_z}-t\right)\right]~, \label{czonka}
\end{eqnarray}
where $P_0=500$ keV is the initial energy modulation,
$\mathcal{E}_0$ is the nominal electron energy,
$\sqrt{\langle(\Delta \mathcal{E})^2\rangle}$ is the local energy
spread of electrons, $z$ is the longitudinal coordinate, $v_z$ is
the longitudinal velocity of electrons and $t$ is the time.
Moreover $J_n$ indicates the Bessel function of the first kind of
order $n$ and, as before, $R_{56}$ is the momentum compaction
factor. Note that the current is, in general a function of the
time, i.e. $I_0=I_0(t)$. However, throughout this paper we will
make use of the adiabatic approximation, because the electron
bunch is much longer than the modulation wavelength. As a result,
we can consider $I_0$ as a local parameter, and discuss about
local amplitude and phase of the density modulation.

As one can see from Eq. (\ref{czonka}), the microbunching depends
on the choice of the dispersion section strength. In fact,
neglecting the exponential suppression factor in
$\sqrt{\langle(\Delta \mathcal{E})^2\rangle}$, the expression for
the fundamental component of the bunched beam current is $I_1
\propto 2 I_0 J_1(X) \sim I_0 X$ for $X\ll 1$, where $X =P_0
{R_{56}}/({\lambdabar_m \mathcal{E}_0})$, with $\lambdabar_m =
c/\omega_m$, is a small dimensionless quantity known as the
bunching parameter.

One might think that all we have to do is to get the microbunching
amplitude to a maximum by increasing the $R_{56}$ of the
dispersion section, thus increasing the output power. In fact, it
is possible to build a dispersion section with a large $R_{56}$
function. However, one of the main problems in the modulator
operation is preventing the spread of microbunching due to local
energy spread in the electron bunch. In other words, for effective
operation, the value of the suppression factor in the exponential
factor in Eq. (\ref{czonka}) should be close to unity \cite{ORSS}.

\begin{figure}
\begin{center}
\includegraphics*[width=110mm]{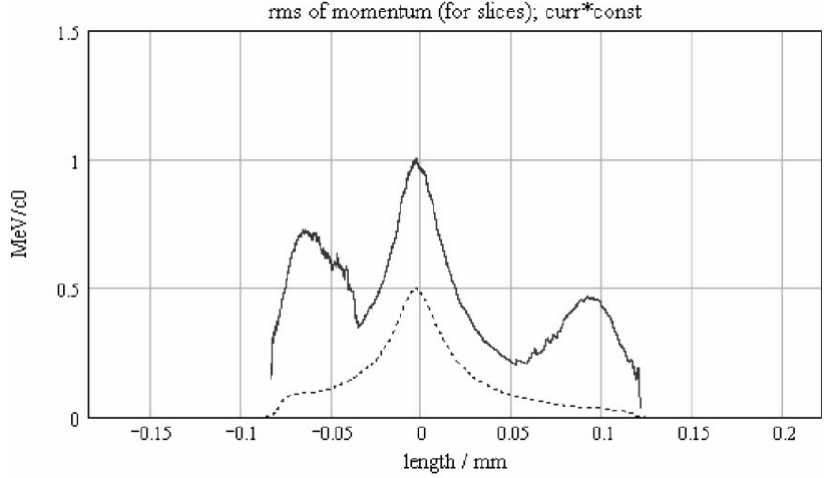}
\caption{\label{Espread}  Energy spread profile (rms, solid line)
at the entrance of the SASE undulator (after XFEL TDR)
\cite{XFEL}). }
\end{center}
\end{figure}

The energy spread is not constant along the electron bunch. For
example, the energy-spread distribution in the case of the
European XFEL is given in Fig. \ref{Espread}, reproduced from
\cite{XFEL}. The maximal energy spread level is about $1$ MeV.
Substituting numbers into the argument of the exponential function
(remember that the chicane of the modulator has dispersion
strength $R_{56} \simeq 50 ~\mu$m) one finds that, for the first
harmonic ($n=1$) the exponential factor is about unity ($\simeq
0.998$) even for $\mathcal{E}_0=2$ GeV, which is the minimal
energy considered. Moreover, the second harmonic ($n=2$) of the
modulation is suppressed by an order of magnitude with respect to
the first, due to the Bessel $J_2$ factor.

As a result, in our case we can approximate Eq. (\ref{czonka})
with

\begin{eqnarray}
&&I = I_0 + I_0  \cdot \frac{P_0 R_{56}}{\lambdabar_m
\mathcal{E}_0} \cos\left[\omega_m
\left(\frac{z}{v_z}-t\right)\right]~, \label{czonka2}
\end{eqnarray}
which ensures that the bunching is uniform along the beam. Calling
$\psi = \omega_m[z/v_z(\gamma_0) - t]$ the modulation phase, the
current can be expressed as $I = I_0[1+ a \cos(\psi)]$, where $a$
is the amplitude of our small ($|a| \ll 1$) density modulation,
taken with its own sign.

Let us now discuss distortions due to self-interaction effects.
Concerning the induced bunching inside the modulator undulator,
perturbations due to collective effects are minimized up to a
negligible level by using a small number of periods ($N_w = 5$).
This optimization is also important in order to increase the
replica resolution and minimize slippage effects ($N_w \lambda/c
\ll \sigma_T$, $\sigma_T$ being the electron bunch duration).

Concerning the effect of LSC interactions, one needs a more
detailed analysis. The propagation of the induced electron bunch
density modulation through the setup is a problem involving
self-interactions.  If the OTR station is placed at a short
distance (say, a few meters) from the modulator, e.g. to align the
bunch formation system, plasma oscillations play no important
effect, as it will be clear after reading the following analysis.
However, the situation changes if one wants to characterize the
electron beam after the linac, to monitor the electron bunch
properties along the machine and, in particular, to monitor
electron trajectories in the undulator. In fact, as the bunch
progresses through the linac, the modulation of the bunch density
produces energy modulation due to the longitudinal impedance
caused by space-charge fields. This process is complicated by the
fact that, due to the presence of energy and density modulation,
plasma oscillations can develop.

As a result, the initial energy and density modulation of the
electron bunch will be modified by the passage through the setup.
In order to study the feasibility of our scheme one needs to
estimate what modifications take place. For a given facility, OTR
screens positioned at a higher electron energy translate into a
longer distance between modulator and OTR screen and, thus, into a
stronger LSC influence. However, the evolution of plasma
oscillations tends to slow down as the energy increases. This
means that LCLS is less affected than the European XFEL, because
the last magnetic bunch compressor at LCLS is positioned at higher
energy ($4.5$ GeV compared to the $2$ GeV of the European XFEL).

\begin{figure}
\begin{center}
\includegraphics*[width=115mm]{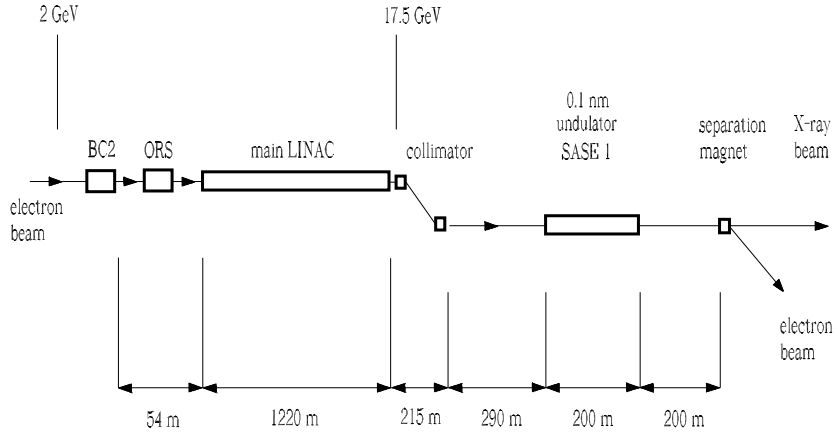}
\caption{\label{Xset}  Schematic diagram of the European XFEL
setup from BC2 through SASE1.}
\end{center}
\end{figure}

\subsection{\label{sub:probl}Distortions of the optical microbunching downstream the main
accelerator}

Let us consider the European XFEL case, where self-interactions
are more important (see Fig. \ref{Xset}). Proceeding as in
\cite{TIME} we find that energy and density modulation along the
accelerator are linked by the following system of differential
equations:

\begin{eqnarray}
\frac{d a}{dz} = \frac{1}{\lambdabar} \frac{\Delta
\gamma(z)}{(\gamma_0+g z)^3}~,\label{Daiacc}
\end{eqnarray}
\begin{eqnarray}
\frac{d(\Delta \gamma)}{dz} \simeq  - \frac{a(z)}{\lambdabar
(\gamma_0+g z)^2} \frac{I_0}{I_A} {\exp\left[\frac{\epsilon_n
\beta}{(\gamma_0+g z)^3 \lambdabar^2}\right]
~\Gamma\left[0,\frac{\epsilon_n \beta}{(\gamma_0+g z)^3
\lambdabar^2}\right]}~, \label{modul2}
\end{eqnarray}
where we assumed that $\gamma(z) = \gamma_0+g z$, with $g$ the
acceleration gradient and $\gamma_0 \simeq 4\cdot 10^3$. The
density modulation $a$ has been defined above, and the energy
modulation is given by $\Delta \gamma \sin(\psi)$. Here $I_A
\simeq 17$ kA is the Alfven current, $\epsilon_n$ is the
normalized emittance, $\beta$ is the average betatron function and
$\Gamma(\alpha,\beta)$ is the incomplete gamma function.

Usually, calculating self-interaction effects is rather
challenging. In our case, however, we used the straightforward 1D
model in Eq. (\ref{Daiacc}) and Eq. (\ref{modul2}). Remarkably,
such model is not a rough approximation of reality, but it rather
constitutes a quantitative approach. In fact, in our case study we
deal with a very particular range of parameters where four
asymptotic limits can be simultaneously exploited. First,
retardation effects can be neglected due to a small current $I \ll
\gamma I_A$, and to the adiabatic acceleration limit\footnote{The
adiabatic acceleration limit can always be used in our case.
Assuming a constant acceleration gradient $d\gamma/dz \equiv g
\simeq 25 ~\mathrm{m}^{-1}$ (see \cite{XFEL}), we have $2
\lambdabar \gamma (d \gamma/dz) \sim 10^{-2}$ for $\gamma = 4\cdot
10^3$, corresponding to the lowest energy of $2$ GeV, and $2
\lambdabar \gamma (d \gamma/dz) \sim 10^{-1}$ for $\gamma = 3.5
\cdot 10^4$, corresponding to the highest energy of $17.5$ GeV.
Note that the largest effects due to longitudinal impedance are
expected in the first part of the acceleration process.}
$\lambdabar (d\gamma^2/dz)\ll 1$. Second, the adiabatic
approximation applies, i.e. $ \omega \sigma_T \gg 1$. Third, a
pencil beam (1D) approximation is valid because\footnote{For
$\gamma = 4\cdot 10^{3}$ we have, for parameters specified here,
$\epsilon_n \beta/(\gamma^3 \lambdabar^2) \sim 3 \cdot 10^{-2}$.}
$\sigma_r \ll \gamma \lambdabar$, with $\sigma_r$ the bunch
transverse dimension. Finally, fourth, we can neglect the
interaction of space-charge fields with material structures
because $\gamma \lambdabar \ll b$, $b$ being the characteristic
dimension of the vacuum chamber. As a result, when dealing with an
XFEL setup and discussing about optical microbunching, we have a
unique situation. If any of the conditions above ceases to be
valid, the model in Eq. (\ref{Daiacc}) and Eq. (\ref{modul2})
ceases to be valid too, as it would be the case e.g. for
calculations of impedance at lower frequencies.

\begin{figure}
\begin{center}
\includegraphics*[width=110mm]{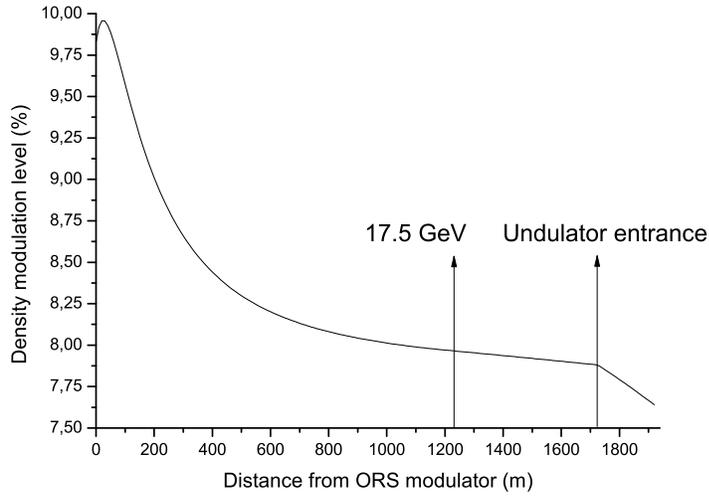}
\caption{\label{modul} Evolution of the electron bunch density
modulation as a function of the distance from the modulator.}
\end{center}
\end{figure}
Let us then use Eq. (\ref{Daiacc}) and Eq. (\ref{modul2}) assuming
an average betatron function of about $\beta = 25$ m along the
main accelerator and a normalized emittance $\epsilon_n = 1.4$
mm$\cdot$mrad (see \cite{XFEL}). We set the acceleration length in
the main linac equal to $d_a \simeq 1220$ m. The undulator does
not follow immediately the main linac, as it can be seen from Fig.
\ref{Xset}. For example, the SASE $1$ undulator is preceded by a
$290$ meters long straight section and by a collimation system,
which is $215$ meters long. According to \cite{XFEL}, the
compaction factor of the collimator can be set to any value from
$R^{(c)}_{56} = - 2 ~$mm to $R^{(c)}_{56} = 2 ~$mm, with
possibility of fine tuning (not in real time) of about $\pm 100
~\mu$m.

In practical situations, one is interested in inserting OTR
screens just after the bunch compression chicane (at $2$ GeV) or
along the undulator at $17.5$ GeV. The situation where the
influence of self-interactions is most important is obviously at
$17.5$ GeV along the undulator. As shown in \cite{OURX}, the
longitudinal Lorentz factor $\gamma_z=\gamma/\sqrt{1+K^2/2}$
should be used in the undulator instead of $\gamma$. Since $K =
3.3$, $\gamma^2$ and $\gamma_z^2$ differ of about an order of
magnitude, hence a different influence of the LSC in the undulator
compared to the straight section. In the undulator, Eq.
(\ref{Daiacc}) and Eq. (\ref{modul2}) are modified to

\begin{eqnarray}
\frac{d a}{dz} =  \frac{1}{\lambdabar} \frac{\Delta \gamma}{\gamma
\gamma_z^2}~,~~~~~~ \frac{d\Delta \gamma}{dz} \simeq -
\frac{a(z)}{\lambdabar \gamma_z^2} \frac{I_0}{I_A}
{\exp\left[\frac{\epsilon_n \beta}{\lambdabar^2 \gamma
\gamma_z^2}\right] ~\Gamma\left[0,\frac{\epsilon_n
\beta}{\lambdabar^2 \gamma \gamma_z^2}\right]}~, \label{modul2X}
\end{eqnarray}
where now $\gamma = 3.5 \cdot 10^4$. Even for a simple case study
where we set $R^{(c)}_{56} = 0$, numerical analysis shows that our
initial conditions $a_i \simeq 0.1$ and $(\Delta \gamma)_i \simeq
1$ yield an unwanted evolution of $a(z)$ as summarized in Fig.
\ref{modul}.

\begin{figure}
\begin{center}
\includegraphics*[width=115mm]{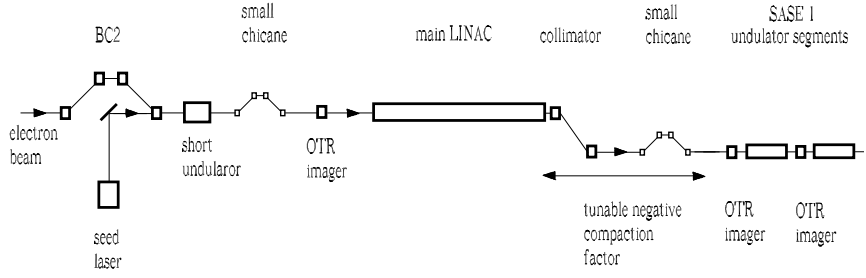}
\caption{\label{compe} Installation of a small dispersive element
(chicane) after the collimation system will allow to compensate
positional-dependent variations of the electron bunch density
modulation level.}
\end{center}
\end{figure}
As one may see, the electron bunch density modulation is
diminished by at most $\Delta a/a_i = 25 \%$ at highest energies
and at the end of the undulator. This constitutes a detrimental
effect concerning our imaging techniques. In fact, we calculated
$a(z)$ assuming a fixed peak-current $I_0 = 5 $ kA. However, the
peak-current along the bunch is not constant, and from Eq.
(\ref{Daiacc}) and Eq. (\ref{modul2}) follows that the modulation
level varies along the bunch in a complicated way depending on the
variation of the peak-current level. This effect is unwanted. In
fact, different parts of the electron bunch should be given the
right weight as concerns their contribution to radiation emission,
meaning that the (absolute) charge modulation in each point of the
electron bunch should be proportional to the charge density
distribution of the unmodulated bunch. This can only be realized
when the bunching factor is uniform along the bunch i.e. when it
does not depend on the charge density distribution nor on the
energy spread distribution.

\begin{figure}
\begin{center}
\includegraphics*[width=120mm]{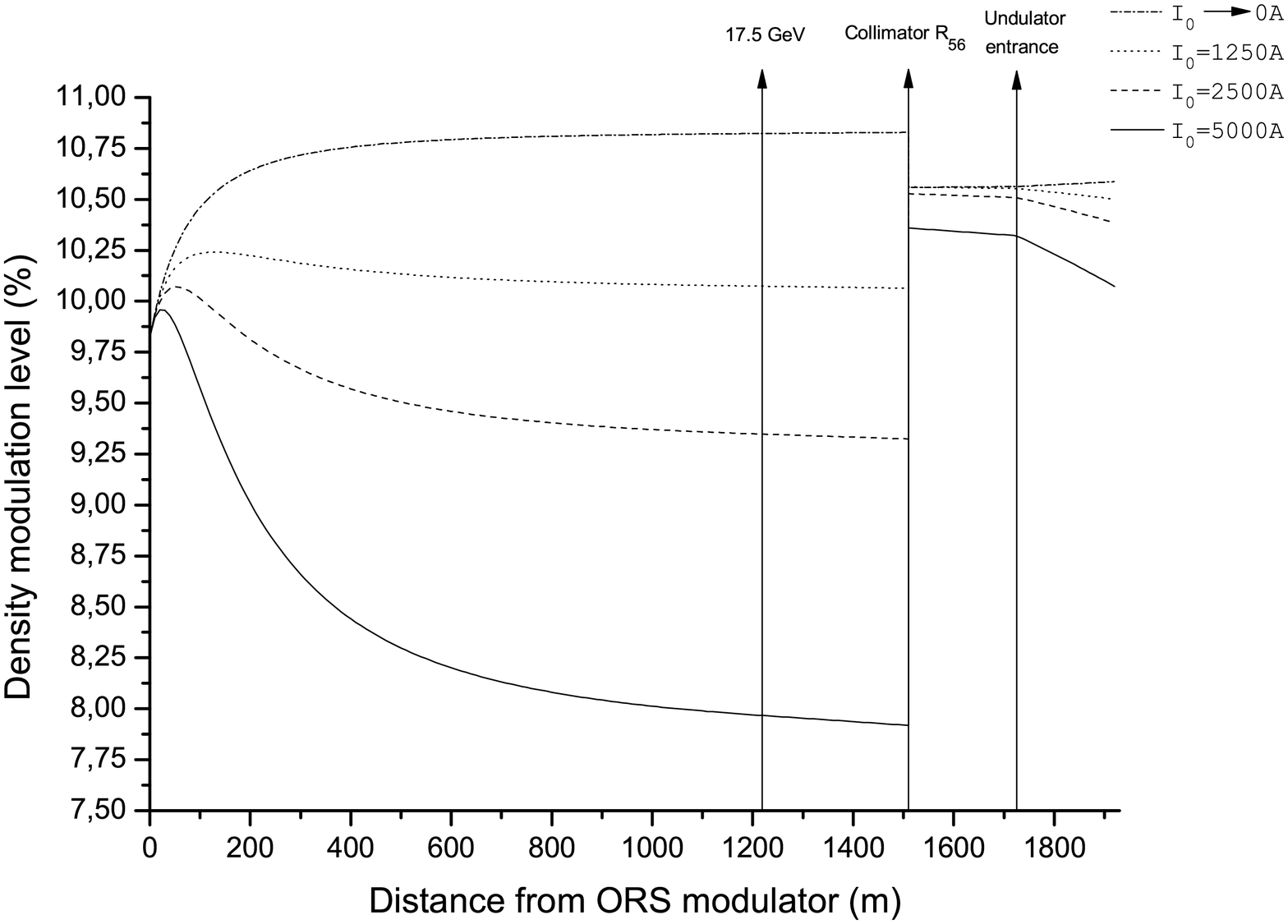}
\caption{\label{modul2fig} Evolution of the level of the electron
bunch density modulation for different values of the peak-current
$I_0$. Here we account for the presence of a dispersive element
$R_{56}^{(c)} \simeq 12 \mu$m at the collimation system.}
\end{center}
\end{figure}
Yet, note that the change in the modulation level is small at any
value of $z$, i.e. $\Delta a(z) \ll a_i$. As a result, from Eq.
(\ref{Daiacc}) and Eq. (\ref{modul2}) follows that, in first
approximation, the current-dependent terms in both $a(z)$ and
$\Delta \gamma$ depend on $I_0$ linearly. Moreover, the design
value for $R_{56}^{(c)}$ is negative. Therefore, one may think of
installing a small chicane as in Fig. \ref{compe} to organize a
tunable negative compaction factor. Calling with $R^{(t)}_{56}$
the total negative dispersion strength of collimator and chicane,
we can compensate the current-dependent term in $a$ at the
position of the extra-chicane by requiring:

\begin{eqnarray}
a-a_i = -\frac{|R^{(t)}_{56}| }{\lambdabar} \frac{\Delta
\gamma}{\gamma}~. \label{R56c}
\end{eqnarray}
Note that from the differential equation for $\Delta \gamma(z)$
and $a(z)$ we obtain

\begin{eqnarray}
&&a(z) = a_i + f(z) \Delta \gamma_i  + g(z) I_0 \cr && \cr
&&\Delta \gamma(z) = \Delta \gamma_i + h(z) I_0 ~, \label{gea}
\end{eqnarray}
where $f(z)$, $g(z)$ and $h(z)$ follow from the integration of Eq.
(\ref{Daiacc}) and Eq. (\ref{modul2}), assuming $a(z)=a_i$ in that
equation. A proper choice of $R^{(t)}_{56}$ can compensate for the
part of the electron bunch density modulation proportional to
$I_0$, but cannot compensate for the term in $\Delta \gamma_i$.
However, such term does not depend on the position inside the
bunch and is not detrimental.

With this in mind, the easiest way to find the proper
$R^{(t)}_{56}$ is to set $a(z) = a_i$ and $\Delta \gamma_i = 0$ in
Eq. (\ref{modul2}), solve it for $\Delta \gamma(z)$ and find
$a(z)$ from Eq. (\ref{Daiacc}). Then, substitution of $\Delta
\gamma(z)$ and $a(z)$ in Eq. (\ref{R56c}) allows one to obtain,
for our parameter choice, $R^{(t)}_{56} = - 12 \mu$m.

The final step is to find the evolution of $a(z)$ accounting for
the presence of $R_{56}^{(t)}$. Note that here we implicitly
modelled the collimator and the small chicane after it as a
$215$-meter-long straight section followed by a single, localized
dispersive element. Since the actual setup is more complicated
than that in our simple  model, results reported here are for the
sake of illustration only. However, detailed calculations would
not present novel effects due to the influence of space-charge, so
that exploitation of the tunability of $R_{56}^{(t)}$ is always
possible, and our simple estimations are qualitatively correct.

Results of numerical calculations for different values of the
peak-current $I_0$ are shown in Fig. \ref{modul2fig}. The
compensation effect of $R_{56}^{(t)}$ can be seen from the spread
of the modulation levels before and after the dispersive element.
In fact, from Fig. \ref{modul2fig} one can see that such spread is
reduced from about $\Delta a/a_i =30 \%$ to about $\Delta a/a_i =2
\%$ after the compensation chicane, to increase up to $\Delta
a/a_i =5 \%$  at the end of the undulator. One concludes that LSC
interactions do not introduce any undesired current-dependent
modification to the amplitude of the level of the electron bunch
density modulation with an accuracy $\Delta a/a_i$ of a few
percent.

We should stress that numbers considered in these examples are for
the worst possible influence of LSC. As said before, at LCLS these
effects would be even less important, as one starts from higher
energies after the second bunch compression chicane (at $4$ GeV).
One can take advantage of coherent OTR emission also at that
facility, which will allow the electron bunch image not to be in
the shadow of parasitic coherent OTR emission \cite{LOO1}. In
fact, with a laser-heater \cite{LASH}, the induced uncorrelated
energy spread is expected to limit the parasitic modulation of the
bunch after the last compressor chicane to less than $1\%$, which
is much smaller compared to the $10\%$ modulation at optical
wavelength induced on purpose in our scheme\footnote{It should
also be realized that in our methods we will almost always use a
narrow bandpass filter with relative bandwidth $0.1 \%$. In this
case, the influence of parasitic microbunching is strongly
reduced.}.

\section{\label{sec:OTRsetup} Characteristics of the OTR source}

\subsection{\label{sub:quali}Qualitative description}

\subsubsection{Parameter space of the problem}

As mentioned before, in this paper we deal with two situations of
practical interest, both common to the European XFEL and LCLS. In
the first, the OTR screen is positioned at low electron beam
energy, just after the modulator. In the second, it is positioned
at high electron beam energy, after the main linac and the
collimator system, possibly within the undulator line. These two
cases, shown in Fig. \ref{compe}, have differences, but the
overall qualitative picture is similar.


Let us define the slowly varying envelope of the field in the
space-frequency domain as $\vec{\widetilde{E}} = \vec{\bar{E}}
\exp{[-i \omega z/c]}$. We will refer to this quantity simply as
"the field". Here $\vec{\bar{E}}(\omega, \vec{r},z)$ is the
Fourier transform of the electric field $\vec{{E}}(t, \vec{r},z)$
in the space-time domain, according to the convention:
$\vec{\bar{E}}(\omega, \vec{r},z) = \int_{-\infty}^{\infty}
\vec{{E}}(t, \vec{r},z) \exp[i \omega t] d t$. Note that $\vec{r}$
indicates the transverse position vector.

We will be interested in the OTR emission from an electron bunch
modulated by the ORS. The typical longitudinal bunch dimension is
in the order of $50 \mu$m (FWHM), while the modulation wavelength
$\lambdabar_m \sim 0.1 \mu$m. Then, the adiabatic approximation
applies, and coherent OTR emission is automatically characterized
by a narrow bandwidth around $\lambdabar = \lambdabar_m$. This
means that we can use the expression for the OTR field from a
single electron in the space-frequency domain convolved with the
instantaneous charge density distribution as a good representation
of the instantaneous OTR field. 
The OTR field from a single electron in space-frequency domain is
usually approximated as\footnote{Further on, when the longitudinal
coordinate $z$ and the frequency $\omega$ will have a given, fixed
value, we will not always include them in the
arguments of the field.}

\begin{eqnarray}
\vec{\widetilde{E}}^{(1)}(\vec{r}) \propto
\frac{\omega(\vec{r}-\vec{r}_e)}{c \gamma
\left|\vec{r}-\vec{r}_e\right|} K_1\left(\frac{\omega
\left|\vec{r}-\vec{r}_e\right|}{c \gamma}\right) ~,
\label{proptoff0}
\end{eqnarray}
where $K_1$ is the first order modified Bessel function of the
second kind, $\vec{r}$ is the observation position on the screen
and $\vec{r}_e$ is the electron position. Eq. (\ref{proptoff0}) is
the result of the Ginzburg-Frank theory \cite{GINZ}.

We will see that the field distribution at the OTR screen is
characterized by two scales of interest. One is associated with
the transverse size of the electron bunch, $\sigma_r \sim 30
~\mu$m. The other is the typical size of the single-particle OTR
spatial distribution, which can be estimated from Eq.
(\ref{proptoff0}) in the order  of $\gamma \lambdabar \sim 1$ mm
for the $2$ GeV case, where we introduced the reduced wavelength
$\lambdabar = c/\omega$. In our case study of interest $\sigma_r
\ll \gamma \lambdabar$, and Eq. (\ref{proptoff0}) can be
approximated as

\begin{eqnarray}
\vec{\widetilde{E}}^{(1)}(\vec{r}) \propto
\frac{\vec{r}-\vec{r}_e}{\left|\vec{r}-\vec{r}_e\right|^2}~.
\label{proptoff}
\end{eqnarray}
This fact will be exploited through all our paper.

We should stress here the vectorial nature of the electric field,
which always exhibits space-variant polarization. This suggests an
electrostatic 2-D analogy with the electric field generated by an
uniformly charged wire. For the parameters of our problem, this
analogy is valid whenever $r \ll \gamma \lambdabar$. This
includes, in particular the range $r \lesssim \sigma_r \ll \gamma
\lambdabar$.

\begin{figure}
\begin{center}
\includegraphics*[width=150mm]{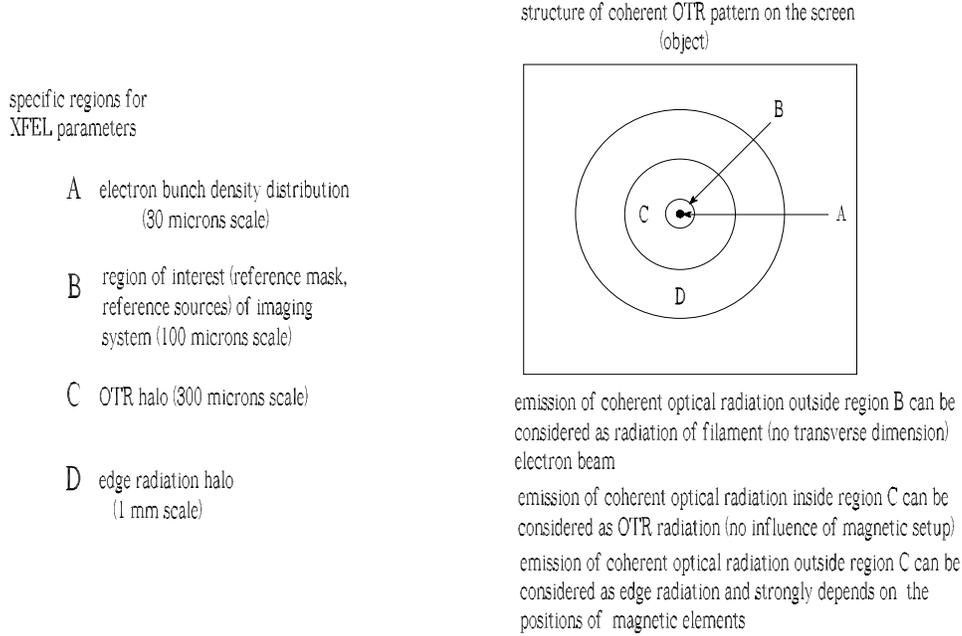}
\caption{Scheme of the coherent OTR radiation pattern as observed
in the object plane for the low electron beam energy case of $2$
GeV.\label{halo} }
\end{center}
\end{figure}

Information about the electron bunch will be shown to be included
in a small region of size $\sigma_r\sim 30 \mu$m corresponding to
the region of nonzero electron density on the OTR screen, which we
will call the "bunch" region, region A in Fig. \ref{halo}. The
region of interest of the imaging system is characterized by $r
\lesssim 100~\mu$m, region B in Fig. \ref{halo}.

The field distribution for $100 ~\mu \mathrm{m} \lesssim r
\lesssim 300~\mu$m does not depend on the transverse size of the
electron beam. In other words, a filament beam approximation
applies. We will refer to this region as the "halo" region, region
C in Fig. \ref{halo}.  Note that in the halo region, information
about the peak-current distribution is encoded in the dependence
of field amplitude versus time.

The regions A, B and C are not influenced by the position of
magnetic structures. These become relevant at larger distances $r
\gtrsim 300~\mu \mathrm{m} \sim \gamma \lambdabar$, region D in
Fig. \ref{halo}. Note that the field in region D remains
independent of emittance effects, i.e. the filament beam
approximation is still valid. Therefore, information about the
peak-current distribution can be extracted from region D exactly
as from region C.

It should be remarked that numbers given here refer to the
relatively low electron beam energy case of $2$ GeV, and should be
multiplied by a corresponding factor when these considerations are
extended to the high energy case of $17.5$ GeV.

As mentioned before, as a result of the adiabatic approximation,
the field distribution seen on the screen actually depends on the
instantaneous transverse charge density distribution. In the bunch
region, the field distribution evolves due to the dependence of
the electron bunch transverse size on the longitudinal coordinate,
while in the halo region the field distribution is constant,
although its amplitude depends on the peak-current.

In this paper we will be mainly interested in the field within the
bunch region. This is important for imaging purposes. However, we
will occasionally need to characterize the halo region too. For
example, as we will see, there is a possibility of simplifying the
Optical Replica setup for peak-current profile measurement by
taking advantage of an OTR screen as a radiator. In this case,
characterization of the halo region is needed.

Outside of the bunch region an exact characterization of the field
becomes more complicated, because the approximation of the
single-electron field considered up to now (proportional to $K_1$)
fails when $r$ approaches $\gamma \lambdabar$. In fact, the
influence of magnetic structures upstream of the OTR screen starts
to be important. Fortunately, not the halo region, nor the D
region for $r \gtrsim \gamma \lambdabar$ will be involved in the
imaging process. In fact, for the halo region the filament-beam
asymptote holds: in this region only information about the
peak-current is
encoded. 
%

\subsubsection{Similarity techniques}

Up to now we gave a picture of our problem introducing the
parameter space of interest. Based on the adiabatic approximation,
we mentioned that the coherent OTR field can be written as a
convolution between the charge density distribution and the
single-electron field. In particular, we used the Ginzburg-Frank
formula to describe the field produced by a single electron on the
OTR screen (proportional to $K_1$).

With the help of similarity techniques it is possible to present
qualitative arguments to explain why the Ginzburg-Frank formula
can be used in our case study, and why we can neglect the presence
of bending magnet and straight section in our setup. Let us
consider a setup composed by a bend, a straight section, and a
target (the OTR screen). Radiation characteristics were shown to
depend on two dimensionless parameters \cite{OUREDGE}:

\begin{eqnarray}
\delta \equiv \frac{\sqrt[3]{R^2 \lambdabar}}{L}~,~~~~\phi \equiv
\frac{L}{\gamma^2 \lambdabar} ~, \label{param}
\end{eqnarray}
where $R$ is the radius of the bend, the reduced wavelength of the
radiation $\lambdabar$ coincides, as before, with the reduced
modulation wavelength $\lambdabar_m$, and $L$ is the length of the
straight section.

By definition, $1/\delta$ is a measure of the straight section
length $L$ in units of the synchrotron radiation formation length
$\sqrt[3]{R^2 \lambdabar}$. The parameter $\phi$, instead, is a
measure of the straight section length $L$ in units of the
characteristic length $\gamma^2 \lambdabar$. In the case of a
straight section of length $L$, the apparent distance travelled by
an electron as seen by an observer is equal to $L/(2 \gamma^2)$.
Since it does not make sense to distinguish between points within
the apparent electron trajectory such that $L/(2\gamma^2) \lesssim
\lambdabar$, one obtains a critical length of interest $\sim
\gamma^2 \lambdabar$.

It is known (see e.g. \cite{OUREDGE}) that the Ginzburg-Frank
theory used above is a limiting case of the more general Edge
Radiation (ER) theory\footnote{Which, in its turn, is a limiting
case of the more general Synchrotron Radiation theory.}. While the
theory of Edge Radiation accounts for the presence (but not for
the detailed structure) of magnetic elements, the Ginzburg-Frank
theory does not. In the theory by Ginzburg and Frank, an electron
coming from an infinitely long straight line ($L \longrightarrow
\infty$) and crossing an interface between vacuum and an ideal
conductor is considered. As a consequence of the crossing,
time-varying currents are induced at the boundary. These currents
are responsible for Transition Radiation. The metallic mirror,
that is treated as the source of Transition Radiation, is usually
modelled with the help of a Physical Optics approach. This is a
well-known high-frequency approximation technique, often used in
the analysis of electromagnetic waves scattered from large
metallic objects. Surface current entering as the source term in
the propagation equations of the scattered field are calculated by
assuming that the magnetic field induced on the surface of the
object can be characterized using Geometrical Optics, i.e.
assuming that the surface is locally replaced, at each point, by
its tangent plane. One may talk of Transition Radiation in the
sense by Ginzburg and Frank \cite{GINZ} if $\delta \cdot \phi \ll
1$, $\delta \ll 1$ and, additionally, $\phi \gg 1$.

In fact, when $\phi \gg 1$ one can neglect the presence of the
upstream bend in the setup, and consider only an electron crossing
an interface between vacuum and an ideal conductor. Condition
$\delta \cdot \phi \ll 1$ means that the critical wavelength of
Synchrotron Radiation from the bend is much shorter than the
(optical) radiation wavelength we are interested in, i.e.
$\lambdabar \gg R/\gamma^3$. Condition $\delta \ll 1$ means that
Synchrotron Radiation from bending magnet radiation is
characterized by a significantly larger opening angle, compared to
that of Edge Radiation. From the viewpoint of electromagnetic
sources (i.e. harmonics of the current and charge density),
$\delta \ll 1$ means that a sharp-edge approximation can be
enforced, i.e. one can neglect the way electromagnetic sources
begin or cease to exist. 

\begin{figure}
\begin{center}
\includegraphics*[width=110mm]{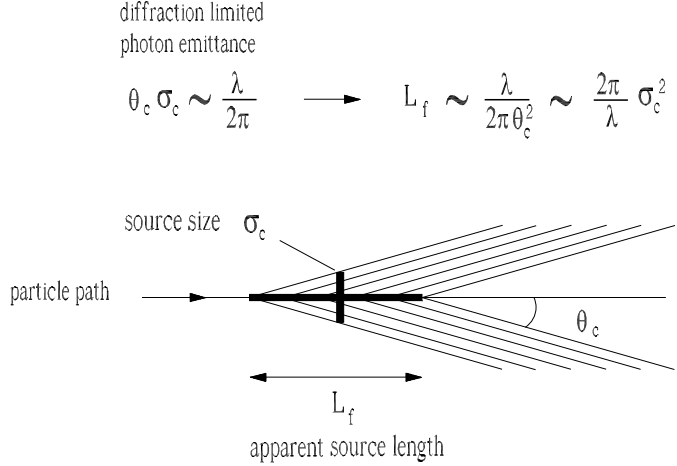}
\caption{Characteristic transverse size $\sigma_c$ and divergence
$\theta_c$ of a diffraction-limited source, and its relation with
the formation length $L_f$, which is interpreted as the apparent
length of the source, adapted from \cite{WIDE}.\label{diffrlim} }
\end{center}
\end{figure}

In our cases of interest, analysis shows that $\phi \gtrsim 1$ for
the low energy case, $\phi \sim 1$ for the high energy case, and
in both cases $\delta \ll 1$, $\delta \cdot \phi \ll 1$.
Although\footnote{In contrast, for example, with the case of
Optical Diffraction Radiation imagers with impact parameters of
order $\gamma\lambdabar$ as discussed in \cite{OUREDGE}.} $\phi
\sim 1$, when $r \ll \gamma \lambdabar$ (i.e. within region B of
Fig. \ref{halo}) we can still use the Ginzburg-Frank theory and
describe the field at the OTR screen with a distribution
proportional to the asymptotic expression for $K_1[r/(\gamma
\lambdabar)]/(\gamma \lambdabar)$. In fact, on the one hand the
parameter $\phi$ compares the length of the straight section $L$
to $\gamma^2 \lambdabar$. On the other hand, however, if we are
interested in a region of the OTR screen such that $r \ll \gamma
\lambdabar$, the effective formation length is decreased to about
$L_f(r) \sim r^2/\lambdabar$. 
We can easily estimate the relation between source size and
formation length remembering that radiation from a single electron
is diffraction limited and using for estimation the fact that for
diffraction limited radiation the photon emittance is given by
$\theta_c \sigma_c \sim \lambdabar$, where $\sigma_c$ is the
characteristic transverse size and $\theta_c$ is the
characteristic divergence of the source (see Fig. \ref{diffrlim}).
Since $\sigma_c \sim L_{f} \theta_c$, one has $\theta_c \sim
\sigma_c/L_f$, which yields $L_f(\sigma_c) \sim
\sigma_c^2/\lambdabar$. For example, for region A the effective
formation length is of order of $1$ cm, while for the full region
of interest of our imaging system (region B) we have $L_f \sim 10$
cm. This is much shorter compared to the distance between the OTR
screen and the upstream bending magnet. Thus, within the deep
asymptotic region $r \ll \gamma \lambdabar$, i.e. within regions A
and B of Fig. \ref{halo}, the Ginzburg-Frank formula can be
applied, while it begins to be less and less accurate in region C,
up to $r \sim \gamma \lambdabar$, where $L_f \sim
\gamma^2\lambdabar$ and $\phi \sim 1$. Note that $L_f \sim
\gamma^2\lambdabar$ is the maximal formation length because,
transversely, the radiation intensity is cut off at $r > \gamma
\lambdabar$.

Further complications arise from the fact that the OTR screen can
be positioned within the undulator line. There the longitudinal
velocity of electrons is effectively decreased, resulting in
effectively weaker electromagnetic sources. The typical transverse
size on the OTR screen associated with the electromagnetic field
decreases from $\gamma \lambdabar$ to $\gamma_z \lambdabar$, where
$\gamma_z^2 = \gamma^2/(1+K^2/2)$ and $K = 3.3$ for the SASE1
undulator of the European XFEL. This means that the transverse
size decreases of about a factor $3$, and $\delta$ is effectively
increased. However, here we will be interested in much smaller
transverse scales of order of the transverse size of the electron
bunch $\sigma_r \sim 30 \mu$m, whereas $\gamma \lambdabar \simeq
4.4$ mm and $\gamma_z \lambdabar \simeq 1.7$ mm. Thus, we expect
that the field distribution changes in the peripheral region
between $\gamma_z \lambdabar$ and $\gamma \lambdabar$ in the halo
region C, but not in the bunch region with $r \lesssim \sigma_r$
nor in the region B. Since, as we will see, these size
modifications enter only logarithmically in the calculation of the
number of photons, they may be neglected in first approximation.

\subsection{\label{sub:una}OTR from a single electron}

Consider a setup composed by a bend, a straight section, and a
metallic mirror. In order to solve the problem of field
characterization, a virtual source method described in
\cite{OUREDGE}, based on the more general work \cite{ARTF}, is
taken advantage of. Following that reference, and defining the
electron charge as $(-e)$, we obtain at the mirror
position\footnote{This convention differs for technical reasons
from that adopted in \cite{OUREDGE} where the mirror position is
at $z=L/2$.}, i.e. at $z=0$ :

\begin{eqnarray}
\vec{\widetilde{E}}\left(\vec{r}\right) &=&  - \frac{2 \omega
e}{c^2  \gamma}   \frac{\vec{r}}{r} K_1\left(\frac{\omega r}{c
\gamma }\right) - \frac{2 \omega e}{c^2 \gamma} \frac{\vec{r}}{r}
\exp\left[-\frac{i \omega L}{2 c \gamma^2}\right] \left.
\right.\cr &&\left.\times \exp\left[\frac{i \omega {r}^2 }{2 c
L}\right]\frac{\omega}{c L} \int_0^{\infty} d{r}' {r}'
K_1\left(\frac{\omega r'}{c \gamma}\right) J_1 \left(\frac{\omega
{r}{r}'}{c L}\right) \exp\left[\frac{i \omega {r}^{'2} }{2 c
L}\right] \right.~,\label{fieldtot2}
\end{eqnarray}
which is valid for any value of $\phi$ under conditions $\delta
\ll 1$ and $\delta \cdot \phi \ll 1$. Note that the field is
radially polarized. One may deal with two asymptotes of the theory
for $\phi \gg 1$ and $\phi \ll 1$.

As discussed before, when $\phi \gg 1$ Eq. (\ref{fieldtot2})
yields back the usual Ginzburg-Frank theory

\begin{eqnarray}
\vec{\widetilde{E}}\left(\vec{{r}}\right) &=&  - \frac{2  \omega
e}{c^2 \gamma} \frac{\vec{r}}{{r}} K_1\left(\frac{\omega{r}}{c
\gamma} \right)~.\label{vir4ea}
\end{eqnarray}
%
%
The following result holds for the asymptote $\phi \ll 1$:

\begin{eqnarray}
\vec{\widetilde{E}}\left(\vec{{r}}\right) = - \frac{2 e}{c}
\frac{\vec{r}}{r^2} \exp\left[\frac{i \omega r^2}{2 c
L}\right]~,\label{vir4ec}
\end{eqnarray}
which comes from the second term in Eq. (\ref{fieldtot2}). The
quadratic phase factor in Eq. (\ref{vir4ec}) describes a spherical
wavefront centered on the optical axis at the upstream edge (the
initial bend). The screen positioned at the downstream end of the
setup detects an electric field given by the spherical wavefront
in Eq. (\ref{vir4ec}) from the upstream magnet at any value
$\sqrt[3]{\lambdabar R^2} \ll L \ll \gamma^2 \lambdabar$ and for
$r \ll L \sqrt[3]{\lambdabar/R}$. Note that when, additionally, $r
\ll \sqrt{\lambdabar L}$, the quadratic phase factor in Eq.
(\ref{vir4ec}) can be dropped. 

In our cases of interest $\phi \sim 1$, which is outside the
region of applicability of the two asymptotes in Eq.
(\ref{vir4ea}) and Eq. (\ref{vir4ec}). However, if we are
interested in imaging an electron bunch with transverse size
$\sigma_r \ll \gamma \lambdabar$ we only deal with the bunch
region, i.e. with the deep asymptotic region A in Fig. \ref{halo},
where $r \lesssim \sigma_r \ll \gamma \lambdabar
\sim\sqrt{\lambdabar L}$. In this region, the argument of the
$J_1$ function under the integral sign in Eq. (\ref{fieldtot2}) is
much smaller than unity, because $r' \lesssim \gamma\lambdabar$
due to the presence of the $K_1$ function under the same integral.
As a result we can substitute $J_1[\omega r r'/(cL)]$ with $\omega
r r'/(2 cL)$ inside the integral. Then, the integration can be
performed analytically. Analysis of the result shows that the
magnitude of the second term in Eq. (\ref{fieldtot2}) is much
smaller (of a factor of order $r/(\gamma \lambdabar)$) than the
first term in $K_1$. We conclude that  Eq. (\ref{vir4ea}) holds
within the bunch region A in Fig. \ref{halo}.

Proceeding to larger values of $r$ outside the bunch region A,
i.e. outside the deep asymptotic region $r \ll \gamma \lambdabar$,
and through regions B and C in Fig. \ref{halo}, Eq. (\ref{vir4ea})
begins to be less and less accurate. However, as we will see,
corrections to Eq. (\ref{vir4ea}) enter only logarithmically in
the calculation of the number of photons in the halo. Therefore,
as concerns photon number estimations, we may still use Eq.
(\ref{vir4ea}) in the halo region C with logarithmic accuracy.

Finally, let us discuss OTR emission in the far-zone. Fresnel
propagation can be performed on Eq. (\ref{vir4ea}), yielding the
following far-zone expression for the single-particle field

\begin{eqnarray}
\vec{\widetilde{E}}\left({z},\vec{\xi}\right) =  \frac{2 e
\gamma^2\vec{{\xi}}}{c {z} (\gamma^2 \xi^2 + 1)} \exp\left[\frac{i
\omega  \xi^2}{2 c }z\right] ~, \label{Efarsum3}
\end{eqnarray}
where $\xi=r/z$, and $z$ is the distance between OTR screen and
observation plane.

The energy radiated per unit frequency interval per unit surface
can be calculated as in \cite{JACK}. From the relation

\begin{eqnarray}
\frac{d^2 W^{(1)}}{d\omega d S} = \frac{c}{4\pi^2}
\left|\vec{\widetilde{E}}\right|^2~.\label{radden0}
\end{eqnarray}
we have, in the far-zone:

\begin{eqnarray}
\frac{d^2 W^{(1)}}{d\omega d S} =  \frac{ e^2}{\pi^2 c {z}^2}
\frac{\gamma^4 \xi^2}{(\gamma^2\xi^2+1)^2}~.\label{radden}
\end{eqnarray}
One can write Eq. (\ref{radden}) in terms of number of photons,
giving

\begin{eqnarray}
\frac{d N_\mathrm{ph}^{(1)}}{d\omega d S} = \frac{ \alpha}{\pi^2
\omega {z}^2} \frac{\gamma^4
\xi^2}{(\gamma^2\xi^2+1)^2}~,\label{radden2}
\end{eqnarray}
with $\alpha \equiv e^2/(\hbar c) = 1/137$ the fine structure
constant. Here and in Eq. (\ref{radden}) the superscript $~^{(1)}$
indicates that we are dealing with single particle emission.

\subsection{\label{OTRmod}Coherent OTR from an optically modulated
electron bunch}

Coherent transition radiation has been introduced a long time ago
into the array of beam diagnostics available for measuring the
microbunching induced in SASE FELs in the infrared, visible and
VUV regimes \cite{ROSE}-\cite{LUM4}. This microbunching
diagnostics provides information about the longitudinal dynamics
of the electron beam in the FEL. There are other applications of
coherent OTR diagnostics. For example, recently, coherent OTR
imaging techniques were used for measuring the microbunching
induced by the space-charge instability at different locations at
LCLS \cite{LOO1}. All these applications are based on measuring
the microbunching induced by FEL or space charge interactions,
effects which introduce distortions in the image formation. We
explained how to avoid these effects in Section \ref{sec:setup}.

Having said this, we can apply our knowledge about the OTR from a
single electron to the case of an optically modulated electron
bunch. In order to obtain the field from the electron bunch, a
microscopic approach can be used where the single-electron field
is averaged over the six-dimensional phase-space distribution of
electrons. In this (Lagrangian) approach particles are labelled
with a given index, and the motion of individual charges is
tracked through space. One follows the evolution of each particle
as a function of energy deviation $\delta \gamma$, angular
direction $\vec{\eta}$, position $\vec{l}$ and arrival time $\tau$
at a given longitudinal reference-position. Knowing the evolution
of each particle, individual contributions to the field are
separately calculated and summed up. Due to the high-quality
electron beams produced at XFELs (highly collimated and nearly
monochromatic) we have the simplest possible situation. Namely,
when performing OTR calculations from an optically modulated
electron bunch we can neglect both angular and energy
distribution, and use a model of a cold electron bunch with given
longitudinal, $f_\tau(\tau)$, and transverse, $f_l(\vec{l})$,
charge density distributions. For a modulated electron bunch we
write $f_\tau(t)$ as

\begin{eqnarray}
f_\tau(t) = f_{\tau 0}(t) [1+a_f \cos(\omega_m t)]
~.\label{ftautau}
\end{eqnarray}
Note that in general we have no factorization of the charge
density distribution into longitudinal $f_\tau(\tau)$ and
transverse $f_l(\vec{l})$ factors. However, here we will be
interested in an estimate of the number of available photons and,
with some accuracy, we  can use a model with separable charge
density distribution function. In fact, this assumption is not
related to fundamental principles, and will only lead to a
different numerical factor in the estimation of the number of
photons.

\begin{figure}
\begin{center}
\includegraphics*[width=140mm]{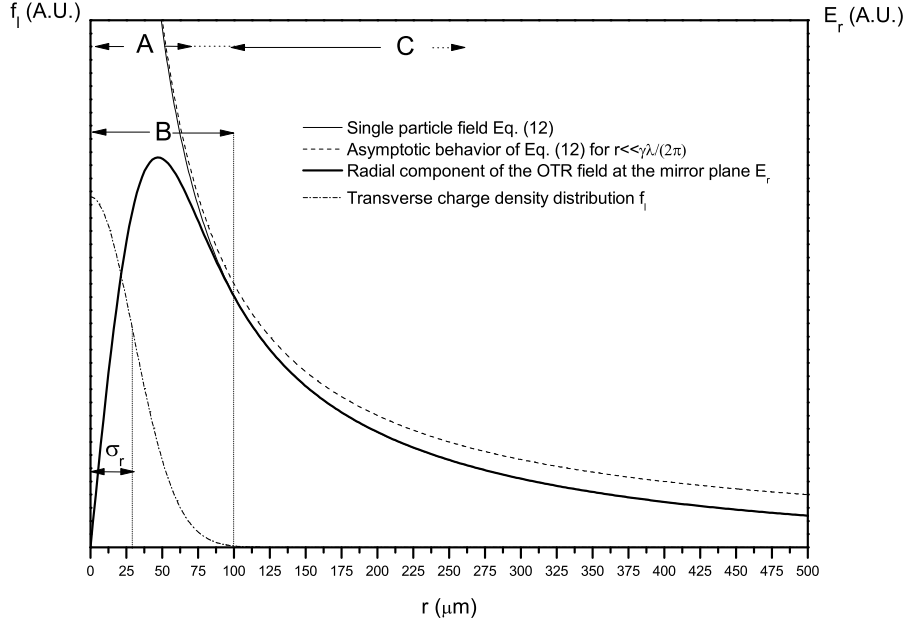}
\caption{\label{field} Radial component of the OTR field in the
mirror plane, Eq. (\ref{totalf2}), transverse (Gaussian) charge
density distribution $f_l$, single particle field, Eq.
(\ref{vir4ea}), scaling as $K_1[r/(\gamma\lambdabar)]/(\gamma
\lambdabar)$, and its asymptote for $r \ll \gamma \lambdabar$
scaling as $1/r$. Here the electron energy is $2$ GeV, the
electron bunch transverse rms size is $\sigma_r = 30~\mu$m, the
modulation wavelength is, as usual $\lambda = 800$ nm.}
\end{center}
\end{figure}
In regions $C$ and $D$ of Fig. \ref{halo} we have $\phi \sim 1$,
and one should use Eq. (\ref{fieldtot2}) to characterize the
field. Only where $r \ll \gamma\lambdabar$ one may use the $K_1$
approximation in Eq. (\ref{vir4ea}), i.e. the Ginzburg-Frank
approximation. However, as mentioned above, we may still use Eq.
(\ref{vir4ea}) in regions $C$ and $D$ with logarithmic accuracy
when dealing with photon number estimations.

The field distribution for the electron bunch at the OTR screen in
the space-frequency domain is essentially a convolution in the
space domain of the temporal Fourier transform of the charge
density distribution and the temporal Fourier transform of the
single-electron field. When the charge density distribution of the
electron bunch can be factorized as product of longitudinal,
${f}_\tau(\tau)$, and transverse, $f_l(\vec{l})$, factors, one
obtains:

%
%
%

\begin{eqnarray}
\vec{\widetilde{E}}(\vec{r}) = N_e \bar{f}_\tau(\omega) \int d
\vec{r'} \vec{\widetilde{E}}^{(1)}(\vec{r'})
f_l(\vec{r}-\vec{r'})~,\label{totalf0}
\end{eqnarray}
where $\bar{f}_\tau(\omega)$ is the Fourier transform of the
temporal charge density distribution ${f}_\tau(\tau)$.

Assuming a Gaussian transverse charge density distribution of the
electron bunch with rms size $\sigma_r$, i.e. $f_l(\vec{l})= (2\pi
\sigma_r^2) \exp[-l^2/(2 \sigma_r^2)]$ and substituting the
single-particle OTR field, Eq. (\ref{vir4ea}) into Eq.
(\ref{totalf0}), we obtain

\begin{eqnarray}
\vec{\widetilde{E}}(\vec{r}) = -\frac{N_e e
\bar{f}_\tau(\omega)}{\pi c \sigma_r^2} \int d \vec{r'}
\frac{\omega}{\gamma c} \frac{\vec{r'}}{r'} K_1\left(\frac{\omega
r'}{\gamma c}\right) \exp\left[-\frac{(\vec{r}-\vec{r'})^2}{2
\sigma_r^2}\right]~.\label{totalf}
\end{eqnarray}
Eq. (\ref{totalf}) can be shown to be equivalent to

\begin{eqnarray}
\vec{\widetilde{E}}(\vec{r}) = -\frac{2 N_e e
\bar{f}_\tau(\omega)}{c \sigma_r^2} \frac{\vec{r}}{r}
\exp\left[-\frac{r^2}{2\sigma_r^2}\right]\int_0^\infty d r'
\frac{\omega r}{\gamma c} K_1\left(\frac{\omega r'}{\gamma
c}\right) I_1\left(\frac{r r'}{\sigma_r^2}\right)
\exp\left[-\frac{r'^2}{2 \sigma_r^2}\right]~,\cr &&\label{totalf2}
\end{eqnarray}
where $I_1$ is the modified Bessel function of the first kind. We
plot the radial component of Eq. (\ref{totalf2}) in Fig.
\ref{field}, together with the charge transverse density
distribution $f_l(\vec{r})$, the single particle field in Eq.
(\ref{vir4ea}) and its asymptote proportional to $1/r$ for $r \ll
\gamma \lambdabar$. For this example, the chosen energy is $2$
GeV, the electron bunch transverse rms size is $\sigma_r =
30~\mu$m and the modulation wavelength is, as usual, $\lambda =
800$ nm. As one can see from Fig. \ref{field}, outside the region
B (i.e. within regions C and D in Fig. \ref{halo}) one has no more
emittance effects, and emission can be considered as radiation
from a filament electron bunch (with no transverse dimensions).
Emittance effects are present within the B region and the A
(bunch) region. They encode information about the transverse
charge density distribution. Note that the outer part of region B
is already fitting well the filament-beam asymptote in the region
$r \ll \gamma \lambdabar$, where it is well approximated by a
$1/r$ behavior. Finally, for values of $r\gtrsim \gamma
\lambdabar$ we expect modifications of the $K_1$ behavior due to
edge radiation contributions.

Similarly to Eq. (\ref{radden}), we can calculate the spectral
distribution of coherent photons on the OTR screen by taking
squared modulus of Eq. (\ref{totalf0}). The field in the
Fraunhofer zone is linked to that at the OTR screen position by a
Fourier transform. Then, it can be seen that the spectral
distribution of photons in the far-zone is equal to the product of
the single-electron result in the far-zone, Eq. (\ref{radden2}),
$N_e^2$ and the squared-modulus of the structure factor

\begin{eqnarray}
\varrho\left(\omega ,\vec{\xi}\right) = \int d\tau f_\tau(\tau)
\exp\left[i {\omega \tau}\right] \int d\vec{l} f_l(\vec{l})
\exp\left[\frac{i \omega }{c} \vec{\xi}\cdot
\vec{l}\right]~,\label{bigf}
\end{eqnarray}
which can be expressed as

\begin{eqnarray}
\varrho\left(\omega ,\vec{\xi}\right) =
\bar{f}_\tau(\omega)\bar{f}_l(\vec{\xi})~,\label{bigf2}
\end{eqnarray}
where $\bar{f}_\tau(\omega)$ and $\bar{f}_l(\vec{\xi})$ are the
Fourier transforms of the temporal and transverse charge density
distributions (and can be identified with the two factors in Eq.
(\ref{bigf})). Thus, one has

\begin{eqnarray}
\frac{d N_\mathrm{ph}}{d\omega d S} = N_e^2\frac{d
N_\mathrm{ph}^{(1)}}{d\omega d S}
\left|\varrho(\omega,-\vec{\xi})\right|^2~.\label{radden3}
\end{eqnarray}
In case of a Gaussian transverse charge density distribution we
have:

\begin{eqnarray} \bar{f}_l(\xi)& = & \exp\left[-\frac{\omega^2 \xi^2 \sigma_r^2}{2 c^2}\right]~, \label{fbarlll}
\end{eqnarray}
A Gaussian temporal electron bunch profile with an rms bunch
duration $\sigma_T$ and an amplitude of density modulation $a_f$
is also assumed. We define the function $f_{\tau 0}(t) = 1/(2\pi
\sigma_T^2)^{-1/2} \cdot \exp[-t^2/(2 \sigma_T^2)]$ and, with the
help of Eq. (\ref{ftautau}) we obtain

\begin{eqnarray}
\bar{f}_\tau(\omega)& = &\frac{
a_f}{2}\left\{\exp\left[-\frac{\sigma_T^2}{2}
\left(\omega-\omega_m\right)^2\right]
+\exp\left[-\frac{\sigma_T^2}{2}
\left(\omega+\omega_m\right)^2\right]\right\} +
\exp\left[-\frac{\sigma_T^2 \omega^2}{2}\right]~, \cr&&
\label{fbarrrr}
\end{eqnarray}
where $\omega_m = 2\pi/\lambda_m$.

\subsubsection{Adiabatic approximation}

We now apply an adiabatic approximation, which relies on the fact
that the bunch duration $\sigma_T$ is much longer than the period
of the modulation. In other words, $\sigma_T \omega_m \gg 1$. This
allows us to simplify Eq. (\ref{fbarrrr}) as

\begin{eqnarray}
\bar{f}_\tau(\omega)& = &\frac{a_f}{2}
\exp\left[-\frac{\sigma_T^2}{2}
\left(\omega-\omega_m\right)^2\right] ~, \label{fbar2}
\end{eqnarray}
which is valid for frequencies $\omega$ around the modulation
frequency $\omega_m$. From Eq. (\ref{fbar2}) follows that
radiation is exponentially suppressed for frequencies outside the
bandwidth $1/\sigma_T$ centered at the modulation frequency
$\omega_m$.

\subsubsection{Angular distribution of coherent OTR photons emitted by an optically modulated  Gaussian electron
bunch}

Since we are interested in coherent emission around the modulation
wavelength, we can consider the wavelength in Eq. (\ref{radden2})
and Eq. (\ref{fbarlll}) fixed. Then, according to Eq.
(\ref{radden3}), the calculation of the angular distribution of
photons emitted around $\omega_m$ amounts to a multiplication of
Eq. (\ref{radden2}) by $|\bar{f}_l|^2$ and by

\begin{eqnarray}
\int_0^\infty d\omega
\left|\bar{f}_\tau(\omega)\right|^2={\sqrt{\pi}  a_f^2}/({4
\sigma_T})~, \label{integrfor}
\end{eqnarray}
leading to the following expression for the number of coherent OTR
photons emitted by an electron bunch per unit solid angle:

\begin{eqnarray}
\frac{d N_\mathrm{ph}}{d \Omega} =  \frac{N_e^2\sqrt{\pi}  a_f^2
\alpha}{4 \pi^2 \omega_m \sigma_T} \frac{\gamma^4
\xi^2}{(\gamma^2\xi^2+1)^2}\exp\left[-\frac{\omega_m^2 \xi^2
\sigma_r^2}{c^2}\right]~,\label{radden4}
\end{eqnarray}
where $d\Omega = \xi d\xi d\phi$, $\phi$ being the azimuthal
angle.

\subsubsection{Total number of coherent OTR photons emitted by an optically modulated  Gaussian electron
bunch}

To estimate the total number of coherent OTR photons it is
sufficient to integrate Eq. (\ref{radden4}) over $d\Omega$, giving

\begin{eqnarray}
{N_\mathrm{ph}}= \frac{N_e^2  a_f^2 \alpha}{4 \sqrt{\pi} \omega_m
\sigma_T}  \left[-1+\exp(N)(1+ N)\Gamma(0,
N)\right]~,\label{radden5}
\end{eqnarray}
where, as before,  $\Gamma(\alpha,\beta)$ is the incomplete gamma
function, and we defined $N=\sigma_r^2/(\lambdabar^2 \gamma^2)$.
The parameter $N$ is analogous to a Fresnel number in diffraction
theory, and is the only (dimensionless) transverse parameter
related to radiation emission in our Gaussian model.

In the high energy case $N \sim 10^{-4}$, whereas in the low
energy case $N \sim 10^{-3}$. Considering $a_f \simeq 0.1$, $N_e
\simeq 6 \cdot 10^9$, i.e. about $1 $ nC of charge, $\sigma_T
\simeq 80$ fs we obtain, for $\lambda = 800$ nm, a total number of
about $10^{13}$ coherent photons per OTR pulse into a $10$-nm
bandwith\footnote{Note, for comparison, that the estimated number
of incoherent photons per OTR pulse is about $10^8$ into a
$100$-nm bandwidth.}.

In the limit for small values of $N$ one obtains the following
asymptotic expression for Eq. (\ref{radden5}):

\begin{eqnarray}
{N_\mathrm{ph}} \simeq \frac{N_e^2  a_f^2 \alpha}{4 \sqrt{\pi}
\omega_m \sigma_T} [-1-\Gamma_E - \ln(N)]  ~,\label{radden6}
\end{eqnarray}
$\Gamma_e \simeq 0.58$ being the Euler constant. The asymptote in
Eq. (\ref{radden6}) can be exploited when dealing with XFEL
setups, because of the extreme high-quality of the electron bunch
(small emittance). As stated before, the electron beam size and
the transverse dimension $\gamma \lambdabar$ enter only
logarithmically in the expression for the number of photons.
Therefore, the number of photons available is almost insensitive
to the value of $N$.


Finally, it is interesting to estimate the number of coherent OTR
photons in the bunch region A, Fig. \ref{halo}. This can be done
substituting Eq. (\ref{totalf}) into Eq. (\ref{radden0}) and
integrating over $\omega$ with the help of Eq. (\ref{integrfor}).
This yields the energy radiated per unit surface on the OTR
screen. Then, integrating over $dS$ and dividing by $\hbar \omega$
yields about $10^{12}$ photons in the bunch region.

\subsubsection{Angular distribution of coherent OTR photons in the case of arbitrary peak-current profile}

Within the adiabatic approximation,  i.e. for $ \sigma_T \omega_m
\gg 1$, it makes sense to introduce an expression for the
instantaneous power as a function of the peak-current $I_0(t)$ and
modulation $a(t)$.

To this purpose we consider, first, a stepped-profile model for
the bunch $f_{\tau 0}(t) = 1/T$ for $-T/2<t<T/2$ and zero
elsewhere. We also suppose $a(t) = a_f$ constant to start with. It
follows that

\begin{eqnarray}
\bar{f}_\tau(\omega ) &=& {a_f } \left\{\frac{\sin\left[T (\omega+
\omega_m)/2\right]}{T(\omega+\omega_m)}+\frac{\sin\left[T (\omega-
\omega_m)/2\right]}{T(\omega-\omega_m)}\right\}~.\cr &&
\label{fbarrrr2}
\end{eqnarray}
Then, in the limit for $T \gg \lambdabar_m/c$ one has
$\int_0^\infty d\omega \left|\bar{f}_\tau(\omega)\right|^2 \simeq
a_f^2 \pi T/2$, and

\begin{eqnarray}
\frac{dP}{dS} &=& \frac{1}{T} \frac{dW}{d S} = \frac{c a_f^2 I_0^2
\pi }{8 \pi^2 e^2} \exp\left[-\frac{\omega_m^2 \xi^2
\sigma_r^2}{c^2}\right] \left|\vec{\widetilde{E}}^{(1)}\right|^2
~, \label{Pinst0}
\end{eqnarray}
where $dP/d S$ power per unit surface. If now the peak-current
$I_0=I_0(t)$ and the modulation level $a=a(t)$ are slowly varying
functions of time on the scale $\lambdabar_m/c$, we can interpret
Eq. (\ref{Pinst0}) as the instantaneous power density at time $t$,
and with the help of Eq. (\ref{Efarsum3}) we obtain the following
expression for photon flux radiated into the unit solid angle:

\begin{eqnarray}
\frac{d \dot{N}_{ph}}{d\Omega} &=&   \frac{\alpha  a(t)^2 I_0(t)^2
 }{2 \pi e^2 \omega_m} \frac{ \gamma^4\xi^2}{
(\gamma^2 \xi^2 + 1)^2}  \exp\left[-\frac{\omega_m^2 \xi^2
\sigma_r^2}{c^2}\right] ~. \label{Pinst01}
\end{eqnarray}
%
Eq. (\ref{Pinst01}) can be used in order to study the general case
of an electron bunch with arbitrary gradient profile and amplitude
of modulation.


\subsubsection{Effect of angular filtering} 

To conclude this Section, it is interesting to consider the effect
of angular filtering \cite{OUREDGE}. For mathematical simplicity,
let us introduce the angular filter profile through an amplitude
transmittance $T(\xi)$. We choose a simple Gaussian model for
$T(\xi)$, and we write:

\begin{eqnarray}
T(\xi) = \exp\left[-\frac{\xi^2}{2 \theta_a^2}\right]~.
\label{anfil}
\end{eqnarray}
%
%

Then, the effect of angular filtering is accounted for by
multiplying Eq. (\ref{radden4}) by $T^2(\xi)$, i.e. the squared of
the amplitude transmittance, and by integrating over angles
$d\vec{\xi}$. One obtains

\begin{eqnarray}
{N_\mathrm{ph}}= \frac{N_e^2  a_f^2 \alpha}{4 \sqrt{\pi} \omega_m
\sigma_T}  \left[-1+\exp(N+A)(1+ N+A)\Gamma(0,
N+A)\right]~,\label{radden5b}
\end{eqnarray}
where $A \equiv 1 /(\gamma \theta_a)^2$. Eq. (\ref{radden5b}) is
analogous to Eq. (\ref{radden5}), where $A$ now plays the same
role of the $N$ parameter and accounts for angular filtering
effects. 


\section{\label{sec:imager}OTR imager}

Having characterized the radiation from an OTR screen for both a
single electron and an optically modulated electron bunch, we will
now consider the physics of the image formation process. In
particular we will discuss the OTR imager setup up to the image
plane where the detector is placed.

\begin{figure}
\begin{center}
\includegraphics*[width=130mm]{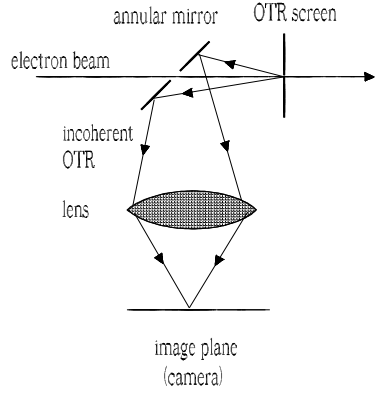}
\caption{\label{incotr} Optical setup for incoherent OTR imager.
For high electron energies, the XFEL case, an important fraction
of the coherent OTR radiation will be lost through the center hole
of the mirror.}
\end{center}
\end{figure}

A simple setup for OTR imaging is schematically shown in Fig.
\ref{incotr}. Radiation is reflected by an annular mirror (which
allows the passage of the electron bunch) and an image is formed
in the image plane with the help of a converging lens. In this
case, the object plane is the OTR screen.

\begin{figure}
\begin{center}
\includegraphics*[width=110mm]{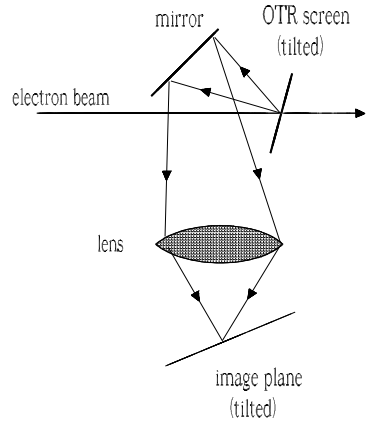}
\caption{\label{noanular} Optical setup for the
near-normal-incident OTR screen. The detector image plane is
slightly tilted to compensate for the tilt of the object plane. }
\end{center}
\end{figure}

The annular-mirror design depicted in Fig. \ref{incotr} is usually
applied in the measurement of transition radiation around the THz
frequency range for longitudinal profile characterization.
However, for optical frequencies ($\lambdabar \sim 0.1 ~\mu$m) an
important fraction of OTR will be lost through the center hole. A
possible solution to this problem is to avoid the use of an
annular mirror as shown in Fig. \ref{noanular}, where a
near-normal-incidence scheme is shown. Note that in order to
compensate for the small tilt-angle of the object plane one needs
to tilt the image plane as well.  The arrangement in Fig.
\ref{noanular} is currently used for incoherent OTR imaging of
electron beams at LCLS (see e.g. \cite{BENG}).

One should therefore consider the mirror design in Fig.
\ref{noanular}, rather than that in Fig. \ref{incotr}. However,
for the sake of simplicity of drawing, we will still refer to the
non-tilted OTR screen design in Fig. \ref{incotr}. This does not
make any difference concerning the description of our methods.

\begin{figure}
\begin{center}
\includegraphics*[width=130mm]{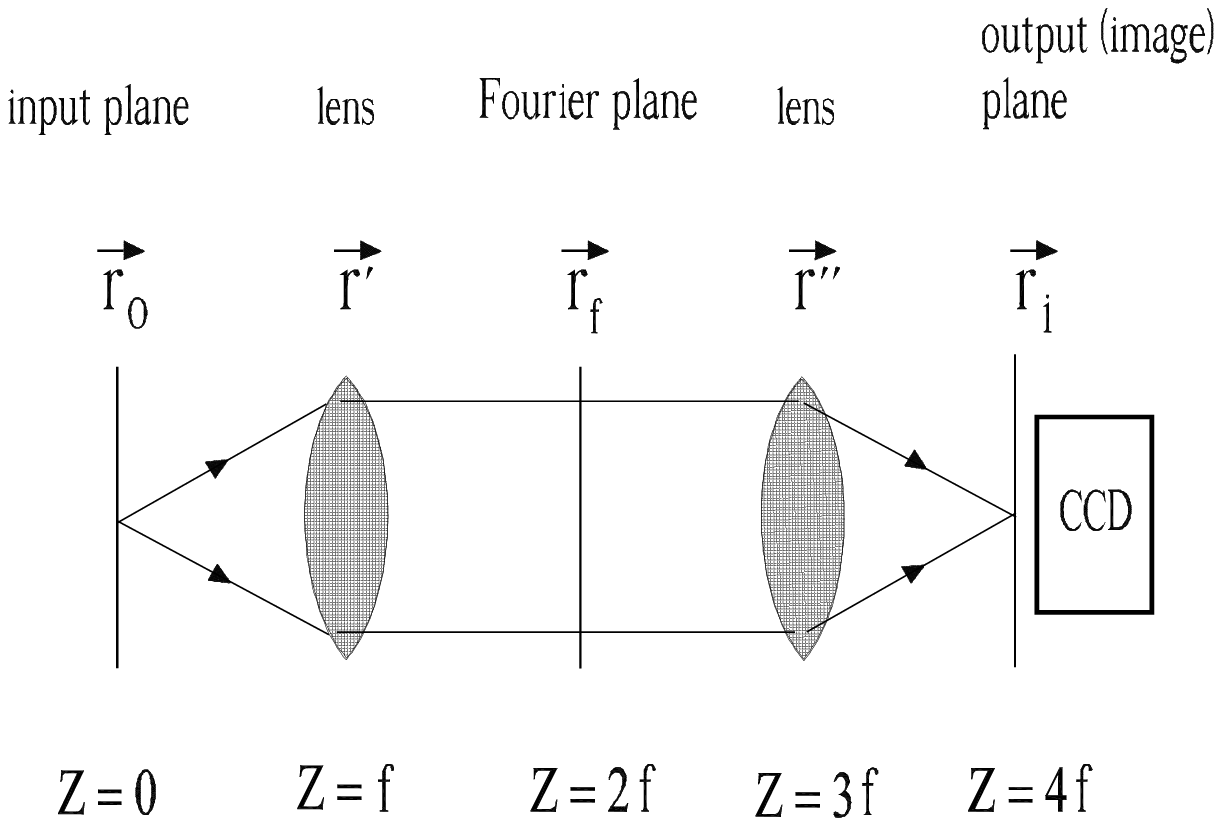}
\caption{\label{4farr}  A two-lens image-processing system. In
order to perform Fourier processing of the input signal a
transparency and a polarization transformer can be inserted in the
Fourier plane to suitably modify the Fourier transform of the
object.}
\end{center}
\end{figure}

Another optical system traditionally used for imaging purposes is
the well-known two-lens image-formation scheme in Fig.
\ref{4farr}. This scheme allows for magnification by changing the
focal distance of the second lens but for simplicity, in the
following we will assume that the two focal distances are the same
(i.e. we consider $1:1$ imaging). This two-lens setup is usually
employed for image-processing purposes, as it can be better used
for image-modification compared to the single-lens system.
Therefore, in the following we will systematically consider the
two-lens scheme in Fig. \ref{4farr} instead of the single-lens
scheme described in Fig. \ref{noanular}.


\subsection{\label{sub:lasttt}Theoretical basis for the analysis of coherent imaging systems}

Given the two-lens setup discussed above, we first consider the
relatively easy problem of characterization of single-particle
radiation in the image plane.

We introduce the lens transmission function

\begin{eqnarray}
{T}\left(\vec{ {r}'}\right) = P\left(\vec{ {r}'}\right)
\exp\left[-  \frac{i \omega{r}'^2}{2 c {f}}\right]~,
\label{transm}
\end{eqnarray}
where $f$ is the focal distance of the lens and $P$ is the pupil
function of the lens, which may account for finite extent of the
lens, apodization, aberrations and is, in general, a complex
function of $\vec{r'}$.

We assume that the two lenses in Fig. \ref{4farr} are identical.
Let us explicitly derive the field in the output (image) plane of
the setup in Fig. \ref{4farr}. In the following, it should be
clear that we discuss about coherent imaging of an extended
object. In fact, as we have seen in the previous Section
\ref{sec:OTRsetup}, the distribution of the OTR field from a
single electron is not point-like, but rather laser-like and in
the XFEL case its extension is macroscopic, in the millimeter
scale. Standard description of image formation by a lens is based
on the assumption that optical systems are space-invariant, or
isoplanatic \cite{GOOD}. Once the space-invariant condition has
been assumed a linear-system treatment can be applied, which
consistently considers a lens as linear filter. We recall that a
linear filter is characterized by convolution equation of the form
$f = h \ast g$, where $g$ is the input function at the object
plane, $h$ the impulse response and $f$ the output function in the
image plane.

Imaging systems that use coherent light are linear in field
amplitude, but are space invariant only under certain conditions
\cite{DUMO,TICH,BRAI}\footnote{As stressed before, here we discuss
about a specific subject, namely coherent imaging of extended
objects. As remarked in \cite{BRAI}, "in general, even
aberration-free thin lenses do not meet isoplanatic condition.
Some spatial phase distortion is unavoidably introduced, which
severely modifies the intensity distribution of the image". Up to
now, this problem attracted little attention of the Optics
community, and the practical importance of this subject was
recognized only very recently.}. This means that the concept of
impulse response or transfer function for coherent imaging has
only a limited use.  In particular, as we will see, a convolution
equation of the form $f = h \ast g$ can be applied to lenses only
after certain quadratic phase factors can be neglected within the
field-propagation equation. Then, the (scaled) pupil function
plays the role of the amplitude transfer function for the system.

With reference to Fig. \ref{4farr} we call with $z$ the coordinate
on the optical axis, $z=0$ being the position of the object plane.
Thus, the position of the Fourier plane is at $z = 2 f$, and that
of the image plane is at $z = 4f$.

Within a paraxial treatment in the space-frequency domain, each
polarization component of the field propagates in free-space from
position $z_s$ to position $z$ according to

\begin{equation}
{ \widetilde{E}}( {z},\vec {r}) = {i \omega\over{2\pi c |z-z_s|}}
\int d \vec{ {r}'}~ {\widetilde{E}}(z_s,\vec{r'}) \exp\left[
i\omega{\mid \vec{r}-\vec{r'}\mid^2\over{2c |z-z_s|}}\right] ~.
\label{fieldpropback}
\end{equation}
For each polarization component, we also introduce the spatial
Fourier transform of a field distribution

\begin{eqnarray}
{F}(z,\vec{u})=\int d \vec{r'} {\widetilde{E}}(z,\vec{r'}) \exp[i
\vec{r'}\cdot \vec{u}]~.\label{fieldpropexpf}
\end{eqnarray}
The propagation equation for the spatial Fourier harmonics of the
field in free-space is given by

\begin{eqnarray}
{F}\left(z, \vec{u}\right) = {F}\left(0, \vec{u}\right)
\exp\left[-\frac{i c z u^2}{2 \omega}\right]~. \label{vecF}
\end{eqnarray}
Note that $F\left(z, \vec{u}\right)$ is the 2D spatial Fourier
transform, with respect to transverse coordinates, of the "field"
${ \widetilde{E}} ( z,\vec{ {r}'})$, which is, in its turn, the
temporal Fourier transform of the electric field in the space-time
domain. As a result, $F\left(z, \vec{u}\right)$ is actually a 3D
Fourier transform of the electric field in the space-time domain,
with respect to time and transverse coordinates. It is, therefore,
a function of the longitudinal coordinate $z$.

The relation between the field distribution on the pupil, $z = f$,
and the field distribution in the focal plane, $z = 2f$, is given
by

\begin{eqnarray}
{ \widetilde{E}}( 2 f,\vec{ {r}}_f) &=& \frac{i \omega}{2 \pi c f}
\exp{\left[ \frac{i \omega | \vec{r}_f |^2}{2 c f} \right]}\int d
\vec{ {r}'}~ { \widetilde{E}} ( f,\vec{ {r}'}) P(\vec{r'})
\exp{\left[- \frac{i \omega (\vec{r}_f\cdot\vec{r'})}{c
f}\right]}~.\label{fieldpropexpf}
\end{eqnarray}
Here ${ \widetilde{E}} ( f,\vec{ {r}'})$ indicates the field
amplitude immediately in front of the lens. Also, $\vec{r}_f $
indicates the transverse position in the focal plane, i.e. at $z=2
f$.  Use of the convolution theorem and Eq. (\ref{vecF}) allow one
to write Eq. (\ref{fieldpropexpf}) as

\begin{eqnarray}
{\widetilde{E}}(2 f,\vec{r}_f) &=& \frac{i\omega}{2\pi c
f}\exp\left[\frac{i \omega |\vec{r}_f|^2}{2 c f} \right]\int d
\vec{u}~ \mathcal{P}\left(\frac{\omega \vec{r}_f}{c
f}-\vec{u}\right)\cdot \exp{\left[- \frac{i c f  {u}^2}{2 \omega}
\right]}F\left(0, - \vec{u}\right)~,\cr && \label{EFP}
\end{eqnarray}
where we defined

\begin{eqnarray}
\mathcal{P}(\vec{u}) = \int d\vec{r'} P(\vec{r'}) \exp\left[- i
\vec{r'}\cdot \vec{u}~\right] ~.\label{apsfs}
\end{eqnarray}
Simple analysis, consisting of a change of variable ($\vec{u}
\longrightarrow -\vec{u}+ {\omega \vec{r}_f}/({c f})$) followed by
expansion of the quadratic phase factor in Eq. (\ref{EFP}) shows
that the field in the focal plane, Eq. (\ref{EFP}), can also be
written as

\begin{eqnarray}
{\widetilde{E}}\left(2 f,\vec{r}_f\right) = \frac{i\omega}{2\pi c
f} \int d \vec{u} ~{F}\left(0, \vec{u}- \frac{\omega \vec{r}_f}{c
f}\right) \exp\left[-\frac{i c u^2 f}{2 \omega}\right]
\mathcal{P}(\vec{u}) \exp[i\vec{u}\cdot \vec{r}_f]~.\cr
&&\label{Fplane}
\end{eqnarray}
Assume now that the pupil simply consists of a finite aperture. In
other words,

\begin{eqnarray}
P(\vec{r}~) = 1  ~~\mathrm{for}~~r < a, ~P(\vec{r}~) = 0 ~~
\mathrm{otherwise}~. \label{PdefP}
\end{eqnarray}
We indicate with $\sigma$ the maximal transverse size of
object\footnote{Not to be confused with the transverse size of the
bunch $\sigma_r$. Here and in the following, $\sigma$ is the
characteristic size of any object in the object plane.}, which is
a characteristic transverse scale of the problem.

We also assume:

\begin{eqnarray}
\sigma \ll a~,~~~~ a^2/(\lambdabar f) \gg 1~. \label{bothc}
\end{eqnarray}
Consider Eq. (\ref{Fplane}). On the one hand, the smallest
structures in the Fourier transform of the field in the object
plane under the integral sign are of the order of $u \sim
1/\sigma$, meaning that we can neglect variations of $F$ for
frequencies $u \ll 1/\sigma$. On the other hand, $\mathcal{P}$
enters the integral sign as well, and has a width of $1/a$. It
follows that, for $\sigma \ll a$, one can neglect variations of
$F$ in $u$, and can simply take $F$ out of the integral sign.
Moreover, $u \lesssim 1/a$ due to the width of the pupil. As a
result, whenever $a^2/(\lambdabar f) \gg 1$ the quadratic phase
factor within the integral sign can be neglected and one obtains

\begin{eqnarray}
{\widetilde{E}}^{}\left(2 f,\vec{r}_f\right) =\frac{i\omega}{2\pi
c f} {F}\left(0, -\frac{\omega \vec{r}_f}{c f}\right) P(\vec{r}_f)
~.\label{Fplane2}
\end{eqnarray}
\begin{figure}
\begin{center}
\includegraphics*[width=110mm]{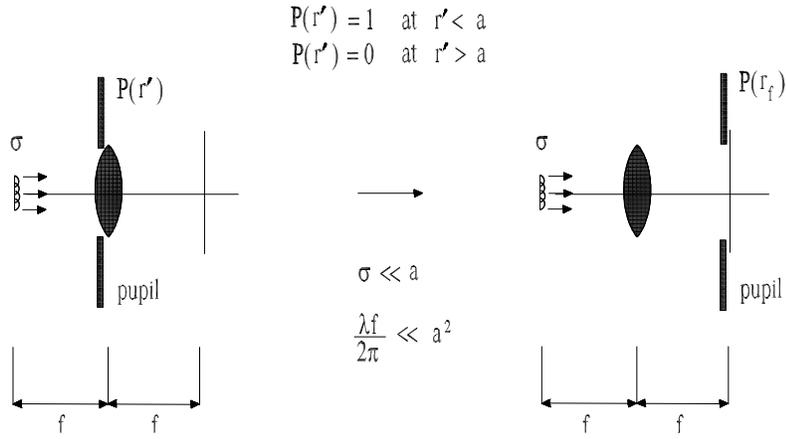}
\caption{\label{sumup}  Left: a coherent object with
characteristic size $\sigma$ is spatially Fourier transformed by a
lens with focal length $f$ and an aperture of radius $a$.  The
pupil aperture is placed directly against the lens. When the pupil
aperture is much larger than a single Fresnel zone, and the object
size is a small fraction of the pupil size, the
diffraction-limited optical system is space-invariant, i.e.
isoplanatic. The concept of amplitude transfer function can be
directly applied to the system, and the relation between pupil and
amplitude transfer function is straightforward: the scaled pupil
function plays the role of the amplitude transfer function. Right:
the pupil aperture is placed behind the lens in the focal plane.
In this case, the pupil plays the role of a transparency, which
sharply limits the range of Fourier components passed by the
system. Under the accepted conditions (\ref{bothc}), left and
right configurations are equivalent.}
\end{center}
\end{figure}
As long as conditions (\ref{bothc}) hold, it does not matter if we
consider the aperture $P$ placed at the lens position, or at any
other position between the lens and the Fourier plane. This can be
understood by inspecting Eq. (\ref{Fplane2}). In fact, Eq.
(\ref{Fplane2}) is the same as for the case when a finite aperture
is placed in the focal plane of a lens with non-limiting aperture.
The situation is summed up in Fig. \ref{sumup}.

The field in the image plane $z = 4f$ is obtained by taking Eq.
(\ref{Fplane2}) as a new object, and propagating the field through
the second lens in Fig. \ref{4farr}. One has

\begin{eqnarray}
{\widetilde{E}}\left(4 f,\vec{r}_i\right) &=& -
 \int d \vec{v} \int d \vec{u} ~{\widetilde{E}}\left(0,
\frac{c f \vec{u}}{\omega}+\frac{c f \vec{v}}{\omega}-
\vec{r}_i\right) \mathcal{P}(\vec{u}) \cr &&
\times\exp\left[-\frac{i c v^2 f}{2 \omega}\right]
\mathcal{P}(\vec{v}) \exp[i\vec{v}\cdot
\vec{r}_i]~.\label{Iplane0}
\end{eqnarray}
Here $\vec{r}_i$ indicates the transverse position in the image
plane, i.e. at $z=4f$. When conditions (\ref{bothc}) hold, one can
show that Eq. (\ref{Iplane0}) can be written as

\begin{eqnarray}
{\widetilde{E}}\left(4 f,\vec{r}_i\right) = -
 \int d \vec{w} ~{\widetilde{E}}\left(0,
\frac{ c f \vec{w}}{\omega}-\vec{r}_i\right)~\cdot \int d \vec{u}~
\mathcal{P}(\vec{u}) \mathcal{P}(\vec{w}-\vec{u})~.\label{Iplane}
\end{eqnarray}
In fact, $u,v \lesssim 1/a$, and the quadratic phase factor in Eq.
(\ref{Iplane0}) can be dropped due to $a^2/(\lambdabar f) \gg 1$.
Moreover, one always has $|\lambdabar f(
\vec{u}+\vec{v})-\vec{r}_i|  \lesssim \sigma$. Therefore $r_i
\lesssim \sigma + 2 \lambdabar f/a$, and also the linear phase
factor $\vec{v} \cdot \vec{r}_i$ can be dropped when conditions
(\ref{bothc}) hold. A further change of integration variable
yields Eq. (\ref{Iplane}).

Now, the autocorrelation integral in $d\vec{u}$ in Eq.
(\ref{Iplane}) is equal to the Fourier transform of
$|P(\vec{u})|^2$ done with respect to the transform variable
$-\vec{w}$. However, for our particular choice of $P$, $P^2 = P$
and the integral in $d\vec{u}$ simply equals
$\mathcal{P}(-\vec{w})$. As a result

\begin{eqnarray}
{\widetilde{E}}^{}\left(4 f,\vec{r}_i\right) =-\int d \vec{u}~
\mathcal{P}\left(\frac{\omega \vec{r}_i}{c f}-\vec{u}\right) \cdot
{\widetilde{E}}^{}\left(0,-\frac{c f \vec{u}}{\omega}\right)
~,\label{Iplane3}
\end{eqnarray}
where we changed again $\vec{w}$ to $\vec{u}$ for notational
reasons.


The previous analysis of the field in the image plane, shows that
the concept of coherent transfer function remains meaningful for
diffraction-limited optical system when the pupil size is much
larger than a single Fresnel zone, and the physical size of the
object is a small fraction of the pupil size (see conditions
(\ref{bothc})).

Note that both conditions (\ref{bothc}) are satisfied in our case
of single-electron imaging, because we assume that the numerical
aperture is $\mathrm{NA} \sim 0.1$, $\lambda \sim 1~\mu$m and $f
\sim 10$ cm. This automatically means that $a$ is in the
centimeter scale, which is much larger than the scale of the
coherent OTR pattern, hence $\sigma \ll a$. Also, one can see that
$a^2/(\lambdabar f) \gg 1$.

\subsection{\label{sub:imsi}Image of a single electron}

Let us now consider the problem of OTR pattern imaging from a
single electron. First, note that Eq. (\ref{Iplane3}) is valid for
each polarization component. Remembering Eq. (\ref{vir4ea}) and
its vectorial character one has

\begin{eqnarray}
\vec{\widetilde{E}}^{(1)}\left(4 f,\vec{r}_i-\vec{r}_0\right) = -
\int d \vec{ u}~ \mathcal{P}\left(\frac{\omega
(\vec{r}_i-\vec{r}_0)}{c f}-\vec{u}\right) \cdot
\left(\frac{2\omega e}{c^2 \gamma}\right) \frac{\vec{u}}{u}
K_1\left(\frac{f u}{\gamma}\right) ~,\label{Iplane4}
\end{eqnarray}
where $\vec{r}_0$ indicates the coordinate in the object plane,
and the superscript $(1)$ indicates the single-particle field. By
virtue of the convolution theorem, Eq. (\ref{Iplane4}) may also be
written as

\begin{eqnarray}
\vec{ \widetilde{E}}^{(1)}( 4 f,\vec{ {r}}_i-\vec{r}_0) &=&
\frac{i \omega
 }{2 \pi c f} 
\int d \vec{ {r}}_f~ \Bigg\{\frac{2 e \gamma^2\vec{{r}}_f
P(\vec{r}_f)}{c  (\gamma^2 {r_f}^2 + {f^2})} \Bigg\}\exp{\left[-
\frac{i \omega
 (\vec{r}_i-\vec{r}_0)\cdot\vec{r}_f}{c f}\right]}~.
\label{fieldpropexpi2}
\end{eqnarray}
%
%
Thus, for our choice of pupil, and when conditions (\ref{bothc})
hold, it does not matter if we consider the pupil aperture $P$
placed at the lens position, or in the Fourier plane.

The energy per unit frequency interval per unit surface in the
image plane is given by

\begin{eqnarray}
I^{(1)}(\vec{r}_i-\vec{r}_0) &\equiv & \frac{d^2 W^{(1)}}{d\omega
d S} = \frac{c}{4\pi^2} \left|\vec{\widetilde{E}}^{(1)}\right|^2 =
\cr &&\frac{c  }{4\pi^2} \left| \int d \vec{ u}~
\mathcal{P}\left(\frac{  \omega (\vec{r}_i-\vec{r}_0)}{c
f}-\vec{u}\right) \cdot \left(\frac{2\omega e}{c^2 \gamma}\right)
\frac{\vec{u}}{u} K_1\left(\frac{f
u}{\gamma}\right)\right|^2~,\label{raddenf}
\end{eqnarray}
or equivalently by

\begin{eqnarray}
I^{(1)}(\vec{r}_i-\vec{r}_0) &\equiv & \frac{d^2 W^{(1)}}{d\omega
d S} = \frac{c}{4\pi^2} \left|\vec{\widetilde{E}}^{(1)}\right|^2 =
\cr && \frac{ \omega^2  }{16 \pi^4 c f^2}\left| \int d \vec{
{r}}_f~ \Bigg\{\frac{2 e \gamma^2\vec{{r}}_f P(\vec{r}_f)}{c
(\gamma^2 {r}_f^2 + {f^2})} \Bigg\}\exp{\left[- \frac{i \omega
 (\vec{r}_i-\vec{r}_0)\cdot\vec{r}_f}{c
f}\right]}\right|^2~.\label{raddenf2}
\end{eqnarray}
Eq. (\ref{raddenf}) (or Eq. (\ref{raddenf2})) is the response to
the single-electron source in the case when the Ginzburg-Frank
formula is valid, when conditions (\ref{bothc}) hold and,
additionally, under the assumption of an ideal lens with a finite
pupil aperture ($P(\vec{r}) = 1$ for $r < a$ and zero otherwise).

In general, the single-particle field in the image plane, Eq.
(\ref{Iplane4}) (or equivalently Eq. (\ref{fieldpropexpi2})), is
non azimuthal symmetric. Note that if $P$ is azimuthal-symmetric
in the cylindrical coordinate system $(r,\phi,z)$, where $z$ is
the optical axis (i.e. $P(\vec{r}) = P(|\vec{r}|)$), Eq.
(\ref{raddenf2}) reduces to

\begin{eqnarray}
I^{(1)}(r_i) =\frac{ \omega^2 e^2}{4\pi^2 c f^2} \left|
\int_0^{\infty} d r_f \frac{ \gamma^2 r_f^2P({r}_f)}{c (\gamma^2
{r}_f^2 + {f^2})} J_1{\left(\frac{ \omega {r_f r_i}}{c
f}\right)}\right|^2~,\label{raddenfsimm}
\end{eqnarray}
where we set $\vec{r}_0=0$ for illustration. For an arbitrary
offset $\vec{r}_0$, a generalization of Eq. (\ref{raddenfsimm})
can be obtained substituting $r_i$ with $\left|\vec{r}_i
-\vec{r}_0\right|$.

Eq. (\ref{raddenf}) (or equivalently Eq. (\ref{raddenf2}))
includes information about the optics, through $P(r_f)$, and about
the way the electron input is converted into electromagnetic
field, through the single-particle field. Because of this, we will
refer to $I^{(1)}(r_i)$, that is the image of OTR produced by a
single electron (i.e. the impulse response for OTR), as the
\textit{particle spread function} of the system\footnote{Sometimes
the particle spread function is called \textit{single particle
function} \cite{CAST}.}. The particle spread function differs from
the standard diffraction pattern from a point source, which is
known in optics as the \textit{point spread function} of the
system and is defined as the squared modulus of the Fourier
transform of the pupil function. Similarly,
$\vec{\widetilde{E}}^{(1)}(\vec{r}_i)$ may be referred to as the
\textit{amplitude particle spread function} of the system.

It should be remarked that the expressions for $I^{(1)}(r_i)$
given above assume that no polarization component is selected.
However, they can be easily modified to deal with such case.

It is also interesting to note that for OTR, the direction of the
electric field depends on the transverse position (because the
field is radially polarized). The fact that a Bessel $J_1$
function enters in Eq. (\ref{raddenfsimm}) (and not a Bessel $J_0$
function as for point-source imaging) is actually due to lack of
azimuthal symmetry of the OTR field. This fact is obviously
responsible for a wider particle spread function. The lack of
azimuthal symmetry can be better appreciated considering Eq.
(\ref{Iplane4}) (or Eq. (\ref{fieldpropexpi2})) rather than Eq.
(\ref{raddenf}) (or Eq. (\ref{raddenf2})). Since we are concerned
with coherent imaging, the former two equations are the ones we
will be actually dealing with.

\subsection{\label{sub:inccoh}Image of an electron bunch. Incoherent and coherent case}

Well defined algorithms exist for calculating the image for a
complicated input signal (i.e. an electron bunch with given
transverse charge density distribution) in the case of incoherent
or coherent radiation. In both cases, resolution of the image is
strictly related to the particle spread function (Eq.
(\ref{raddenf})) and the amplitude particle spread function (Eq.
(\ref{Iplane4})) discussed above.

Let us introduce the electron density distribution of the electron
bunch, $\rho(t,\vec{r})$. Here we are interested, in particular,
in the electron density distribution of a bunch crossing the
bondary between vacuum and OTR screen. Let us also call
$\bar{\rho}(\omega,\vec{r})$ the Fourier transform of
$\rho(t,\vec{r})$ with respect to time.

In the coherent case, the intensity distribution in the image
plane is given by the squared modulus of the convolution of the
amplitude particle spread function, Eq. (\ref{Iplane4}), with the
electron beam transverse profile at frequency $\omega$:

\begin{eqnarray}
I(\omega,\vec{r}_i)&=&\frac{c N_e^2}{4\pi^2}\cr && \times
\left|\int d\vec{r'} \bar{\rho}(\omega,\vec{r'}-\vec{r}_i) \int d
\vec{ u}~ \mathcal{P}\left(\frac{  \omega \vec{r'}}{c
f}-\vec{u}\right) \cdot \left(\frac{2\omega e}{c^2 \gamma}\right)
\frac{\vec{u}}{u} K_1\left(\frac{f u}{\gamma}\right)\right|^2~.\cr
&&\label{totintcoh}
\end{eqnarray}

In the incoherent case, the intensity distribution in the image
plane is given by an ensemble average of independent contributions
of the form of the particle spread function, Eq. (\ref{raddenf})
over the electron density distribution. These contributions can be
ascribed to each electron, and are independent of one another. The
ensemble average procedure amounts to a convolution of the
particle spread function with $\bar{\rho}(0,
\vec{r})=\int_{-\infty}^{\infty} dt ~\rho(t,\vec{r})$ 
Therefore, one obtains:

\begin{eqnarray}
I(\omega,\vec{r}_i)  =\frac{c   N_e}{4\pi^2}  \int d\vec{r'}
\bar{\rho}(0, \vec{r'}-\vec{r}_i)\left| \int d \vec{ u}~
\mathcal{P}\left(\frac{ \omega \vec{r'}}{c f}-\vec{u}\right) \cdot
\left(\frac{2\omega e}{c^2 \gamma}\right) \frac{\vec{u}}{u}
K_1\left(\frac{f u}{\gamma}\right)\right|^2~ .\cr
&&\label{totintinc}
\end{eqnarray}
Note that both Eq. (\ref{totintcoh}) and Eq. (\ref{totintinc})
reduce to Eq. (\ref{raddenf}) in the case of a single electron,
i.e. for $\bar{\rho}(\omega,\vec{r})=
\delta(\vec{r}-\vec{r}_0)\exp[i\omega {t}_0]$, where $\vec{r}_0$
and ${t}_0$ indicate the electron offset and arrival time at a
given longitudinal reference position.

In practical cases of interest we will also introduce a
Fourier-plane mask with amplitude transmittance $T_m (\vec{r}_f)$.
Thus, Eq. (\ref{Fplane2}) is modified to

\begin{eqnarray}
{\widetilde{E}}^{}\left(2 f,\vec{r}_f\right) =\frac{i\omega}{2\pi
c f} {F}\left(0, -\frac{\omega \vec{r}_f}{c f}\right) P(\vec{r}_f)
T_m(\vec{r}_f) ~.\label{Fplane2biss}
\end{eqnarray}
Defining $\mathcal{T}_m(\vec{u}) = \int d\vec{r'} T_m(\vec{r'})
\exp[- i \vec{r'}\cdot \vec{u}]$, one should therefore perform the
substitution

\begin{eqnarray}
\mathcal{P}(\vec{u}) \longrightarrow \int d\vec{u'}
\mathcal{P}(\vec{u}-\vec{u'}) \mathcal{T}_m(\vec{u'})= \int
d\vec{u'} \mathcal{T}_m(\vec{u'}) \frac{2 \pi
a}{|\vec{u}-\vec{u'}|} J_1(a |\vec{u}-\vec{u'}|)~ \label{puppil}
\end{eqnarray}
in Eq. (\ref{totintinc}) and Eq. (\ref{totintcoh}).

In both cases one needs to calculate the integral

\begin{eqnarray}
\vec{\widetilde{E}}^{(1)}_\mathrm{mod}( \vec{r}) = \int d \vec{
u}~ \mathcal{P}\left(\frac{\omega \vec{r}}{c f}-\vec{u}\right)
\cdot \left(-\frac{2\omega e}{c^2 \gamma}\right) \frac{\vec{u}}{u}
K_1\left(\frac{f u}{\gamma}\right)~, \label{Emod}
\end{eqnarray}
which may be interpreted as a modified expression for the field
accounting \textit{ad hoc} for the presence of the
spatial-frequency filter associated with the finite aperture of
the lenses. Eq. (\ref{totintcoh}) and Eq. (\ref{totintinc}) remain
formally identical, but we now substitute the single particle
field with $\vec{\widetilde{E}}^{(1)}_\mathrm{mod}$. Therefore,
Eq. (\ref{totintinc}) can also be written as:

\begin{eqnarray}
I(\omega,\vec{r}_i)  \propto   \int d\vec{r'} \bar{\rho}(0,
\vec{r'}-\vec{r}_i)\left| \int d \vec{ u'}~
\mathcal{T}_m\left(\frac{ \omega \vec{r'}}{c f}-\vec{u'}\right)
\cdot \vec{\widetilde{E}}^{(1)}_\mathrm{mod}\left(\frac{c f
\vec{u'}}{ \omega}\right)\right|^2~ .\label{totintincmod}
\end{eqnarray}
Similarly, Eq. (\ref{totintcoh}) can be presented as

\begin{eqnarray}
I(\omega,\vec{r}_i)  \propto \left|\int
d\vec{r'}\bar{\rho}(\omega,\vec{r'}-\vec{r}_i) \int d \vec{ u'}~
\mathcal{T}_m\left(\frac{  \omega \vec{r'}}{c f}-\vec{u'}\right)
\cdot \vec{\widetilde{E}}^{(1)}_\mathrm{mod}\left(\frac{c f
\vec{u'}}{ \omega}\right)\right|^2~.\label{totintcohmod}
\end{eqnarray}
%

\subsection{OTR particle spread function}

We now want to calculate the particle spread function of the
system, which is essentially the radiation pattern produced by a
single electron in the image plane. This is obtained from Eq.
(\ref{totintincmod}) or Eq. (\ref{totintcohmod}) by substituting
the transverse electron density distribution with a Dirac
$\delta$-function. Discussing, for simplicity, the case $\vec{r}_0
= 0$  one has

\begin{eqnarray}
I^{(1)}(\omega,\vec{r}_i)  &\propto& \left| \int d \vec{u'}~
\mathcal{T}_m\left(\frac{  \omega \vec{r_i}}{c f}-\vec{u'}\right)
\cdot \vec{\widetilde{E}}^{(1)}_\mathrm{mod}\left(\frac{c f
\vec{u'}}{ \omega}\right)\right|^2 ~.\label{PaSF}
\end{eqnarray}
%

Let us consider the simplest case when the Fourier-plane mask is
absent, i.e. when we account for a pupil which sharply limits the
range of the Fourier components passing through the system. When
one deals with imaging properties, one is only interested in
relative distributions of the radiation intensity. For the present
discussion we are only interested in the relative distribution of
the radiation intensity in the image plane. Rewriting Eq.
(\ref{PaSF}), imaging of a single electron by a
diffraction-limited, two-lens optical system gives

\begin{eqnarray}
I^{(1)}(\omega, r_i) \propto A_p^2(\omega, r_i)
=\left|\int_0^{\theta_a} d {\theta} ~
\frac{{{\theta^2}}}{(\gamma^2 {\theta}^2 + {1})} J_1\left(\frac{
\omega \theta {r_i}}{c} \right)\right|^2~. \label{AAA}
\end{eqnarray}
Here $\theta_a=a/f$. The resolution of the imaging system is
related to the width of $A_p^2$. The function $A_p(\omega,
\vec{r}_i)$ is, instead, to be considered as a mathematical
construct, which does not coincide with the field amplitude, being
in fact the square root of the sum of the intensities along
orthogonal polarization directions.

\begin{figure}
\begin{center}
\includegraphics*[width=110mm]{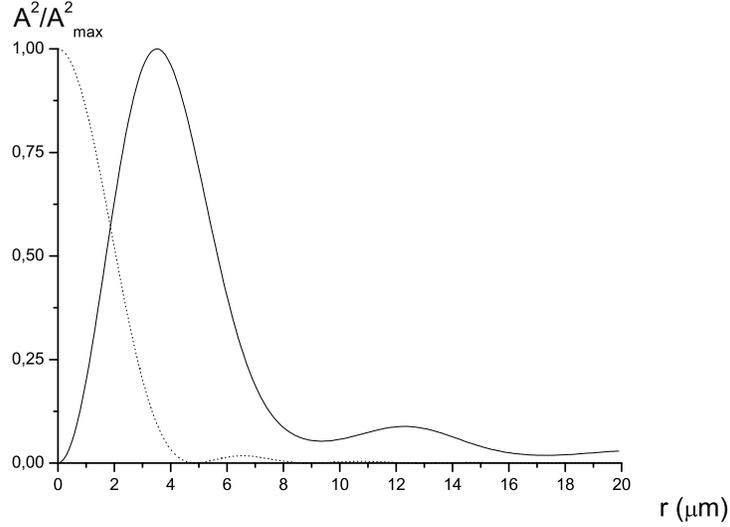}
\caption{\label{A2}  The particle spread function of the system
(solid line), Eq. (\ref{AAA}), for $\lambda = 800$ nm and
$\theta_a=0.1$. The point spread function, for which $A^2 =
\lambdabar^2 J_1^2[ r_i \theta_a/(\lambdabar)]/( r_i \theta_a)^2$,
is shown for comparison (dashed line).}
\end{center}
\end{figure}
An example of the particle spread function $A_p^2$ for $\lambda =
800$ nm and $\theta_a=0.1$ is plotted in Fig. \ref{A2}. In the
region of interest for electron bunch imaging, i.e. $r_i \ll
\gamma \lambdabar$, the result is practically independent on the
choice of $\gamma$ for XFEL setups. In this region, one may
approximate $\gamma^2 \theta^2 \gg 1$. 
As a result one may approximate the right hand side of Eq.
(\ref{AAA}) obtaining \cite{LEBE}

\begin{eqnarray}
A_p(\omega, r_i) = \frac{1}{\gamma^2} \int_0^{\theta_a} d {\theta}
~ J_1\left(\frac{  \omega \theta {r_i}}{c} \right) = \frac{
\theta_a}{\gamma^2}\cdot \frac{1-J_0(\omega  r_i
\theta_a/c)}{\omega r_i \theta_a/c}~, \label{AAA2}
\end{eqnarray}
where, as before, we set $\vec{r}_0 = 0$.

The OTR particle spread function is quite different compared with
the point spread function  $A^2 =\lambdabar^2 J_1^2[  r_i
\theta_a/(\lambdabar)]/( r_i \theta_a)^2$ for a circular pupil. In
particular, as one can see from Fig. \ref{A2}, in the case of the
OTR particle spread function the resolution is worse.

Finally, note that the function $A_p$ presented here does not
account for the peculiar field polarization properties of OTR. The
same reasoning done to derive $A_p$ starting from Eq. (\ref{PaSF})
can be done separately considering the response to a single
electron in the $x$ or in the $y$ polarization direction. In this
case, one obtains the following amplitude particle spread
functions for the orthogonal polarization components of the field
in a Cartesian coordinate system:

\begin{eqnarray}
A_p^{(x)}(\omega, r, \phi)=\frac{\cos(\phi)}{2} A_p(\omega, r)
\label{Apx}
\end{eqnarray}
and

\begin{eqnarray}
A_p^{(y)}(\omega, r, \phi)=\frac{\sin(\phi)}{2} A_p(\omega, r)~.
\label{Apy}
\end{eqnarray}

\subsection{Method for improving the OTR particle spread function}

\begin{figure}
\begin{center}
\includegraphics*[width=140mm]{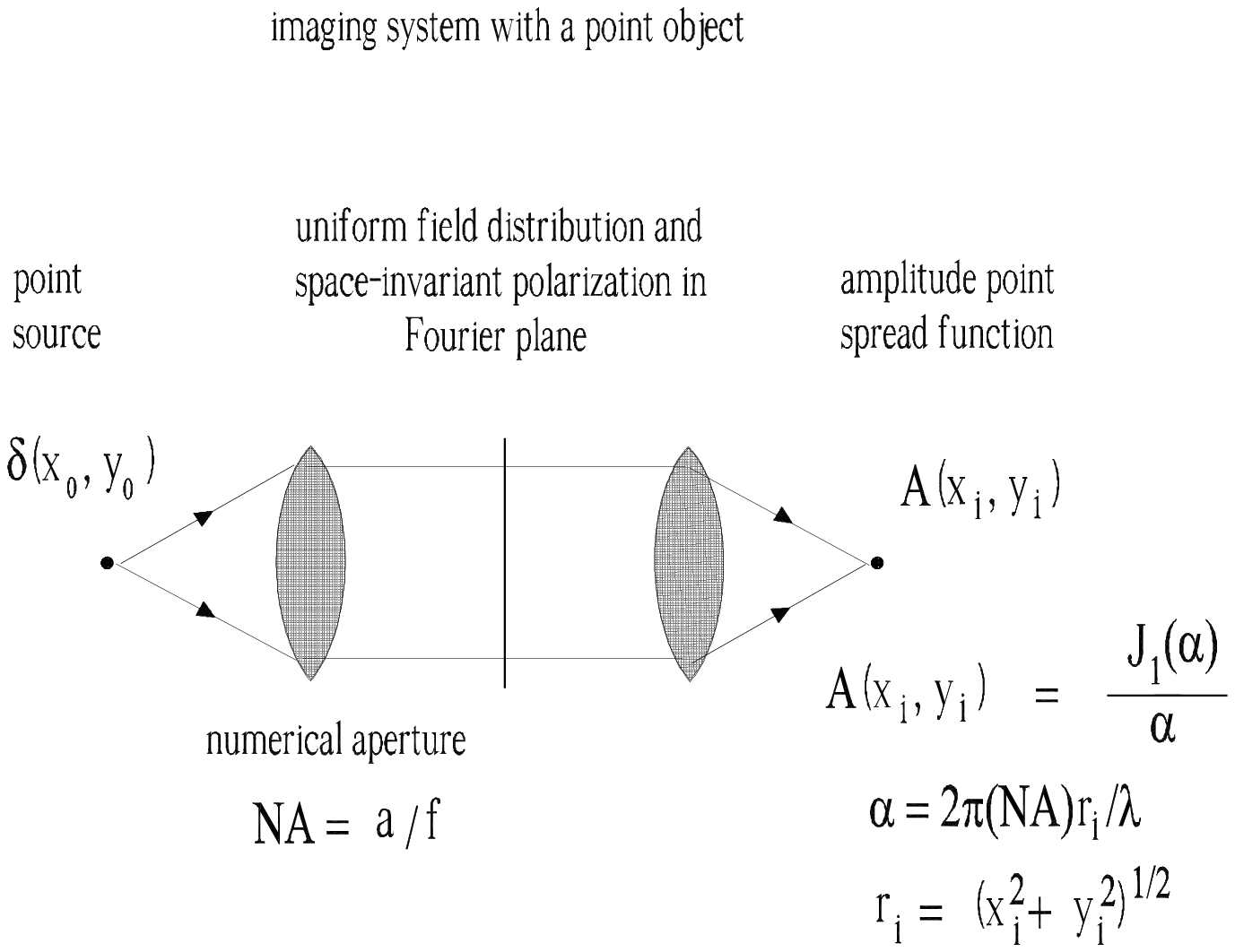}
\caption{\label{imagp} Two-lens imaging system with a point
source.}
\end{center}
\end{figure}

One will have the best resolution of the OTR imaging system when
the width of the particle spread function $A_p^2$ is minimized,
the optimum being the point spread function, the dashed line in
Fig. \ref{imagp}. Such optimal\footnote{Actually, in the coherent
case the resolution can still be improved by introducing pupil
apodization, see e.g. \cite{BAKA}. Here the words "optimal
situation" and "best resolution" are to be understood in a narrow
sense. Namely, as we will see, one can avoid blurring of the OTR
image due to the particular particle spread function of OTR and
reduce the problem to the standard case when point-like sources
are considered.} situation is sketched in Fig. \ref{imagp}. Once
the wavelength is fixed, the resolution only depends on the
numerical aperture of the system.  Note that, with reference to
the system shown in Fig. \ref{imagp}, the field distribution in
the focal plane is uniform, and the polarization is spatially
invariant. As a result, the field in the image plane is azimuthal
symmetric. It is the response of the system to a point source.

\begin{figure}
\begin{center}
\includegraphics*[width=140mm]{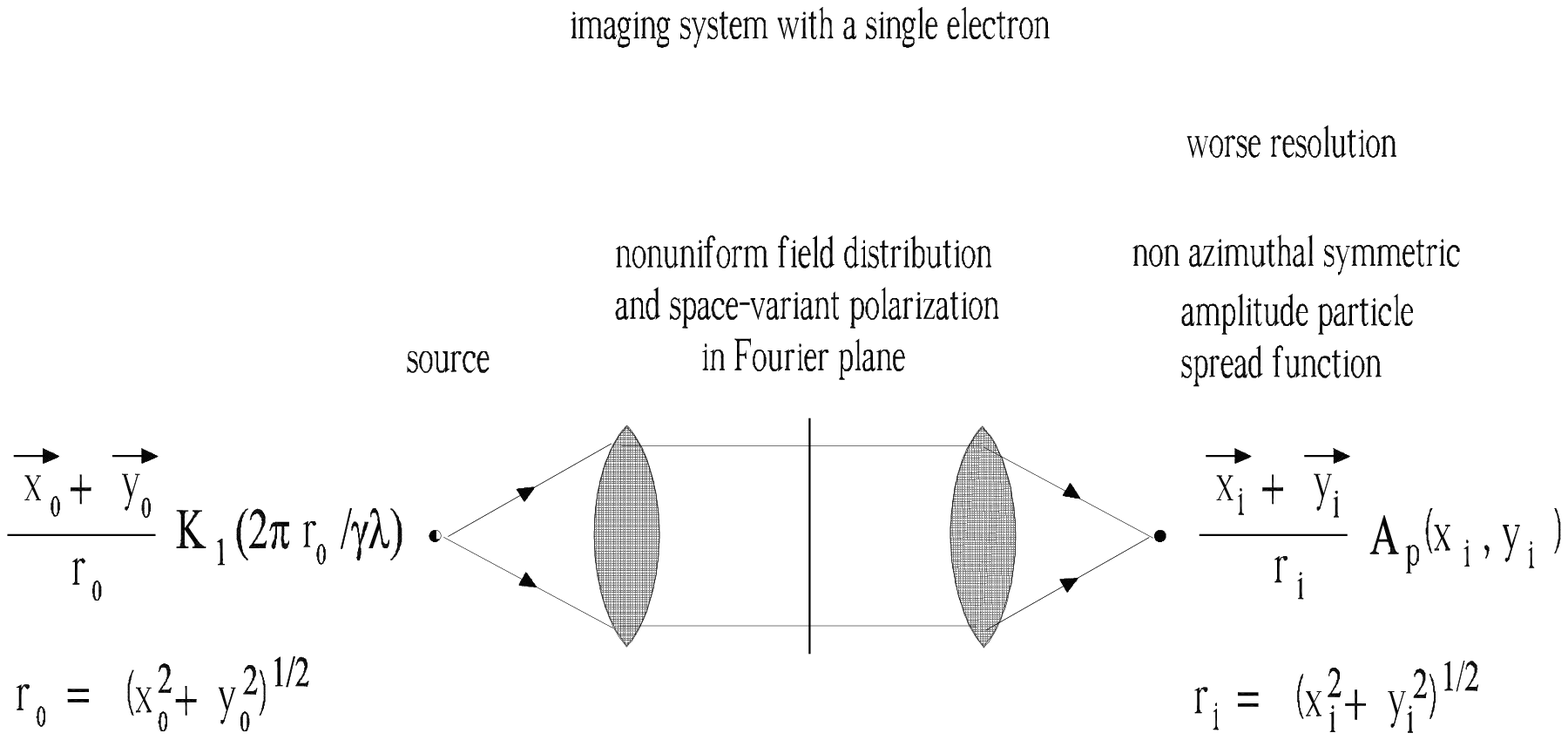}
\caption{The response of the system in the case of Optical
Transition Radiation emitted by a single electron.\label{image}}
\end{center}
\end{figure}

In the OTR imaging case, the situation is different, as one can
see from Fig. \ref{image}. The very specific OTR source is not
point-like but, rather, laser-like with a polarization
singularity. As a result, the field distribution in the focal
plane is not uniform, and the direction of the electric field
varies as a function of the transverse position. Therefore, in our
case we have two separate amplitude particle spread functions for
two orthogonal polarization directions. They are not azimuthal
symmetric, because they depend, respectively, on $x/r$ and $y/r$,
i.e. on cosine and sine of the azimuthal angle. Therefore, in the
image plane one obtains a non-azimuthal symmetric response, which
worsens the resolution. Only if one sums up the \textit{intensity}
patterns referring to the two polarization components for a single
electron, one obtains an azimuthal symmetric distribution, since
$(x^2+y^2)/r^2 = 1$.

\begin{figure}
\begin{center}
\includegraphics*[width=140mm]{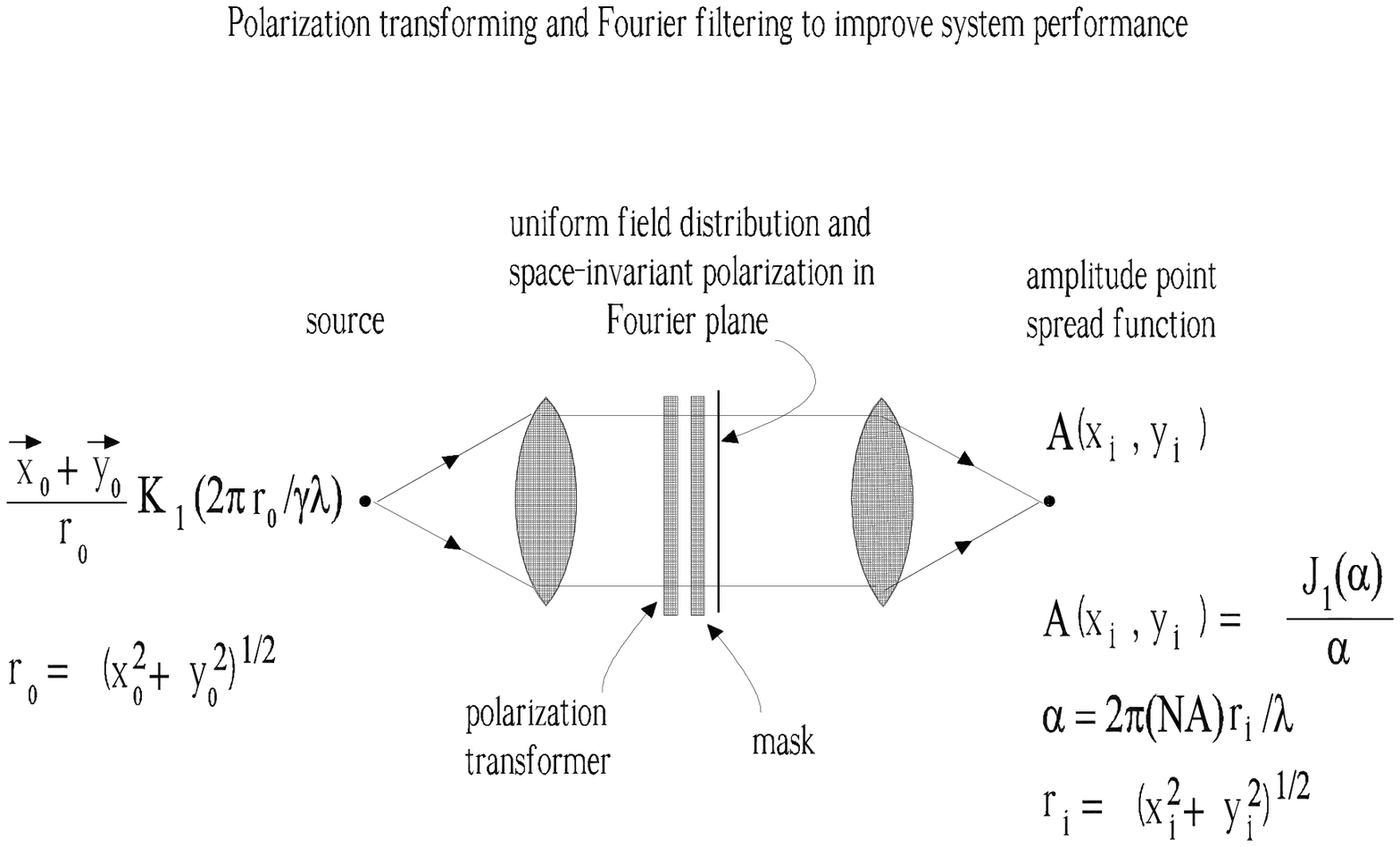}
\caption{\label{ptfilt} OTR particle spread function synthesis by
a Fourier-plane filter and a polarization transformer.}
\end{center}
\end{figure}
\begin{figure}
\begin{center}
\includegraphics*[width=130mm]{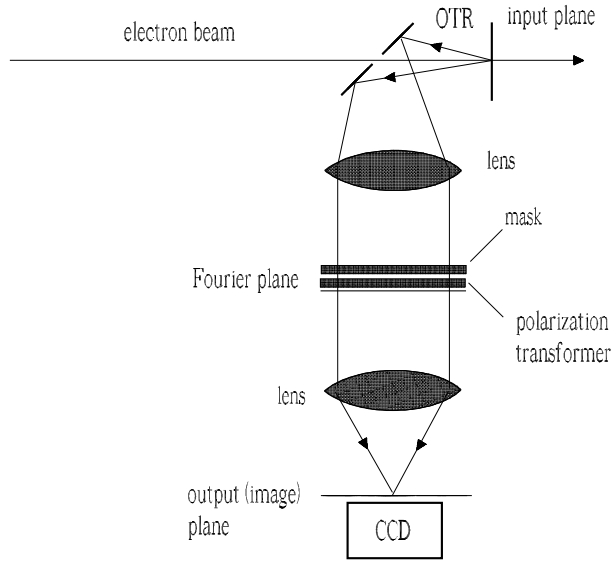}
\caption{\label{4finte} A practical arrangement of the coherent
OTR image-processing system.}
\end{center}
\end{figure}
Methods to improve the particle spread function for OTR imager
have been studied previously, e.g. in \cite{CAST}. The particle
spread function is improved by selecting one polarization
component of the field and by introducing a Fourier filter with
the help of a "round, opaque mask placed in the back focal plane
of the lens to prevent the passage of photons at angles smaller
than the mask angular acceptance" \cite{CAST}. Selection of one
polarization component of the field can improve the resolution in
one projected direction ($x$ or $y$) only, but not in both
simultaneously. Moreover, the other projection yields a much worse
resolution.

Here we propose a method to improve the particle spread function
up to the best possible point spread function. Our method is based
on the introduction of a filter transparency (mask) with a
particular amplitude transmittance and a polarization transformer
in the Fourier plane.

A sketch of the scheme is given in Fig. \ref{ptfilt}. Use of a
Fourier-plane mask and a polarization transformer allows one to
get back to the case shown in Fig. \ref{imagp}, where the field in
the focal plane is uniform, and the polarization is spatially
invariant. As a result, one improves the OTR particle spread
function to obtain the usual point spread function. Our scheme can
be implemented in an OTR imaging system as shown in Fig.
\ref{4finte}. Down here we will consider in more detail the effect
of both polarization transformer and amplitude mask.

\subsubsection{Polarization transformer}

\begin{figure}
\begin{center}
\includegraphics*[width=110mm]{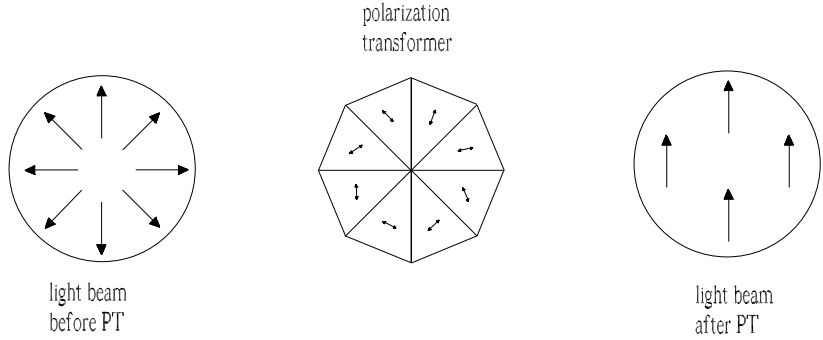}
\caption{\label{trasf1} Schematic plot illustrating the
transformation of a radially polarized beam into a linearly
polarized beam. Instead of a continuous spatially varying
retarder, a sectioned device can be used, where each section has
different orientation of the optical axis of the half-wave plate.}
\end{center}
\end{figure}

The function of the polarization transformer is to change the
radially polarized OTR into linearly polarized radiation as shown
in Fig. \ref{trasf1}. Here we will consider, as an example of
technical realization of a polarization transformer, a device
recently demonstrated in \cite{POLA,POL2}.

When a radially polarized beam passes through the transformer,
each sector rotates the polarization vector by different angle. As
a result, the polarization distribution just after the
polarization transformer is nearly-linear.


We can account for the polarization transformer in our expression
for the field by replacing Eq. (\ref{fieldpropexpi2}) with

%

\begin{eqnarray}
\vec{ \widetilde{E}}^{(1)}( \vec{ {r}}_i) &=& \frac{i \omega
\vec{e}_x}{2 \pi c f}
\int d \vec{ {r}}_f~ \Bigg\{\frac{2 e \gamma^2 r_f P(\vec{r}_f)}{c
(\gamma^2 {r}_f^2 + {f^2})} \Bigg\}\exp{\left[- \frac{i \omega
(\vec{r}_i\cdot\vec{r}_f)}{c f}\right]}~,
\label{fieldpropexpi2mod}
\end{eqnarray}
where we set $\vec{r}_0=0$ as before.

\subsubsection{\label{subsub:fpm}Fourier-plane mask}

Consider now the amplitude particle spread function, Eq.
(\ref{fieldpropexpi2mod}). Suppose that we insert a transparency
in the Fourier plane, whose amplitude transmittance essentially
yields an extra factor $r$. In other words,  we substitute
$P(\vec{r}_f)$ in Eq. (\ref{fieldpropexpi2mod}) with the product
$P(r) T_m(r)$, with $P(r) = \mathrm{circ}_a(r)$ defining the a
limiting aperture for the system. With the help of the cylindrical
coordinate system $(r,\phi)$ we can integrate Eq.
(\ref{fieldpropexpi2mod}) over azimuthal coordinates, thus
obtaining

\begin{eqnarray}
\vec{ \widetilde{E}}^{(1)}( 4 f,{ {r}}_i) &=& \frac{i \omega
\vec{e}_x}{c f}
\int_0^a d {r}_f~r_f^2~ T_m(r_f) \Bigg\{\frac{2 e \gamma^2 }{c
(\gamma^2 {r}_f^2 + {f^2})} \Bigg\} J_0\left( \frac{\omega r_i
r_f}{c f}\right)~. \label{fieldpropexpi2mod0}
\end{eqnarray}
Let us now consider a mask characterized by

\begin{eqnarray}
T_m(r) = T_0 \cdot r/a ~,\label{TMMM}
\end{eqnarray}
where $T_0$ is a common amplitude-attenuation coefficient (equal
to unity or smaller). In this case, assuming $T_0 = 1$ for
simplicity, Eq. (\ref{fieldpropexpi2mod0}) is replaced by

\begin{eqnarray}
\vec{ \widetilde{E}}^{(1)}( 4f,{ {r}}_i) &=& \frac{i \omega
\vec{e}_x}{a c f}
\int_0^a d {r}_f~r_f^3~ \Bigg\{\frac{2 e \gamma^2 }{c (\gamma^2
{r}_f^2 + {f^2})} \Bigg\} J_0\left( \frac{\omega r_i r_f}{c
f}\right)~.\label{fieldpropexpi2mod2}
\end{eqnarray}
As noted before, in the region of interest for electron bunch
imaging, i.e. $r_i \ll \gamma \lambdabar$, we can neglect the term
in $f^2$ in the denominator of the integrand in Eq.
(\ref{fieldpropexpi2mod2}), which gives

\begin{eqnarray}
\vec{ \widetilde{E}}^{(1)}(4 f,{ {r}}_i) &=& \frac{i \omega
\vec{e}_x}{a c f}  \frac{2 e}{c}
\int_0^a d {r}_f~r_f~  J_0\left( \frac{\omega r_i r_f}{c
f}\right)~. \label{fieldpropexpi2mod3}
\end{eqnarray}
Finally\footnote{Note that propagating the field in Eq.
(\ref{vir4ea}) from the object plane ($z=0$) to the Fourier plane
($z=2 f$), one obtains a phase shift $\pi$. This is the phase
shift between "impulse", the OTR field on the screen and
"response", the OTR field in the image plane (the geometrical
phase shift $\exp[-i\omega z/c]$ having already been accounted for
in the relation between $\vec{\widetilde{E}}$ and
$\vec{\bar{E}}$). Such phase shift represents the analogous of the
Gouy phase shift in laser physics. For azimuthal-symmetric beams,
the Gouy phase shift is known to be $\pi/2$. However, this result
is not valid for Eq. (\ref{vir4ea}), where the cartesian
components of the field depend on the azimuthal angle. After
propagating to $z=4 f$, the result is the usual Gouy phase shift
$\pi/2$.},

\begin{eqnarray}
\vec{ \widetilde{E}}^{(1)}( 4 f,{ {r}}_i) &=& \frac{i
\vec{e}_x}{r_i}  \frac{2 e}{c}
J_1\left(\frac{\omega r_i a}{c f}\right)~.
\label{fieldpropexpi2mod4}
\end{eqnarray}
Rewriting Eq. (\ref{fieldpropexpi2mod4}) yields the well-known
expression for the amplitude point spread function:

\begin{eqnarray}
A(\omega,r_i) = \frac{c}{\omega  r_i \theta_a}
J_1\left(\frac{\omega r_i \theta_a}{c}\right)~,\label{AAC}
\end{eqnarray}
which is the diffraction pattern of a circular aperture, where
$\theta_a$, as usual, is the angular dimension of the pupil
aperture.

\begin{figure}
\begin{center}
\includegraphics*[width=120mm]{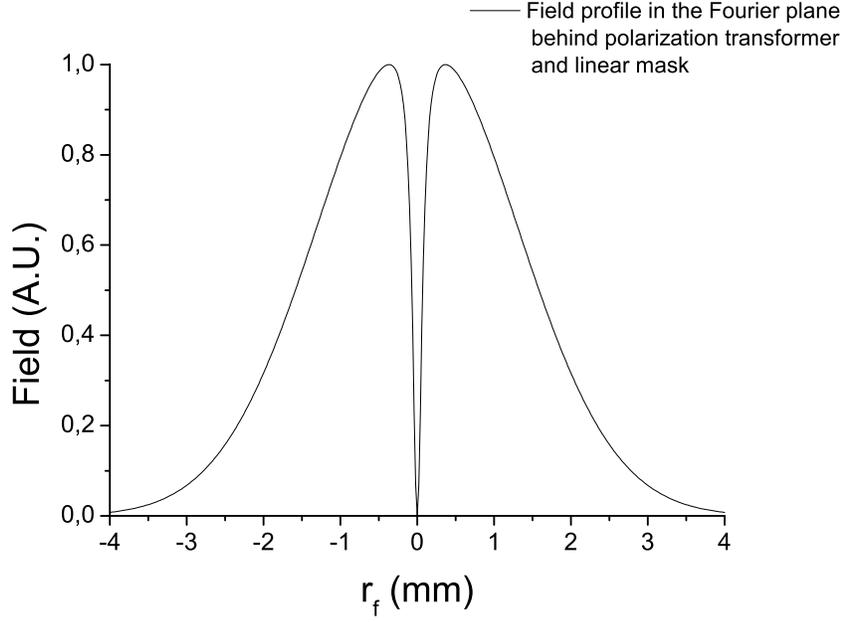}
\caption{\label{FTOTR1} The radial profile of the OTR field from a
Gaussian electron bunch in the Fourier plane, behind polarization
transformer and linear mask.}
\end{center}
\end{figure}
Note that Eq. (\ref{AAC}) is only valid for some range of $r_i$.
In fact, the validity of Eq. (\ref{AAC}) for all values of $r_i$
would mean uniform field distribution amplitude in the Fourier
plane, while on the optical axis we always have zero amplitude.
With our mask profile, Eq. (\ref{TMMM}), the validity region of
Eq. (\ref{AAC}) is for $0< r_i \ll \gamma \lambdabar$. This
corresponds to an improved pattern of the field profile in the
Fourier plane up to frequencies of order $1/(\gamma \lambdabar)$.

\begin{figure}
\begin{center}
\includegraphics*[width=120mm]{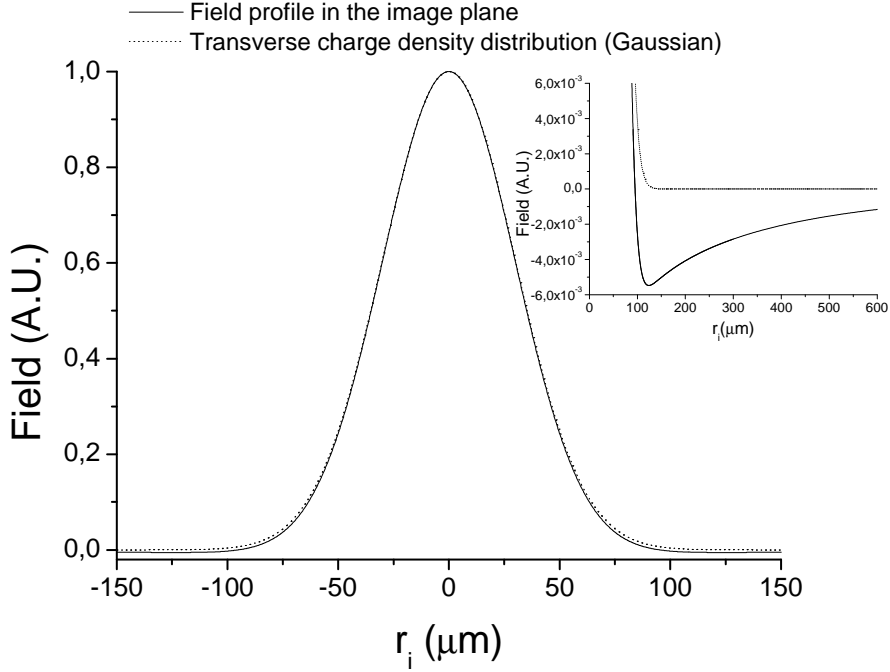}
\caption{\label{FTOTR2}  The radial field profile in the image
plane (solid line) compared with the Gaussian electron charge
density distribution (dotted line). In the inset we show the same
plots on a different scale.}
\end{center}
\end{figure}

In principle, one may use a more complicated mask profile to
improve the pattern in the Fourier plane up to frequencies even
much smaller than $1/(\gamma \lambdabar)$. Yet, we can demonstrate
that for the XFEL case our simple choice of linear transmittance
provides a sufficient accuracy. This fact is shown in Fig.
\ref{FTOTR1} and Fig. \ref{FTOTR2} for the $2$ GeV case, with
$a=3$ cm and $f=30$ cm. Fig. \ref{FTOTR1} shows the field profile
generated by an electron bunch with transverse dimension $\sigma_r
= 30~\mu$m in the focal plane of the two-lens imaging system. Note
that the field has zero amplitude on the optical axis. This fact
corresponds to deviations of the tails of the field distribution
in the image plane relative to the Gaussian shape, as shown in the
inset Fig. \ref{FTOTR2}. However, as one can see, the field
profile in the image plane shown in Fig. \ref{FTOTR2} is
reproducing the bunch profile quite accurately. A Gaussian fit of
the field profile with a Gaussian function yields
$\sigma_r^{(\mathrm{fit})} = 29.86 ~\mu$m. The relative difference
between $\sigma_r$ and $\sigma_r^{(\mathrm{fit})}$ is thus about
$0.5\%$.

\subsection{Resolution analysis for incoherent and coherent imaging}

No conventional resolution criterion can be applied for the
particle spread function in Eq. (\ref{PaSF}). However, with the
help of our method we improved the particle spread function to
obtain the ultimate resolution, so that the amplitude particle
spread function is now equal to the amplitude point spread
function, Eq. (\ref{AAC}).

Instead of dealing with the amplitude particle spread functions in
Eq. (\ref{Apx}) and Eq. (\ref{Apy}), we deal with the solution of
a simple diffraction problem, where the field produced by a point
source is diffracted by a circular pupil. Therefore, the
resolution analysis is reduced now to the theoretical framework of
standard incoherent or coherent imaging theory (see e.g.
\cite{BOO1}). One can take advantage of well-known resolution
criteria like the Rayleigh or Sparrow criteria, or alternatively,
of a formulation of the resolution problem in terms of Optical
Transfer Function (OTF), incoherent case or Coherent Transfer
Function (CTF), coherent case.


In principle, our technique could be applied both for incoherent
and coherent radiation. 
However, in the incoherent case, while spatial resolution improves
towards the point source, limitations arise due to the small
number of photons available. As a result, the method described
above is better applied for the coherent imaging case.

When we deal with coherent OTR imaging, limitations of techniques
to improve the particle spread function due to a small number of
photons can be avoided, given the much larger available photon
flux.


As seen before, processing the spatial spectrum and polarization
in the Fourier plane allows us to reduce a vectorial problem to a
scalar one, so that, instead of dealing with the amplitude
particle spread functions in Eq. (\ref{Apx}) and Eq. (\ref{Apy})
we consider the scalar diffraction pattern in Eq. (\ref{AAC}).
Thus, the imaging problem is reduced to an imaging problem for
coherently radiating point sources. 

We note with \cite{BOO1} that the Rayleigh criterion is not
suitable for defining the resolution in the case of coherent
imaging, because it tends to underestimate the minimal distance
where resolving two point sources is still possible. An acceptable
resolution criterion turns out to be, instead, the Sparrow
criterion, stating that the minimal separation of two points on
the OTR screen is given by $d \simeq 4.6\cdot
\lambdabar/\theta_a$. This value may be compared with that given
by the Sparrow criterion for incoherent radiation, i.e. $d \simeq
3.0 \cdot \lambdabar/\theta_a$. Such comparison suggests that
incoherent imaging techniques would yield (in the case of point
emitters) an ultimate resolution about $1.5$ better than that in
the coherent case. However, as explained before, techniques to
improve the particle spread function in the incoherent case are
limited by the number of photons available.

Finally, it should be clear that the definition of resolution is
somewhat arbitrary, as it consists in a single number, a figure of
merit to be extracted according to some criterion or algorithm
from the actual image. A more quantitative approach to the
resolution problem would be to use the CTF representation, where
the point spread function is given in terms of its spatial Fourier
transform.

Since one has

\begin{eqnarray}
|~\mathrm{CTF}~|\left({u}\right) \propto \left|\int d \vec{r}_i~
A(\omega, r_i) \exp[i \vec{u}\cdot \vec{r}_i]\right| \propto
\mathrm{circ}\left(\frac{u c}{\omega
\theta_a}\right)~,\label{AACCC}
\end{eqnarray}
one recovers the result that the pupil behaves like a spatial
filter. Spatial frequencies up to $ \theta_a/\lambdabar$ are
allowed to pass through the aperture without distortion, while
higher frequencies are blocked. The resolution of the image of the
electron bunch reduces to the standard diffraction-limited
resolution of a point source.

Let us consider, as a numerical example, our particular case of an
XFEL with an ORS setup. Here we are dealing with coherent imaging.
As shown above, an acceptable resolution criterion for the
coherent imaging case is the Sparrow criterion, which defines the
resolution as $d \simeq 4.6\cdot \lambdabar/\theta_a$. According
to the Sparrow criterion, for our case, i.e. for $\lambda = 800$
nm and $\theta_a = 0.1$, we obtain $d \simeq 6~ \mu$m.

It should be noted that both Sparrow and Rayleigh criteria can
only be applied in the case of a Fraunhofer diffraction pattern
from a lens aperture. They cannot be directly applied when one
deals with a complicated OTR particle spread function like that in
Eq. (\ref{AAA2}). Therefore, up to now, the diffraction-limited
resolution for OTR imaging has been based on the estimation of an
"effective width" of the OTR particle spread function.

For example, in order to estimate the diffraction-limited
resolution for the incoherent OTR imager, authors of \cite{BENG}
fit the particle spread function, Eq. (\ref{AAA2}), to a Gaussian
function and calculate the $rms$ width of the Gaussian fit.  They
conclude that, for $\lambda = 800$ nm and $\theta_a=0.1$, the
particle-spread function given by Eq. (\ref{AAA2}) has a width of
about $6~\mu$m, and they consider this as a measure of the
diffraction-limited resolution for their system. However, using
their method, one concludes that the resolution for a point source
is about three times better than what is predicted by the Rayleigh
criterion, which is given by $d \simeq 3.8 \cdot
\lambdabar/\theta_a$. In fact, a resolution $d \simeq 1.2 \cdot
\lambdabar/\theta_a$ was reported in \cite{BENG} for the
incoherent case and for a  point-source. 
For example, following \cite{BENG}, one would obtain a resolution
$d \simeq 2 ~\mu$m , while according to the Rayleigh criterion the
resolution is $d \simeq 5 ~\mu$m. The difference is due to the use
of the width of the point spread function to define the resolution
of the system in \cite{BENG}.

The main result reached in this Section is an optimization of the
particle spread function of the system, so that the imaging
problem can be reduced to the usual (coherent or incoherent)
imaging theory for point-like radiators. This result is summarized
in Eq. (\ref{AAC}), presenting the amplitude particle spread
function of the system after processing, which is equivalent to
the amplitude point spread function.

\subsection{Comparison of OTR and undulator sources for electron bunch imaging}

We have seen that OTR pulses are intrinsically difficult to deal
with, due to the presence of the polarization singularity of the
OTR  field from a single electron, resulting in the space-variant
polarization and in the singular behavior of the field amplitude.
These difficulties can be avoided for instance if one uses
coherent radiation emitted by a modulated electron bunch in the
radiator-undulator of the ORS, where the polarization is
space-invariant, and the field amplitude has no singular behavior.
Yet, coherent undulator radiation is not suitable for imaging
applications, while coherent OTR is. In fact, coherent undulator
radiation is highly collimated within an angle
$\sqrt{\lambdabar/L_w}$, $L_w$ being the undulator length, which
in our case is smaller than a milliradian. Therefore, effectively,
the numerical aperture of the imaging system is limited by the
divergence of the undulator radiation, defining the resolution of
the system. As a result, an electron bunch of a tens-micron-size
cannot be resolved with the help of undulator radiation. In
contrast to this, the OTR radiation pulse from a single electron
has a non-limited angular distribution, and the resolution will
only be limited by the numerical aperture of the optical system.
This basic characteristics of OTR plays an important role in both
coherent and incoherent imaging.

\section{\label{sec:ft}Coherent OTR imaging for characterizing electron bunches for XFELs}

In the previous Section we have shown that, due to the particular
features of the OTR source, the OTR image of a single electron
significantly deviates from the image of a point source. This fact
causes problems concerning the measure of the charge density
distribution. We demonstrated that the resolution can be improved
with the help of a polarization transformer and a specific
transparency in the Fourier plane at the cost of reduction in
photon density. Techniques described in the previous Section can
improve the OTR image from a single electron up to the standard
diffraction pattern from a point source. We shall now discuss some
practical aspects which are specifically related to the
characterization of electron bunches in an XFEL system, such as 3D
electron bunch structure, signal-to-noise ratio, aberrations and
alignment of mask and polarization transformer.

\subsection{Electron bunches with spatio-temporal coupling}

\begin{figure}
\begin{center}
\includegraphics*[width=110mm]{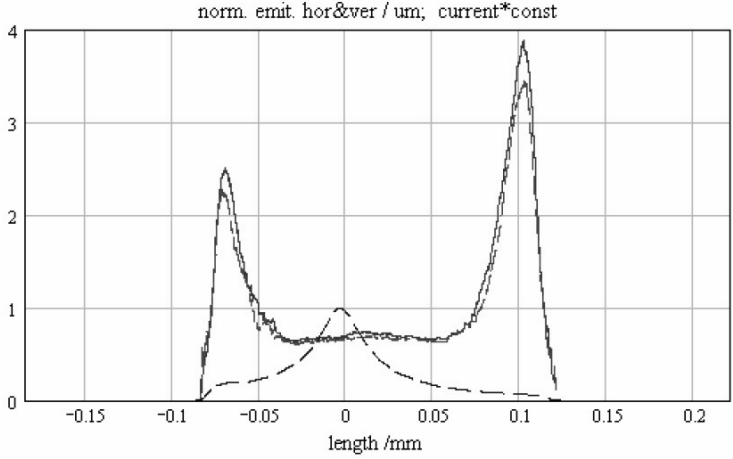}
\caption{\label{emittance} Normalized horizontal and vertical
emittance as a function of the longitudinal position within the
electron bunch taken from \cite{XFEL}. The dashed line indicates
the peak-current profile.}
\end{center}
\end{figure}
\begin{figure}
\begin{center}
\includegraphics*[width=140mm]{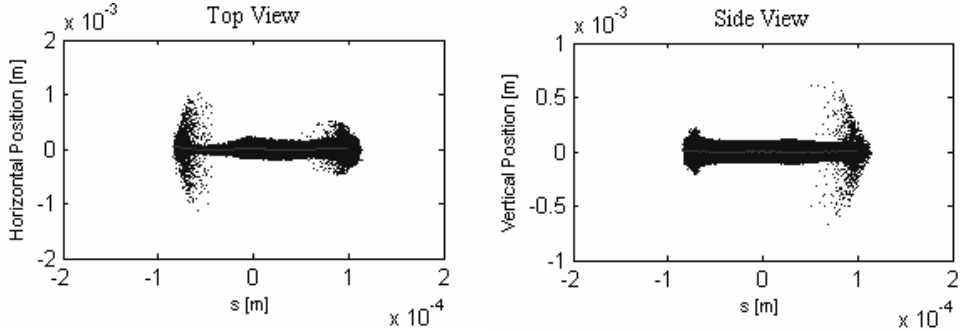}
\caption{Vertical (left) and horizontal (right) projections of the
electron bunch density distribution, \cite{BEAM}. \label{BLBL} }
\end{center}
\end{figure}
The intensity profile in the image plane is readily calculated. In
our case of interest radiation is coherent. Therefore, we must
convolve, in the space-frequency domain, the field given in Eq.
(\ref{fieldpropexpi2mod4}) with the Fourier transform of the
electron density distribution $\bar{\rho}(\omega,\vec{r})$, and
take the squared modulus to get the energy radiated per unit
spectral interval per unit surface

\begin{eqnarray}
\frac{d^2W}{d\omega dS}(\omega,\vec{r}_i)= \frac{e^2 N_e^2}{\pi^2
c} \left| \int \d \vec{r'} \bar{\rho}(\omega, \vec{r'}-\vec{r}_i)
\frac{1}{r'} J_1\left(\frac{\omega r' a}{c f}\right)\right|^2~.
\label{INTBA}
\end{eqnarray}
%
%
%
%
When dealing with XFEL applications, it is important to realize
that the electron-bunch slice emittance changes along the bunch,
as one can see from the example shown in Fig. \ref{emittance},
which refers to the European XFEL \cite{XFEL}. Note that even if
the slice emittance were constant, some mismatch is always
present, leading to a dependence of the electron bunch transverse
size on the longitudinal coordinate (see Fig. \ref{BLBL}). A
comparison of Fig. \ref{emittance} with Fig. \ref{BLBL} shows that
even in the part of the electron bunch where the emittance does
not vary much there are significant changes in the transverse
size. As a matter of fact, XFEL beam formation systems will
generate electron bunches that exhibit a coupling between space
and time even in the case of nominal operation. As a result,
optical replica pulses will also exhibit coupling between space
and time, i.e. the electric field components in the
space-frequency domain cannot be factorized as a product
$\widetilde{E}(\omega,\vec{r}) = f(\vec{r}) g(\omega)$.
%


\begin{figure}
\begin{center}
\includegraphics*[width=140mm]{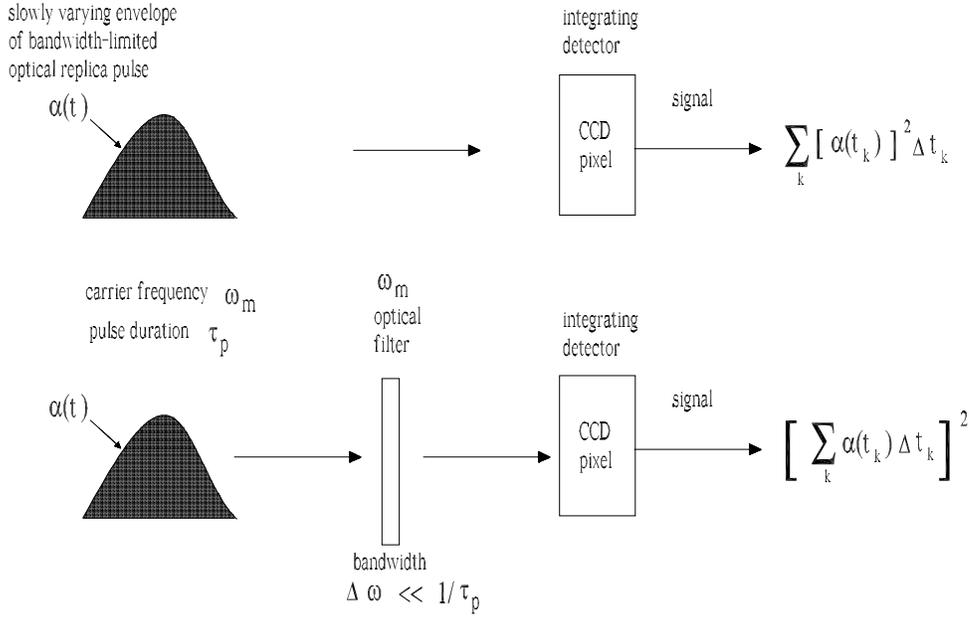}
\caption{\label{bpfil} Effect of the bandpass optical filter on
recording the bandwidth-limited optical replica pulse by a single
detector pixel.}
\end{center}
\end{figure}

\begin{figure}
\begin{center}
\includegraphics*[width=140mm]{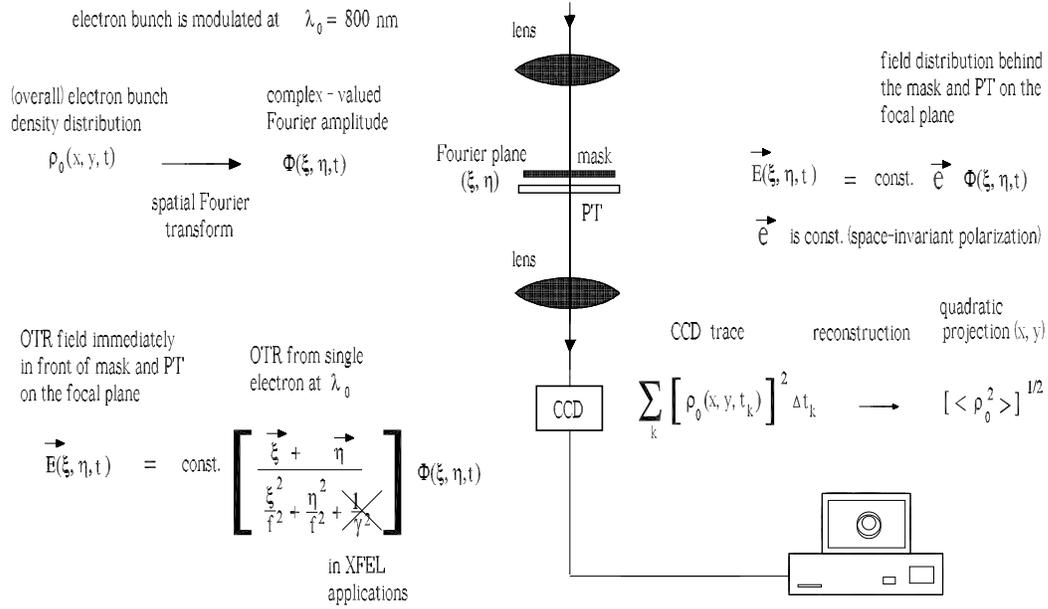}
\caption{\label{coima1} Schematic diagram of coherent OTR imaging
of electron bunch without bandpass optical filter.}
\end{center}
\end{figure}

\begin{figure}
\begin{center}
\includegraphics*[width=140mm]{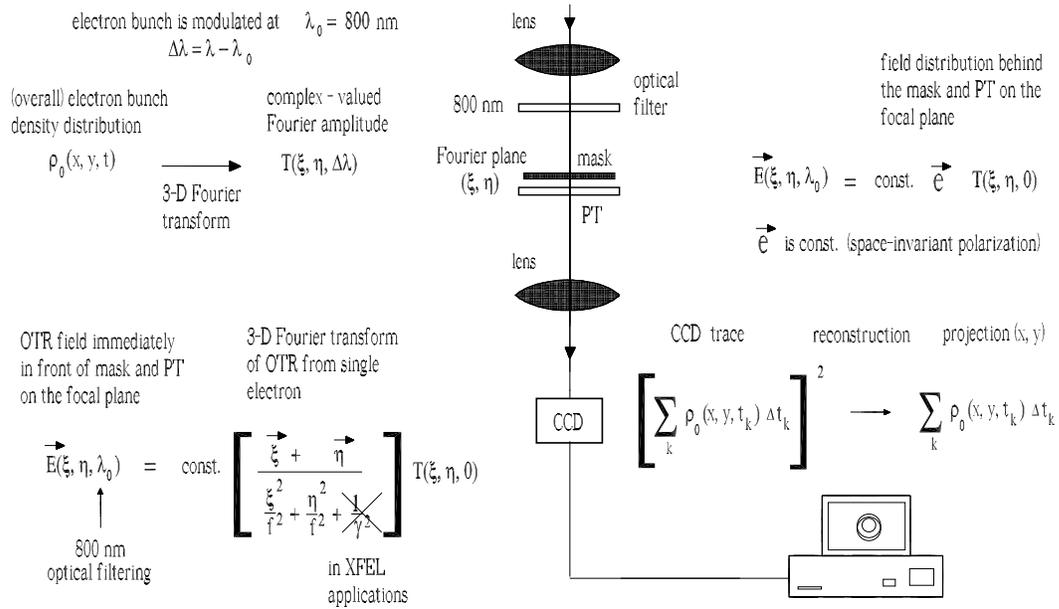}
\caption{\label{coima2} Schematic diagram of coherent OTR imaging
of electron bunch with bandpass optical filter. The setup with
bandpass optical filtering can be used to reconstruct the (x,y)
electron bunch projection.}
\end{center}
\end{figure}
\begin{figure}
\begin{center}
\includegraphics*[width=120mm]{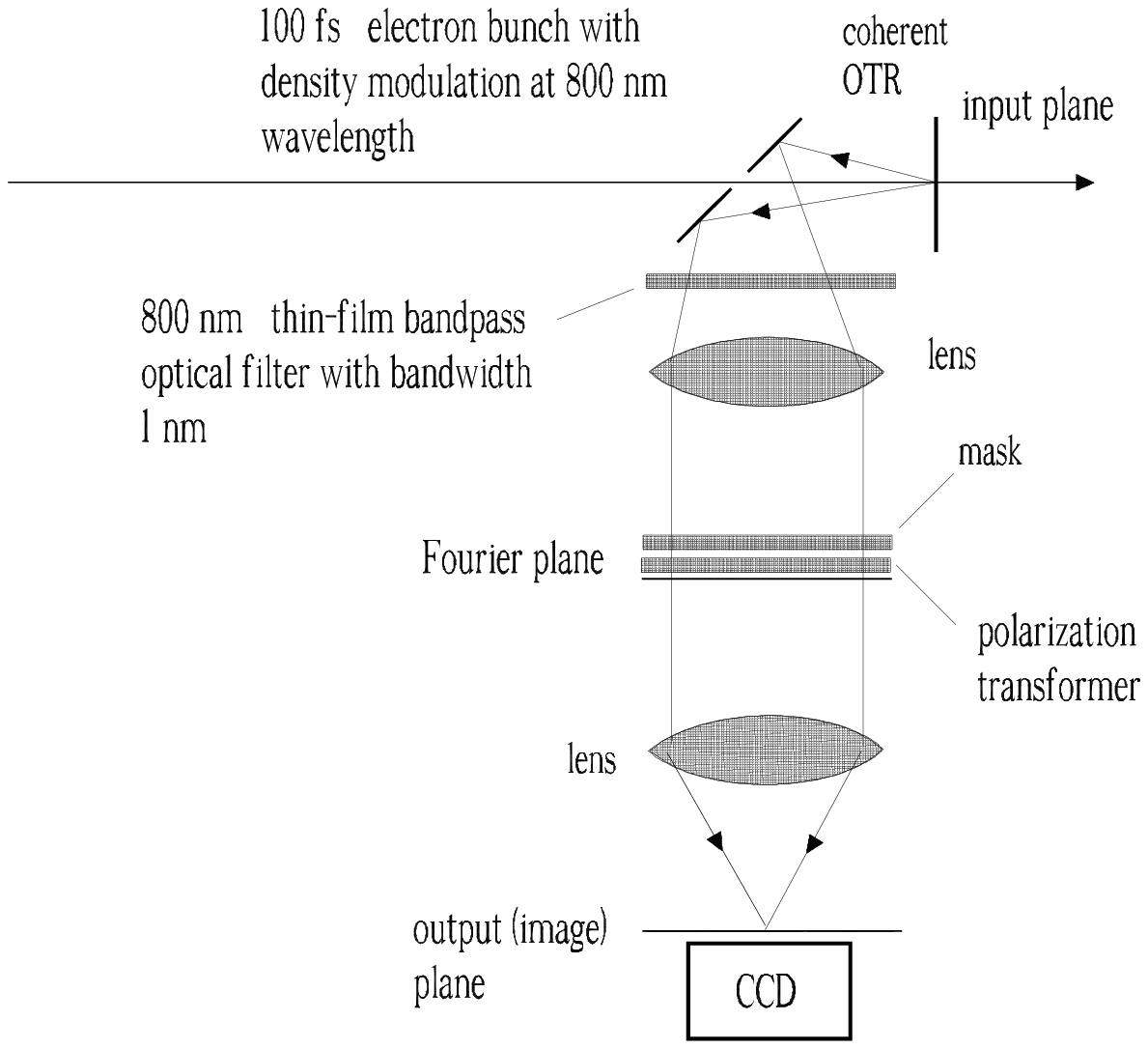}
\caption{\label{schemecoh}  Setup of high-resolution coherent OTR
imager with bandpass optical filter to reconstruct the (x,y)
electron bunch projection.}
\end{center}
\end{figure}
The detector automatically performs an integration over time of
the intensity $I(t,\vec{r}_i)$ in the space-time domain, which can
be extended over all times, because the detector is not fast
enough to resolve a single radiation pulse.  Such integration over
times is equivalent, due to Parseval's theorem, to an integration
over all frequencies of Eq. (\ref{INTBA}). Therefore, the detector
will record the energy per unit surface

\begin{eqnarray}
\frac{d W_\mathrm{det}}{d S}(\vec{r}_i) \propto
\int_{-\infty}^{\infty} d \omega \left| \int \d \vec{r'}
\bar{\rho}(\omega, \vec{r'}-\vec{r}_i) A(\omega,r') \right|^2~.
\label{INTBA22}
\end{eqnarray}
Let us define ${\rho}_0(t, \vec{r})$, as the electron density
distribution of the unmodulated electron bunch. In the case of a
bunch modulated with frequency $\omega_m$, ${\rho}_0(t, \vec{r})$
and ${\rho}(t, \vec{r})$ are linked by

\begin{eqnarray}
{\rho}(t, \vec{r}) = {\rho}_0(t, \vec{r}) [1+a_f \cos(\omega_m
t)]~, \label{duerho0}
\end{eqnarray}
where $a_f$ is the modulation level. Let us also call
$\bar{\rho}_0(\omega,\vec{r})$ the Fourier transform of
$\rho_0(t,\vec{r})$   with respect to time.

Since the adiabatic approximation applies, the function
$\bar{\rho}(\omega, \vec{r})$ has a narrow bandwidth $\Delta
\omega/\omega \ll 1$. Thus, we can write $\bar{\rho}(\omega,
\vec{r}) = \bar{\rho}_0(\omega - \omega_m,\vec{r}) a_f$, $a_f$
being the modulation amplitude. Note that here we assume $a_f =
\mathrm{const}$, an assumption justified in Section
\ref{sec:setup}, where we discussed that the main goal of the ORS
setup is to produce an exact optical replica of the electron
bunch. Since $\rho_0(t,\vec{r})$  is a positive function and $a_f$
is constant, the optical pulse is automatically bandwidth
limited\footnote{In order to ensure a bandwidth-limited optical
pulse we need that the seed laser be bandwidth-limited along the
electron-bunch pulse. However, even a bandwidth-limited seed pulse
can yield, in practice, to a non bandwidth-limited electron
density distribution. This is the case e.g. when there is a given
energy chirp within the electron bunch. We will discuss this
situation in detail later on. For now, we put attention that Eq.
(\ref{booo0}) is actually based on the use of a weaker assumption
than $a_f = \mathrm{const}$, namely $|a_f| =\mathrm{const}$.}.

%
%
The situation is described\footnote{Notations in figures are
slightly different compared to those in the text for editing
reasons. However, all relevant definitions are included.} in the
time domain by Fig. \ref{bpfil} (top) and Fig. \ref{coima1}. The
fact that $\bar{\rho}(\omega, \vec{r}) = \bar{\rho}_0(\Delta
\omega,\vec{r}) a_f~$, where $\Delta \omega = \omega - \omega_m$,
is implied by the validity of Eq. (\ref{duerho0}) with
$a_f=\mathrm{const}$. Note that if $a_f$ were not constant,
$\bar{\rho}(\omega, \vec{r})$ would be a convolution of
$\bar{\rho}_0(\Delta \omega,\vec{r})$ with the Fourier transform
of $a_f$.

In Fig. \ref{bpfil}, the envelope of the optical replica pulse
seen by a single pixel of the detector is shown as a function of
time. This function is integrated by the detector pixel over a
given temporal interval, which is longer than the duration of the
pulse. As shown in Fig. \ref{bpfil}, the result is proportional to
the time-integrated squared modulus of the amplitude of the pulse
envelope. In its turn, the pulse envelope can be written as a
convolution between $\rho_0(t,\vec{r})$ and the point spread
function $A(\omega_m,r)$, where the point spread function is
calculated at $\omega = \omega_m$ because the adiabatic
approximation holds. The detector thus records

\begin{eqnarray}
\frac{d W_\mathrm{det}}{d S}(\vec{r}_i) \propto \int dt \left|\int
dr' \rho_0(t,\vec{r'}-\vec{r}_i) A(\omega_m,r')\right|^2~,
\label{booo0}
\end{eqnarray}
which is equivalent to Eq. (\ref{INTBA22}) by means of Parseval's
theorem. Neglecting the blurring one simply obtains

\begin{eqnarray}
\frac{d W_\mathrm{det}}{d S}(\vec{r}_i)
\propto\left\langle\rho_0^2\right\rangle = \int dt
\left|\rho_0(t,-r_i)\right|^2 ~.\label{booo}
\end{eqnarray}
%

The right hand side of Eq. (\ref{booo}) is an integral in time of
the \textit{squared modulus} of the charge density distribution,
i.e. it is a \textit{quadratic} projection, along the temporal
axis, of the charge density distribution.

Note that in the case of incoherent OTR imaging (still neglecting
the blurring) one would sum the intensity of all electrons
independently. As a result, in order to obtain the energy absorbed
by a detector pixel at a given transverse position $\vec{r}_i$, we
need to sum over all the electron arrival times. Therefore, the
signal from each pixel will be proportional to the integral in
time of $\rho_0(t,\vec{r}_i)$, i.e. it is a \textit{linear}
projection of $\rho_0(t,\vec{r}_i)$ along the temporal axis.

Both linear and quadratic projections can be taken advantage of.
They can be used for comparison with results of numerical
simulations and they are both sensitive to perturbations of the
transverse size of the bunch, which we need to monitor. 

However, also in the coherent OTR imaging case it is possible to
obtain the linear projection of the charge density distribution,
instead of the quadratic projection, by inserting a narrow-band
optical filter with bandwidth much smaller than the inverse
duration of the optical pulse, and centered at the modulation
frequency $\omega_m$. Calling with $\mathcal{G}(\omega-\omega_m)$
the filter transfer function, one obtains the following
modification to Eq. (\ref{INTBA22})

\begin{eqnarray}
\frac{d W_\mathrm{det}}{d S}(\vec{r}_i) \propto
\int_{-\infty}^{\infty} d \omega \left| \int \d \vec{r'}
\bar{\rho}(\omega, \vec{r'}-\vec{r}_i)
\mathcal{G}(\omega-\omega_m) A(\omega,r') \right|^2~. \cr
&&\label{INTBA3}
\end{eqnarray}
%
%
%

Due to the narrow-band nature of $\mathcal{G}$, and remembering
that due to the adiabatic approximation $\bar{\rho}(\omega,
\vec{r}) = \bar{\rho}_0(\omega - \omega_m,\vec{r}) a_f$, both
$\bar{\rho}$ and $J_1$ can be taken out of the integral in
$d\omega$, thus yielding

\begin{eqnarray}
\frac{d W_\mathrm{det}}{d S}(\vec{r}_i) &\propto & a_f^2
\int_{-\infty}^{\infty} d \omega
\left|\mathcal{G}(\omega-\omega_m) \right|^2 \cdot \left| \int \d
\vec{r'}  \bar{\rho}_0(0, \vec{r'}-\vec{r}_i) A(\omega_m,r')
\right|^2~. \cr &&\label{INTBA4}
\end{eqnarray}
In the time domain, the envelope of the optical replica pulse (see
also Fig. \ref{bpfil}) coincides with the non-modulated electron
density distribution, $\rho_0(t,\vec{r})$. One has

\begin{eqnarray}
\bar{\rho}_0(0,\vec{r}_i) = \int_{-\infty}^{\infty} dt
\rho_0(t,\vec{r}_i) ~. \label{INTBA4b}
\end{eqnarray}
Note that $\bar{\rho}_0(0,\vec{r}_i)$ is a positive function, in
contrast with the complex-valued function $\bar{\rho}_0(\Delta
\omega,\vec{r}_i)$ for $\Delta \omega \neq 0$. One sees that the
detector just records the squared modulus of linear projection (in
time) of the charge density distribution.

The same conclusion can be reached directly in the time domain by
inspecting Fig. \ref{bpfil} (bottom) and Fig. \ref{coima2}. The
optical-replica pulse is stretched by the optical filter to a
duration of $1/(\Delta \omega) \gg \tau_p$, with $\tau_p$ the
duration of the original optical replica pulse. The result of this
operation is another pulse whose envelope is given by the
convolution between the envelope $\rho_0(t)$ and the Fourier
Transform of $\mathcal{G}$. Since $1/(\Delta \omega) \gg \tau_p$,
the detector pixel sees the integrated intensity, that is just
given by Eq. (\ref{INTBA4b}). A possible setup for high-resolution
OTR imaging for XFEL setups including a bandpass filter is shown
in Fig. \ref{schemecoh}.

In closing, it is interesting to note that in the incoherent case
one sums up intensities from single particle contributions. For
example, in the case of the usual incoherent OTR setup in Fig.
\ref{incotr} (with no image processing), the energy per unit
surface in the image plane is given by:

\begin{eqnarray}
\frac{d W^\mathrm{incoh}_\mathrm{det}}{d S}(\vec{r}_i) \propto
\int_{-\infty}^{\infty} d \omega   \int \d \vec{r'} \bar{\rho}(0,
\vec{r'}-\vec{r}_i) \left| A_p(\omega, r') \right|^2~,
\label{INTBA2}
\end{eqnarray}
which already presents a linear projection of the charge density
distribution. From Eq. (\ref{INTBA2}) we see that, in the
incoherent case, the spectral energy density per unit surface is
proportional to $\bar{\rho}(0, \vec{r}_i)$ (neglecting the
blurring), i.e. to the Fourier transform of the charge density
distribution at zero frequency. This follows from an average over
ensembles. 
In Eq. (\ref{INTBA2}), $\bar{\rho}(0, \vec{r}_i)$ may also be
presented as $\bar{\rho}(0, \vec{r}_i)=\int \rho(t, \vec{r}) dt$.
Then, based on Eq. (\ref{duerho0}), which is valid in general, due
to adiabatic approximation we have $\int \rho(t,\vec{r})dt =\int
\rho_0(t,\vec{r})dt$, for arbitrary modulation strength
$a_f(t,\vec{r})$.

\subsection{\label{sub:abee}Effect of aberrations on the performance of coherent optical systems}

When an optical system acts as a linear, space-invariant filter
for a complex field amplitude, it is referred to as
isoplanatic\footnote{Note that in \cite{TICH}, conditions for
isoplanatism are discussed for a simple case of an aberration-free
thin lens with a finite extent. We did not find, in literature,
any discussion about conditions for isoplanatism for lenses with
aberrations and apodization. }. When a coherent optical system is
isoplanatic, the effect of aberrations on its performance can be
treated within the framework of a standard theory \cite{MILL}.
However, in general, even aberration-free optical systems are not
isoplanatic. Some spatial phase distortion is unavoidably
introduced, which severely modifies the intensity distribution of
the image.

In section \ref{sec:imager} we discussed the effect of
non-isoplanatism on coherent image formation.  We demonstrated
that thin aberration-free lenses can only be considered as
isoplanatic system if limitations (\ref{bothc}) are satisfied.
Here we discuss the effect of non-isoplanatism on coherent imaging
of extended objects when the thin lenses composing the optical
system are no more aberration-free, and we discuss conditions when
the standard isoplanatic theory can be applied.

\begin{figure}
\begin{center}
\includegraphics*[width=140mm]{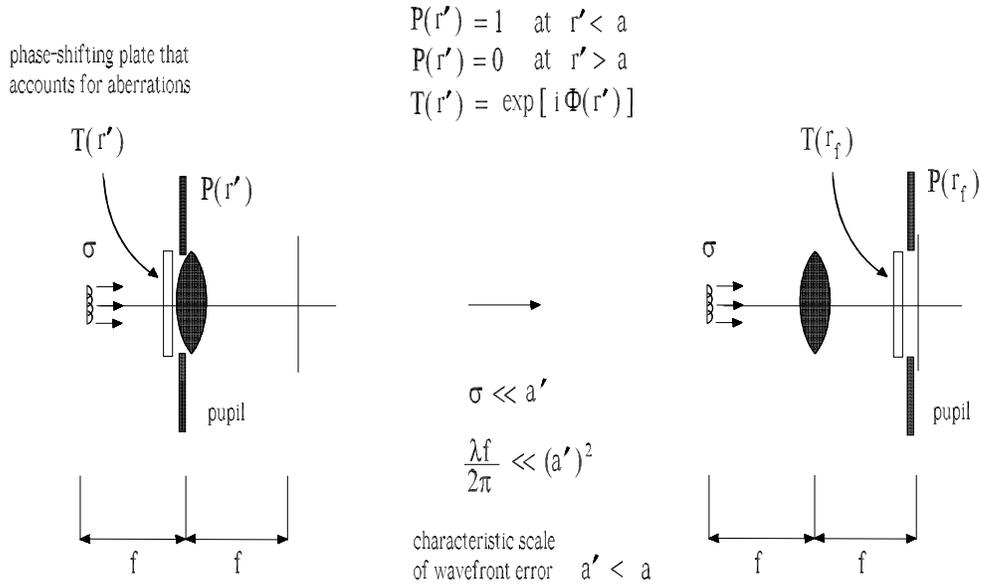}
\caption{\label{Aberration} Isoplanatic system in case of
aberrations. When wavefront errors exist, we can imagine that a
phase-shifting plate is placed at the aperture, thus deforming the
wavefront. When the characteristic scale of the wavefront errors
$a'$ is much larger than a single Fresnel zone, and the object
scale $\sigma$ is much smaller than $a'$, the optical system with
aberrations is isoplanatic. The amplitude transfer function
concept can be applied to such system and the generalized pupil
function plays the role of the amplitude transfer function. In
this case, left and right configurations are equivalent.}
\end{center}
\end{figure}
Consider Fig. \ref{Aberration}. Aberrations, i.e. wavefront
errors, can be imagined as equivalent to a phase-shifting plate
located at the aperture, which deforms the wavefront. The complex
amplitude transmittance $T(\vec{r}')$ of the phase-shifting plate
is given by

\begin{eqnarray}
T(\vec{r}') = \exp[i \Phi(\vec{r}')] ~,\label{trabe}
\end{eqnarray}
where $\Phi(\vec{r}')$ characterizes the aberration. As before, we
indicate with $P(r')$ the pupil function characterizing the lens
aperture. The complex function $P(r')T(\vec{r}')$ is referred to
as the generalized pupil function.

When an aberrated optical system is isoplanatic, the amplitude
point spread function of the system is space-invariant, and is
simply given by the Fraunhofer diffraction pattern of the
generalized pupil function $P(r')T(\vec{r}')$, which is
essentially its Fourier transform. The generalized pupil function
itself plays the role of the amplitude transfer function for the
system. In this case, the spatial bandwidth of the amplitude
transfer function is limited by the finite pupil aperture $P(r')$.
The effect of aberrations consists in the introduction of phase
distortions within the bandwidth of the amplitude transfer
function. These phase distortions can have a severe effect on the
fidelity of the imaging system.

Conditions for an aberrated optical system to be isoplanatic are
easily derived on the basis of Section \ref{sub:imsi}. Let us call
with $a'$ the characteristic scale of the wavefront error. Suppose
now that $a'$ is much larger than a single Fresnel zone, and that
the object scale $\sigma$ is much smaller than $a'$. In this case,
the reasoning in Section \ref{sub:imsi} can be repeated replacing
conditions (\ref{bothc}) with

\begin{eqnarray}
\sigma \ll a'~,~~~~ a'^2/(\lambdabar f) \gg 1~. \label{bothc2}
\end{eqnarray}
Then, both Eq. (\ref{Fplane}) and Eq. (\ref{Iplane3}) can be
recovered, and the optical system is isoplanatic.

More generally, geometrical aberrations are described by the
Siedel wavefront representation \cite{BORN}:

\begin{eqnarray}
\Phi(r',\theta) = b_{20} \left(\frac{r'}{a}\right)^2 + b_{40}
\left(\frac{r'}{a}\right)^4 + b_{31} \left(\frac{r'}{a}\right)^3
\cos(\theta) + b_{22} \left(\frac{r'}{a}\right)^2
\cos^2(\theta)~,\label{siedel}
\end{eqnarray}
where $b_{20}$, $b_{40}$, $b_{31}$ and $b_{22}$ are the amounts of
defocus, spherical aberration, coma, and astigmatism. The
quantities $r'$ and $\theta$ are defined through

\begin{eqnarray}
r'^2 \equiv (x')^2+(y')^2~,~~~~~ \theta \equiv \tan^{-1}(x'/y')~.
\label{uao}
\end{eqnarray}
For the sake of illustration, it is interesting to discuss a
particular example for the case of defocusing aberrations. In this
case it is customary to introduce a quadratic phase error
$\Phi(r')$ in the form

\begin{eqnarray}
\Phi(r') = b_{20} \left(\frac{r'}{a}\right)^2 ~,\label{Phir2}
\end{eqnarray}
In Eq. (\ref{Phir2}), the characteristic scale $a'$ is linked to
the aberration strength by $a'^2 \sim a^2/b_{20}$. Therefore, from
the second condition in (\ref{bothc2}), one has $b_{20} \ll
a^2/(\lambdabar f)$, which means that $b_{20}$ should be much
smaller than the number of Fresnel zones on the aperture.
Similarly, from the first condition in (\ref{bothc2}) one has
$b_{20} \ll a^2/\sigma^2$, meaning that the aberration strength
should be much smaller than the ratio between the aperture area
and the object area too. Note that these conditions actually set
criteria for the system to be isoplanatic in terms of aberration
strength, diffraction, and geometrical dimension of the source
compared with the pupil aperture $a$. Finally, note that in
general, for an aberration of order $n$, the aberration strength
should satisfy the conditions $b_{nm}^{2/n}\ll a^2/(\lambdabar f)$
and $b_{nm}^{2/n}\ll a^2/\sigma^2$.

In our case of coherent OTR imaging, the impact of chromatic
aberrations on the system resolution can be avoided by taking
advantage of the narrow bandwidth of the optical replica radiation
pulse.

\subsection{Sensitivity to displacement of mask and polarization transformer away from the focal
plane}

Up to now we discussed two cases:  a pupil consisting of a finite
aperture only, and a pupil consisting of a finite aperture with
aberrations.

In the case of a pupil without aberrations we saw, Fig.
\ref{sumup}, that conditions (\ref{bothc})  guarantee that the
system is isoplanatic. The field in the Fourier plane is the
product of the Fourier transform of the field at the object plane
and the pupil function (see Eq. (\ref{Fplane2})). Based on Eq.
(\ref{Fplane2}), we demonstrated that the position of the aperture
can be located anywhere between lens and focal plane.
%

\begin{figure}
\begin{center}
\includegraphics*[width=120mm]{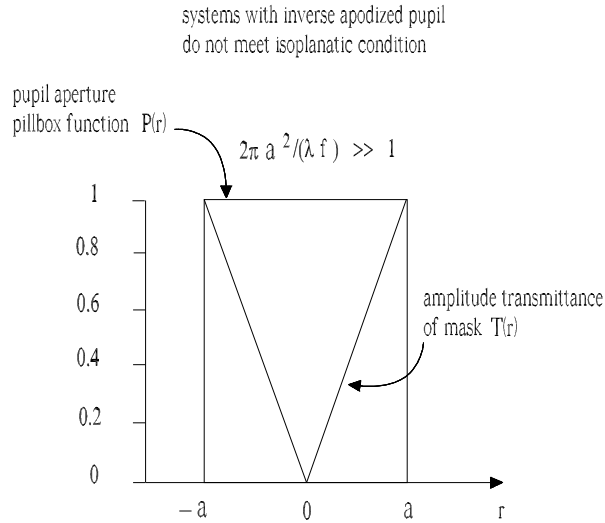}
\caption{\label{DOF1} Introduction of a mask in the pupil. The
problem is related to the so called inverse apodization. In this
case the generalized pupil function $P(r)T_m(r)$ has no
characteristic size and, generally speaking, the optical system is
non-isoplanatic. For coherent imaging of extended object some
spatial phase distortions is introduced, which severely modifies
the low (spatial) frequency components of the object spectrum.}
\end{center}
\end{figure}
This reasoning was extended to the case of aberrations in the
previous Section \ref{sub:abee}. There we saw, Fig.
\ref{Aberration}, that the same conclusions can be drawn whenever
conditions (\ref{bothc2}) hold, which have the same mathematical
structure of conditions (\ref{bothc}), with the only difference
that they now include the characteristic scale of the wavefront
error $a'$ and not the size of the pupil aperture $a$.

Both the problems that we discussed are related to coherent image
formation theory. This knowledge can be applied to treat related
technical issues.

The goal of this subsection is to deal with the problem of
alignment of optical elements along the optical axis. As was shown
in Section \ref{sub:lasttt}, Fig. \ref{sumup}, a lowpass spatial
frequency filter under conditions (\ref{bothc}) is insensitive to
the position of the filter (which can be located anywhere between
lens and focal plane positions).

It is now a good time to generalize our considerations, and to
examine the sensitivity to displacement of a Fourier mask of
particular interest. The mask that we want to consider has a
profile given in Eq. (\ref{TMMM}), Fig. \ref{DOF1}. This profile
changes the shape of the optical transmission of the optical
system. Such changing is known in literature as apodization. Our
case is actually known as inverse apodization (see e.g.
\cite{GOOD}) because it suppresses the central part, and not the
periphery of the optical signal.


Considerations made above will lead us to find isoplanatic
conditions for a pupil with inverse apodization. The pupil
function in this case given by the product $P(r)T_m(r)$.

We start with Eq. (\ref{Fplane}) assuming that the second of
conditions (\ref{bothc}) is satisfied. This time, we analyze Eq.
(\ref{Fplane}) in a different way.  Expressing ${F}(0, \vec{u})$
in terms of a Fourier integral, neglecting the quadratic phase
factor in $u^2$ because of the second of conditions (\ref{bothc}),
and integrating in $d \vec{u}$ we obtain

\begin{eqnarray}
{\widetilde{E}}\left(2 f,\vec{r}_f\right) = \frac{i\omega}{2\pi c
f} \int d \vec{r'} ~\widetilde{E}\left(0, \vec{r'}\right)
P(\vec{r}_f-\vec{r'}) T_m(\vec{r}_f-\vec{r'}) \exp\left[-\frac{i
\omega \vec{r'}\cdot \vec{r}_f}{c f}\right]~.\cr
&&\label{Fplaneee}
\end{eqnarray}
Now $\widetilde{E}\left(0, \vec{r'}\right)$ is our object, whose
extension we estimate with the characteristic size $\sigma$.

We begin by analyzing the case without mask, i.e. $T_m(r) = 1$,
using Eq. (\ref{Fplaneee}). If also the first of conditions
(\ref{bothc}) is verified, $\sigma \ll a$, then we can have
$\sigma^2/(\lambdabar f) \sim 1$, meaning that the pupil is in the
near zone. Since $r' \lesssim \sigma$ (otherwise the field at
$z=0$ under the integral vanishes), due to the exponential factor
in Eq. (\ref{Fplaneee}), one is actually limited to the range $r_f
\lesssim \sigma$. Then, $P(\vec{r_f}-\vec{r'}) = 1$, because
$|\vec{r_f}-\vec{r'}| \lesssim \sigma \ll a$. Therefore, the pupil
function $P$ can be taken out of the integral sign in Eq.
(\ref{Fplaneee}), meaning that it does not perturb the image of
the object. In the case $\sigma^2/(\lambdabar f) \ll 1$, since $r'
\lesssim \sigma$, the exponential term in Eq. (\ref{Fplaneee}) is
of order unity for maximal values of $r_f \sim \lambdabar f/\sigma
\gg \sigma$. Now the pupil operates as a cutoff filter,
introducing a little blurring of the sharp edge on a scale $\sigma
\ll a$.

Now let us turn back to the case a mask with inverse apodization
is present. We can appreciate the difference compared to the
previous case. When $\sigma^2/(\lambdabar f) \sim 1$ we cannot
take the generalized pupil function $P(r) T_m(r)$ out of the
integral sign anymore, because on the scale $r_f \sim \sigma$
(determined by imposing that the exponential term in Eq.
(\ref{Fplaneee}) be at most of order unity) we have significant
variation of the generalized pupil function through $T_m(r)$,
which does not present any characteristic scale of interest. When
instead $\sigma^2/(\lambdabar f) \ll 1$, we have the
characteristic scale $r_f \sim \lambdabar f/\sigma \gg \sigma$.
Then, with accuracy $\sigma^2/(\lambdabar f)$ we can take the
generalized pupil out of the integral sign, because for these
large values of $r_f$ the result of the integration changes little
within excursions of $r' \lesssim \sigma$.

A qualitatively different result is reached, because low
frequencies are always perturbed, while in the case of a simple
pupil $P(\sigma) =1$ for $\sigma \ll a$. It is therefore
impossible to apply the concept of amplitude transfer function to
the system in Fig. \ref{DOF1}. In other words, the system is not
isoplanatic.

Note that inverse apodization is discussed in literature only in
relation with incoherent imaging. If we were imaging an incoherent
collection of point sources the problem of isoplanatism would not
exist at all, and the autocorrelation function of the pupil
actually would serve as transfer function of the system. However,
this observation applies to an object consisting of an incoherent
collection of point sources only, and not to the case of
incoherent OTR imaging\footnote{This example is not unique, think
e.g. to the case of Synchrotron Radiation sources. However, the
OTR case is more complicated due to the polarization singularity
of the source.}, where we deal with an incoherent collection of
laser-like sources. In this case, each source has a finite
transverse extent, and the problem of isoplanatism exists exactly
as in the coherent case.

\begin{figure}
\begin{center}
\includegraphics*[width=120mm]{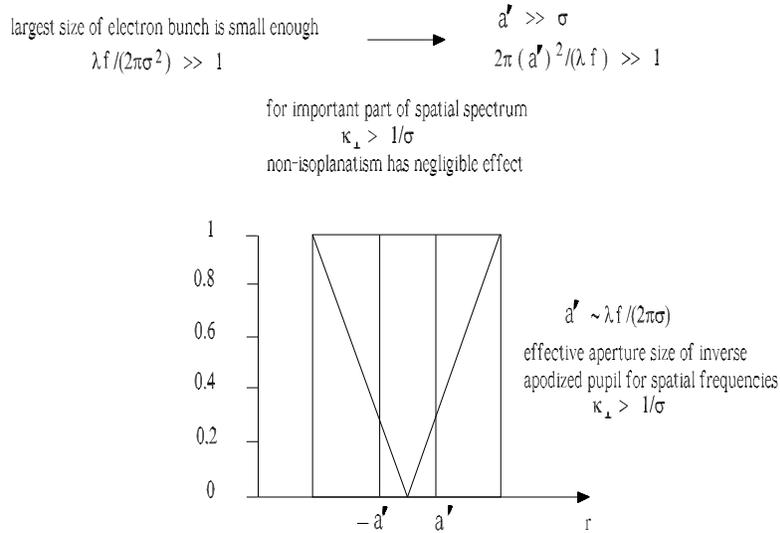}
\caption{\label{DOF2} In our case of interest, the largest
transverse electron-bunch size is small enough for its Fraunhofer
diffraction pattern to be formed at the pupil plane. As a result,
the most important part of the object spectrum  (mid-range and
high spatial frequencies) meet isoplanatic condition. For this
range the generalized pupil function plays the role of amplitude
transfer function. Perturbations of low frequency components of
the object spatial spectrum can only modify the intensity
distribution on a scale of much larger size compared to the bunch
size.}
\end{center}
\end{figure}
\begin{figure}
\begin{center}
\includegraphics*[width=110mm]{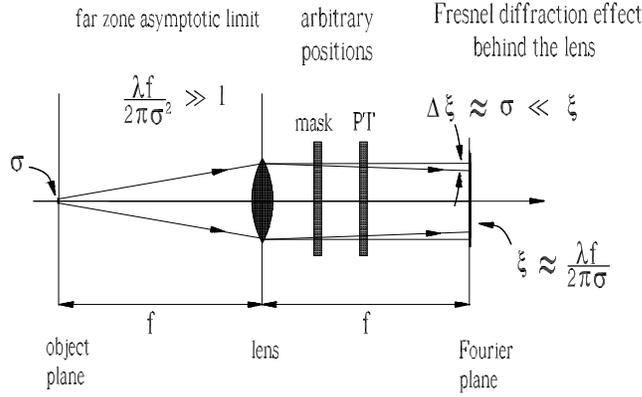}
\caption{\label{DOF} When the transverse size of the electron
bunch is small enough for its Fraunhofer diffraction pattern to be
formed at the pupil plane, mask and polarization transformer can
be installed at arbitrary position within the focal distance.}
\end{center}
\end{figure}
Clearly, as soon as we consider inverse apodization for coherent
imaging, the situation becomes complicated. Nevertheless the
isoplanatic conditions (\ref{bothc2}) can be satisfied for
mid-range and high spatial frequencies of the object spectrum.
Fortunately, in the XFEL case perturbations of the low spatial
frequency components lead to the introduction of some background
only, but not distortions in the electron-bunch image (see Fig.
\ref{FTOTR2}).

Considering the specifics of our problems as illustrated in Fig.
\ref{DOF2}, we are interested in spatial frequencies higher than
$1/\sigma$, where $\sigma$ is the largest size of interest in our
object. Thus, we are not interested in spatial frequencies lower
than $1/\sigma$, which are modified by a shift in the mask
position. This reasoning yields an effective scale $a'$ on the
inverse apodized pupil of order $a' \sim \lambdabar f/\sigma$.
Such scale can be used in conditions (\ref{bothc2}), which are now
equivalent to

\begin{eqnarray}
\frac{\sigma^2}{\lambdabar f} \ll 1~. \label{bothc3}
\end{eqnarray}
When condition (\ref{bothc3}) is satisfied, i.e. when the lens is
in the far zone with respect to the object structure of size
$\sigma$, conditions (\ref{bothc2}) are satisfied too, and one can
shift the mask from the focal plane without appreciable effects.

The far-zone condition (\ref{bothc3}) means that the transverse
size of the object is much smaller than the size of the Fourier
pattern. Intuitively, as it can be seen from Fig. \ref{DOF}, when
placed in the far-zone, the lens transforms a spherical wavefront
from a point source into a plane wave. When the object has some
transverse dimension $\sigma$, such that $\sigma^2/\lambdabar f
\ll 1$, one can neglect Fresnel diffraction effects at the mask
placed anywhere between pupil and Fourier plane. Thus, with
accuracy $\sigma^2/\lambdabar f$, the wavefront behind the lens
can be still considered plane and does not change at any position
between lens and focal plane.

As a result, in the far zone, there is a very low sensitivity to
mask and polarization-transformer displacement from the focal
plane.

\subsection{Sensitivity to transverse displacement for mask and Polarization Transformer}

With the help of a polarization transformer and a Fourier-plane
mask, as schematically illustrated in Fig. \ref{4finte}, we have
seen in the previous Section that spatial spectrum and
polarization of OTR radiation can be conveniently processed.

\begin{figure}
\begin{center}
\includegraphics*[width=120mm]{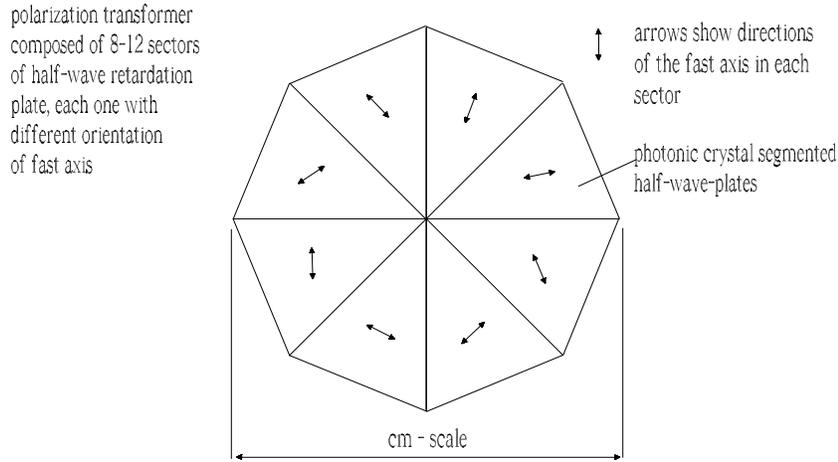}
\caption{\label{trasf2} Segmented half-wave plate with octonary sectors. Arrows schematically show the direction of the fast axis of the crystal in each sector.} 
\end{center}
\end{figure}

A schematic representation of the polarization transformer is
shown in Fig. \ref{trasf2}.  The polarization transformer
demonstrated in \cite{POLA,POL2} is composed of eight \cite{POLA}
up to twelve \cite{POL2} sectors of half-wave retardation plates,
each one with different orientation of the crystal's "fast" axis.
The linear dimension of the device is of the order of a
centimeter. A photonic crystal segmented half-wave-plate is the
perfect radial-to-linear polarization converter for coherent OTR
imager. For a twelve segments device, the linear polarization
purity can be as high as $99\%$, with transformation efficiency of
about $92\%$. Due to the good quality of the photonic crystal
segmented half-wave-plates, the incident wavefront is negligibly
perturbed.

%
Both mask and polarization transformer are endowed with rotational
invariance. Their centers of rotation should coincide with the
position of the optical axis, where the center of the Fourier
transform is located. The accuracy needed in the alignment of the
Fourier-plane mask and the polarization transformer is related to
the fact that we wish to obtain a proper image within region B in
Fig. \ref{halo}, i.e. $r_i \lesssim 100~\mu$m in the image plane
for a focal distance $f \simeq 30$ cm. This maximal value of $r_i$
corresponds to a region in the focal plane given by $r_f \gtrsim
\lambdabar f/r_i \simeq 400~\mu$m. Within this accuracy, the
electron bunch image will not be sensitive to transverse
displacements of the center of the mask in the Fourier-plane and
of the polarization transformer with respect to the optical axis.
This estimation also specifies the required manufacturing
accuracy, close to the center of rotation, of the polarization
transformer and of the mask in the Fourier-plane, which should be
in the range of about $100~\mu$m.


\subsection{\label{sub:SN}Signal-to-noise Ratio}

The signal-to-noise ratio of our system is strictly related to the
practical resolution achievable. In fact, one has to make sure
that the number of photons available are enough to allow
implementation of the imaging scheme described above.

As an example, let us consider the case when the detector is a CCD
camera. Supposing that the CCD charge capacity is about $2 \div 4
\cdot 10^4$ electrons per pixel, and assuming a quantum efficiency
of about $50 \%$, one concludes that the CCD camera will saturate
at about $10^5$ photons per pixel or more.

For our particular case of coherent OTR imaging, the estimated
resolution according to the Sparrow criterion is about $6 ~\mu$m.
Because of the Nyquist sampling theorem, the CCD pixel size should
be about $3~\mu$m. A typical CCD pixel size is about $10 \times 10
~\mu\mathrm{m}^2$. Thus, to resolve $6 ~\mu$m features in the
electron bunch, the magnification of the optical system should be
about $1:3$ or more. Choosing for simplicity the magnification $1
: 3$, our zone of interest (region B in Fig. \ref{halo}) will
result in an image size of $300 \times 300 ~\mu\mathrm{m}^2$ in
the image (detector) plane, i.e. $30 \times 30$ pixels. Similarly,
the electron bunch density distribution (region A in Fig.
\ref{halo}) of order of $30 ~\mu$m is imaged into an area of about
$100 \times 100 ~\mu\mathrm{m}^2$, i.e. $10 \times 10$ pixels.

Assuming $10^{12}$ photons in the region of interest of the
imaging system (region B in Fig. \ref{halo}),  we obtain about
$10^9$ photons per pixel. The difference of four orders of
magnitude beyond the saturation level of the CCD can be used to
implement the particle-spread function improvement described in
the previous Section, which may be quite expensive in terms of
photon budget.

Obviously, in any case, we should make sure that no pixel on the
CCD camera gets more than $10^5$ photons. From Poisson statistics
follows that the shot-noise level at saturation is about a
fraction of a percent of the signal, i.e. the signal-to-noise
ratio is a few hundreds. Shot noise is the dominant cause of noise
at high photon-level conditions. However, while the signal
decreases, other sources of noise become important. As the photon
level drops, also the shot noise level diminishes, and the CCD
camera tends to become read-out-noise limited. Assuming a read-out
noise e.g. of order of $10$ electrons, a certain pixel operates in
read-out-noise limited regime when the number of photons incident
on that pixel is of order $10^2$ or less. At this photon level the
shot-noise and the read-out noise are comparable. The
signal-to-noise ratio is of order $10$, and quickly decreases as
the number of photons decreases. The dynamical range of our device
is therefore of about $4$ orders of magnitude, i.e. the ratio
between the full-well capacity and the read-out noise.


\section{\label{sec:diffite}3D electron-bunch imaging from diffraction intensity measurements by use of iterative phase-retrieval methods}

Up to now we proposed to process the particle spread function of
the OTR imager in order to improve the overall resolution of the
system. We proposed to use coherently generated radiation but,
from a fundamental viewpoint, techniques described in the previous
Section maybe implemented for incoherent radiation as well. The
only problem using incoherent radiation is linked with the smaller
number of available photons, compared to the coherent case, which
makes the discussed technique unpractical. In other words, up to
now we took advantage of the coherent nature of OTR in the sense
that we exploited the large number of available photons.

The fact that we are dealing with coherent radiation, however,
opens up new imaging possibility. Namely, the detector can be
directly installed in the Fourier plane, and imaging from
diffraction intensity measurements can be enforced. Once the
Fourier transform of the object is known, the recovery of the
image is reduced to the calculation of an inverse Fourier
transform. However, one can only measure the modulus of the
Fourier transform, and a phase-retrieval problem arises. With
their algorithm \cite{GERC}, Gerchberg and Saxton demonstrated
that numerical solutions to the two-dimensional phase retrieval
problem is possible. A number of phase-retrieval algorithms were
developed based on the original Gerchberg-Saxton algorithm. For
the interested reader we refer to extensive reviews on the subject
\cite{FINN}-\cite{LUKE}.

Two issues need to be mentioned, which may affect the usefulness
of the iterative phase-recovery method for our case. The first is
the problem of uniqueness of the reconstructed object, the second
is the speed of the algorithm, which may not be fast enough to
perform coherent OTR imaging at an acceptable repetition rate.

In the following we will study two cases when object and
propagated field are related by a Fourier transform. In the first
case, the detector is placed in the focal plane of a lens. In the
second, it is placed in the far (Fraunhofer) zone of the
propagated field. The lens setup allows one to control the image
parameters, but it can potentially introduce aberrations. The
lensless setup instead, is defined by the position of the detector
only.

\subsection{\label{sub:HR}Lens-based setup for diffractive imaging}

\subsubsection{Deconvolution by Fourier-plane mask}

\begin{figure}
\begin{center}
\includegraphics*[width=120mm]{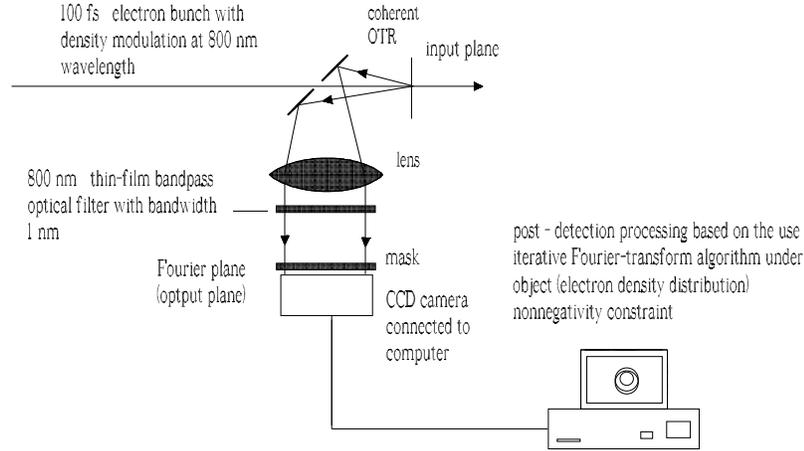}
\caption{\label{recon}  Lens-based setup for diffractive imaging
of an electron bunch.}
\end{center}
\end{figure}
Let us consider the setup in Fig. \ref{recon}. A lens is used to
obtain the Fourier transform of the object in the focal plane.
Once the squared modulus of the Fourier transform of the object is
recorded, the object can be reconstructed with the help of an
iterative phase-retrieval algorithm.

Let the object be placed in the front focal plane of the
lens\footnote{If not, a phase correction needs to be accounted
for.}. The field across the back focal plane is given by Eq.
(\ref{Fplane2}), where we neglect lens imperfections, and assume
that conditions (\ref{bothc}) are satisfied. By means of the
convolution theorem, the (spatial) Fourier transform of the field
in the object plane, which enters Eq. (\ref{Fplane2}), is given by
the product of the Fourier transform of the single-particle field
and the Fourier transform of the charge density distribution,
$\varrho(\omega ,\vec{u})$. Note that here we are discussing a 3D
Fourier transform, both with respect to time and space. In other
words, $\varrho(\omega ,\vec{u})$ is the 3D Fourier transform of
$\rho(t,\vec{r})$:

\begin{eqnarray}
\varrho(\omega, \vec{u}~)= \int d \vec{r} d
t~\rho(t,\vec{r})\exp[i \vec{r}\cdot \vec{u}~]\exp[i\omega t]~.
\label{varrhorhodef}
\end{eqnarray}
Accounting for the vectorial nature of the OTR field, Eq.
(\ref{Fplane2}) can be written as

\begin{eqnarray}
\vec{\widetilde{E}}^{}\left(2 f,\vec{r}_f\right) = \frac{2 N_e e
\gamma^2 \vec{r}_f}{c(\gamma^2 r_f^2+f^2)} \varrho\left(\omega,
-\frac{\omega \vec{r}_f}{c f}\right) P(\vec{r}_f) ~.
\label{Fplane2mody0}
\end{eqnarray}
%

%

%

\begin{figure}
\begin{center}
\includegraphics*[width=140mm]{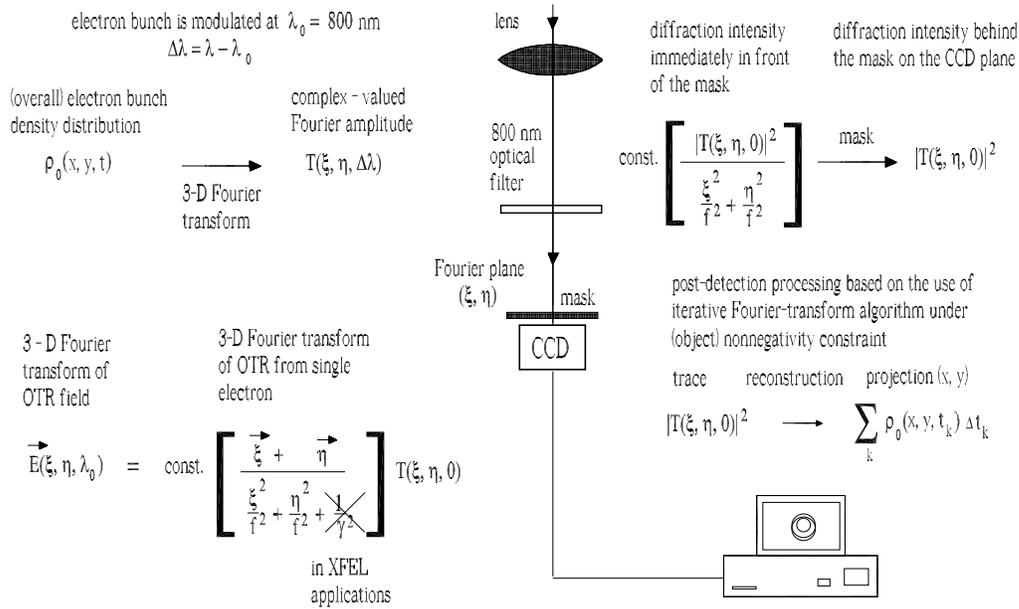}
\caption{\label{xy-proj} Reconstruction of an object from the
modulus of its Fourier transform by using an iterative
phase-retrieval method. For the electron-beam imager problem the
a-priory constraint is the object nonnegativity. For this support
phase retrieval is much easier and almost always unique.  The
method will open up the possibility of real-time,
wavelength-limited optical imaging of electron bunches.}
\end{center}
\end{figure}
\begin{figure}
\begin{center}
\includegraphics*[width=130mm]{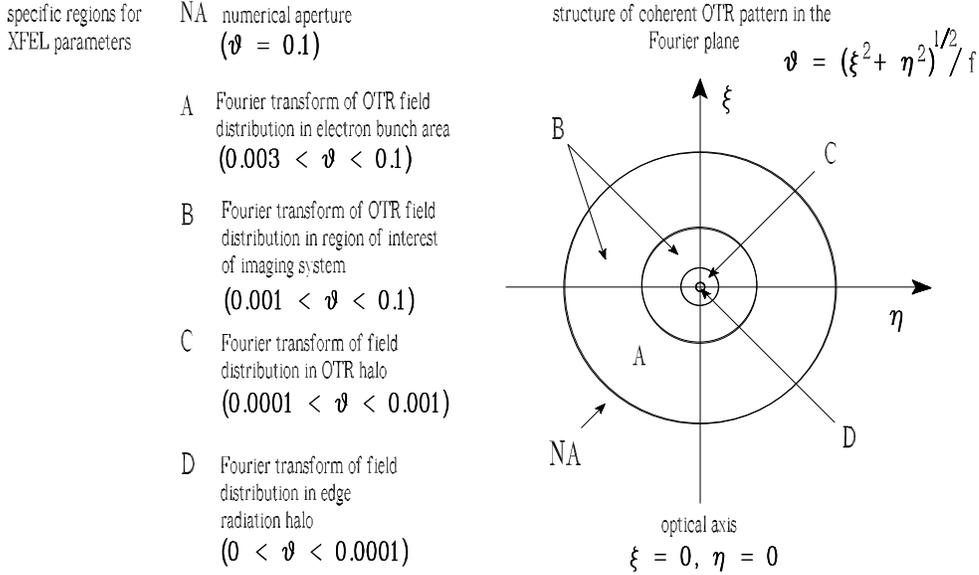}
\caption{\label{Zonerec} Structure of the coherent OTR radiation
pattern as observed in the Fourier plane. Parameters are as in
Fig. \ref{halo}.}
\end{center}
\end{figure}
The intensity in the Fourier plane is thus given by

\begin{eqnarray}
I(\omega,\vec{r}_f) &=& \frac{d^2 W}{d \omega d S}= \frac{c} {4
\pi^2}\left|\vec{\widetilde{E}}^{}\left(2
f,\vec{r}_f\right)\right|^2 \cr &=& \frac{N_e^2 e^2 \gamma^4
r_f^2}{\pi^2 c(\gamma^2 r_f^2+f^2)^2} \left|\varrho\left(\omega,
-\frac{\omega \vec{r}_f}{c f}\right)\right|^2
\left|P(\vec{r}_f)\right|^2 ~. \cr && \label{Fplane2mody}
\end{eqnarray}
The processing of the intensity is better explained in Fig.
\ref{xy-proj}, where the part of the scheme after the lens is
zoomed-in. The detector is installed in the Fourier plane.

The structure of the coherent OTR radiation pattern as observed in
the Fourier plane is shown in Fig. \ref{Zonerec}. We are
interested in the intensity distribution in the Fourier plane, in
contrast to the previously discussed cases when we were interested
in the intensity distribution in the image plane. The situation is
much simpler: in fact, Eq. (\ref{Fplane2mody}) consists of the
product of a single-particle factor, the squared electron bunch
structure factor, and the squared pupil function. This is due to
the fact that the polarization of the coherent OTR field in the
Fourier plane is radial\footnote{For this to be the case, we need
a uniform screen, i.e. our mirror has an aperture larger than
$\gamma \lambdabar$.}, with respect to the optical axis, exactly
as the polarization of a single-electron OTR field. This property
greatly simplifies the image processing, because we do not need a
polarization transformer to deconvolve the electron bunch
structure factor in the Fourier plane. 

Separation of the squared modulus of the spatial Fourier transform
of the charge density distribution from the squared modulus of the
spatial Fourier transform of the amplitude particle spread
function may be done at software level, by taking the ratio of the
experimental data with the (known) squared modulus of the spatial
Fourier transform of the amplitude particle spread function. An
alternative possibility is to insert a filter in the Fourier
plane, as shown in Fig. \ref{recon}. The mask to be inserted has
the same characteristics discussed in the previous Section
\ref{subsub:fpm}. Note that here we are considering a range of
$r_f$ such that $r_f/f \gg 1/\gamma$. This is our range of
interest because the electron beam dimension $\sigma_r \ll \gamma
\lambdabar$ (see Section \ref{sec:OTRsetup}) is reciprocally
related to the high spatial-frequency limit, so that the
relativistic factor $\gamma$ cancels out in Eq.
(\ref{Fplane2mody}).  Therefore, Eq. (\ref{Fplane2mody}) is
independent of $\gamma$, and the mask transmittance is given by
$T_m(r)$ in Eq. (\ref{TMMM}).

Note that here we implicitly assumed that a bandpass filter is
used, as discussed in the previous Section, to generate a linear
integral of the signal.  Since such filter is tuned at $\omega_m$,
the intensity in the detector plane behind the mask and the
bandpass filter is

\begin{eqnarray}
I(\omega_m,\vec{r}_f) &=& \mathrm{const} \cdot
\left|\varrho\left(\omega_m, -\frac{\omega_m \vec{r}_f}{c
f}\right)\right|^2 \left|P(\vec{r}_f)\right|^2 \cr &=&
\mathrm{const} \cdot a_f^2 \left|\varrho_0\left(0, -\frac{\omega_m
\vec{r}_f}{c f}\right)\right|^2 \left|P(\vec{r}_f)\right|^2 ~, \cr
&& \label{Fplane2mody2}
\end{eqnarray}
where (see Eq. \ref{duerho0}) $\varrho(\omega_m,\vec{u}) = a_f
\varrho_0(0,\vec{u}) = a_f \int_{-\infty}^{\infty} dt
\Phi(t,\vec{u})$, where $\Phi$ is the spatial Fourier transform of
$\rho_0(t, \vec{r})$. The main difference to be remarked here,
with respect to the discussion in the previous Section, is that
for diffractive imaging one needs a bandpass filter, while in the
previous Section one could choose between a linear or a quadratic
integral projection of the signal. In fact, without a filter, one
would record $\int d (\Delta \omega) |\varrho_0(\Delta
\omega,\vec{u)}|^2$, and it would be problematic to reconstruct
the structure of the electron bunch in real space. On the
contrary, once $|\varrho_0(0, {\omega_m \vec{r}_f}/({c f}))|^2$ is
measured, the solution of the phase retrieval problem yields back
$\bar{\rho}_0(0,\vec{r})=\int_{-\infty}^{\infty} dt
\rho_0(t,\vec{r})$. Note that the filter should have a bandwidth
smaller than the bandwidth of $\bar{\rho}_0(\Delta \omega,
\vec{r})$.

\subsubsection{Phase-retrieval problem}

Once the squared modulus of the spatial Fourier transform of the
electron-bunch projection along the time axis, $|\varrho_0(0,
\omega_m \vec{r}_f/({c f}))|^2$, is obtained, one deals with a
typical phase-retrieval problem. In fact, in order to reconstruct
the object one needs to know both amplitude and phase of the
diffraction field. However, the only directly-measurable quantity
in the Fourier plane is the intensity, which is proportional to
the squared modulus of the field. Determination of the missing
phase from the observed intensity is known as phase-retrieval. In
our case, $\varrho_0$ must be obtained from the knowledge of
$|\varrho_0|^2$. Only when the phase of $\varrho_0$ is known one
can retrieve the object $\bar{\rho}_0(0,\vec{r})$.

The phase retrieval is typically based on an Iterative Transform
Algorithm (ITA) \cite{GERC}, \cite{FINN}-\cite{LUKE}. This cycles
between real and reciprocal space, enforcing known constraints. In
the reciprocal space, the known constraint is, naturally, the
modulus of the Fourier transform of the object.

In the real space, some generic constraints are usually granted
too. First, any physical object is square-integrable. Moreover, in
real space, a finite support can be defined for any optical
system. The support is defined as the set of points over which the
object is nonzero, a finite support meaning that the object is
nonzero only within a finite region of space. The shape of the
object is usually required to be known in order to set the support
constraint in the ITA. This support is typically larger than the
actual boundary of the object, in which case it is said to be
loose. A method for finding an estimate of the loose support from
the support of the autocorrelation function may be used. In fact,
the autocorrelation of the object can be calculated from the
intensity distribution in the Fourier plane. Hence, the support of
the autocorrelation is known. Then, from the support of the
autocorrelation one can determine upper bounds on the support of
the object, i.e. the loose support.

Once the loose support is constructed, the initial phase input to
the iterative algorithm can be a random phase set. Iteration by
iteration, non-zero electron density outside the finite support is
gradually pushed to zero. In the reciprocal space, the measured
magnitude of the Fourier transform of the object is enforced at
each iteration.

Since only the modulus of the Fourier transform is measured, the
uniqueness of the solution of the ITA is a central question. When
the object is complex, reconstruction is not trivial. Although,
usually, the ITA gives good results, the uniqueness of the
solution is not granted. However, when an object is endowed with
the property of nonnegativity, the application of iterative
methods is drastically simplified.

Note that when our object is a temporal projection of $\rho_0$,
i.e. $\bar{\rho}_0(0,\vec{r}) = \int_{-\infty}^{\infty} dt
\rho_0(t,\vec{r})$, we are dealing with a positive object, because
$\rho_0$ is obviously a real positive function.

Besides the non-zero electron density outside the finite support,
also the negative electron-density inside the support is pushed to
zero, iteration after iteration. 
The ITA will easily converge to the original object in the case of
a two-dimensional phase-retrieval problem for positive images.


\begin{figure}
\begin{center}
\includegraphics*[width=140mm]{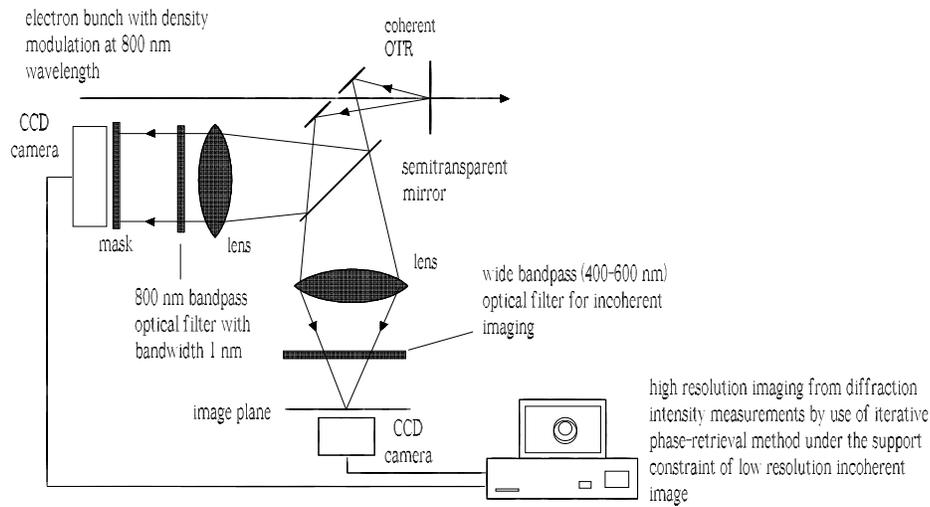}
\caption{\label{recon2}  Reconstruction of an object from the
modulus of its Fourier transform using the phase-retrieval method.
The object domain constraint is a low-resolution optical image.
Under such support, the method will open up the possibility of
real-time, wavelength-limited imaging of electron bunches.}
\end{center}
\end{figure}
When more a-priori information about the object is known,
reconstruction becomes easier. In case a low-resolution image of
the object is available, that image can be used as a constraint
for reconstruction, and allows the ITA \cite{FIN2} to work much
faster.

This suggests the idea of combining, as illustrated in Fig.
\ref{recon2}, incoherent OTR imaging and diffractive imaging in a
single technique. One may register, as before, the squared modulus
of the Fourier transform of the charge density distribution with a
coherent imager while, at the same time, a low-resolution image of
the bunch may be taken with an incoherent imager. Then, the
low-resolution image can be used as a-priori information in the
reconstruction process, allowing real-time, diffraction-limited
resolution.


\subsection{Lensless setup for diffractive imaging}

An alternative application of diffractive imaging is a lensless
setup. Actually, lensless imaging is one of the most promising
techniques for microscale imaging of an electron bunch. The
high-momentum vectors available in the Fraunhoffer diffraction
plane can yield, in principle, wavelength-limited resolution
without limitations which apply to lenses.

\begin{figure}
\begin{center}
\includegraphics*[width=110mm]{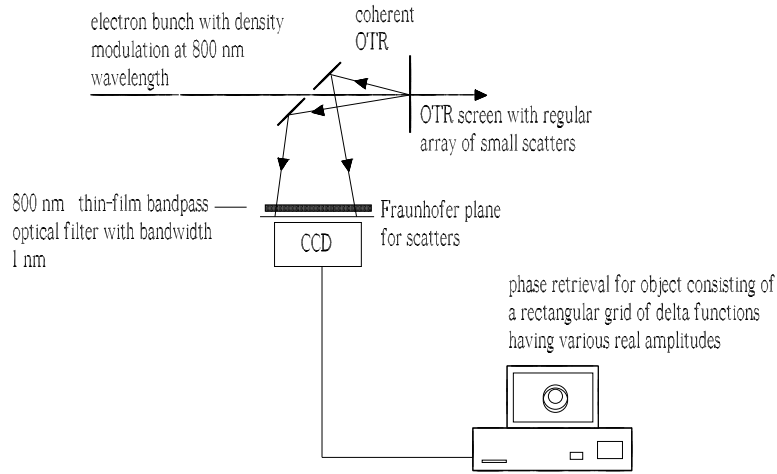}
\caption{\label{diffim1} Simplest lensless setup for diffractive
imaging of the electron bunch when the detector is placed in the
Fraunhofer diffraction plane. }
\end{center}
\end{figure}

\begin{figure}
\begin{center}
\includegraphics*[width=110mm]{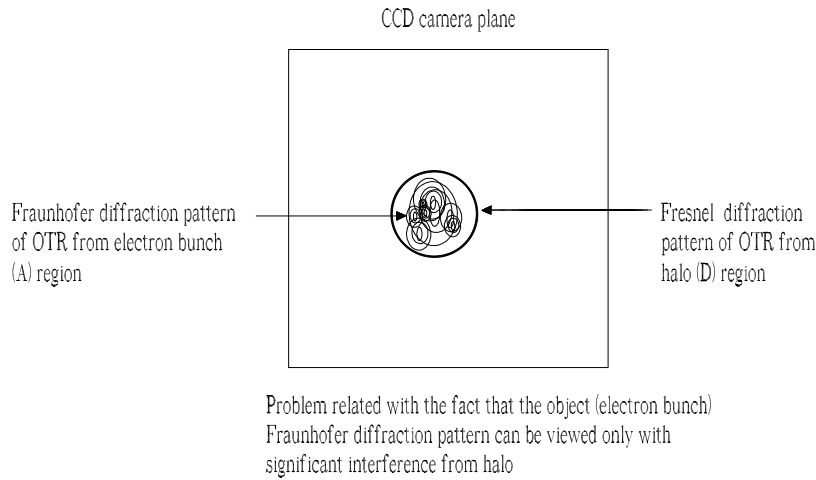}
\caption{\label{diffim2r} Radiation intensity pattern in the
detector plane for a lensless diffractive imaging setup with a
metallic mirror as an OTR screen. The problem is related to the
fact that the OTR halo is not small enough for its Fraunhofer
diffraction pattern to be formed in the detector plane. As a
result, the Fraunhofer diffraction pattern of the object can only
be viewed with significant interference from the OTR halo.}
\end{center}
\end{figure}
\begin{figure}
\begin{center}
\includegraphics*[width=150mm]{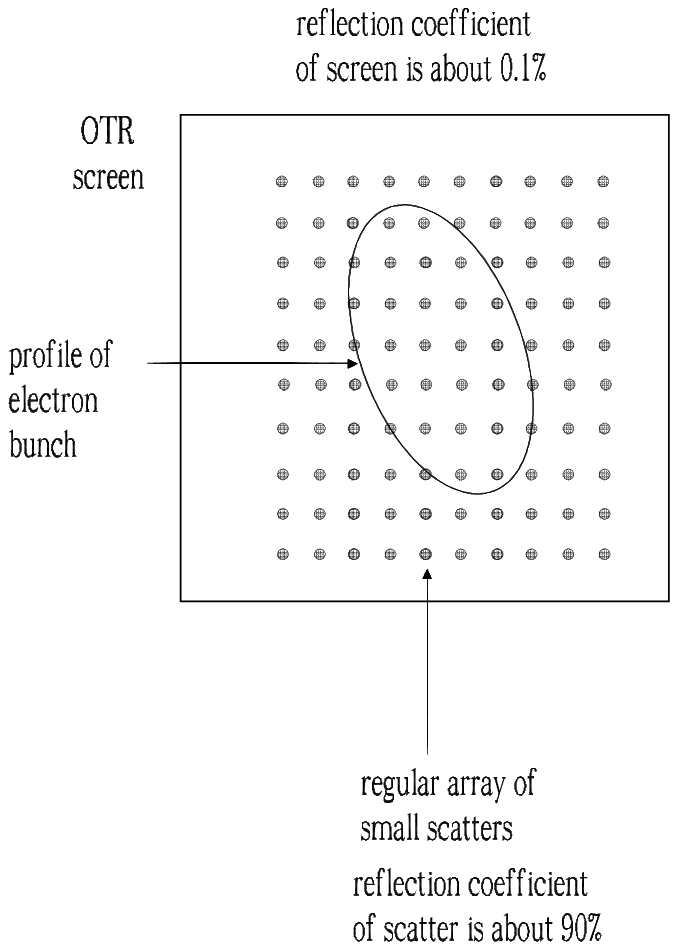}
\caption{\label{diffim2l} Image processing with an object plane
mask for lensless diffractive imaging setup as possible solution
of OTR halo problem. }
\end{center}
\end{figure}

\subsubsection{Validity of the Fraunhoffer approximation}

In Fig. \ref{diffim1} we describe a setup where one takes
advantage of free-space propagation of the electric field to the
object plane. Ideally, detection should be performed in the
Fraunhofer diffraction region. However, in practical cases, the
diffraction pattern from the charge density distribution overlaps
with the OTR halo, whose diffraction pattern may still be in the
near zone, at the position of the detector plane.

With reference to Fig. \ref{halo}, specific regions on the OTR
screen are related to characteristic scales from tens of microns
up to a millimeter. The far zone condition reads as
$(\lambdabar/\sigma^2) z \gg 1$, where $\sigma$ is the
characteristic object size and $z$ the position of the observation
plane. For small features of $\sigma \simeq 10 ~\mu$m in region A,
within the electron density distribution, the far-zone condition
requires $z \gg 1$ mm, but for the largest features in the halo
region D we have $\sigma \simeq \gamma \lambdabar$, and for
$E=17.5$ GeV one needs $z \gg 100$ m to reach the far zone, which
is obviously unfeasible. In other words, when one deals with the
diffraction pattern from region D, the observation plane is in the
very near zone.

This fact leads to overlapping of the Fraunhofer diffraction
patterns of regions A and D, as shown in Fig. \ref{diffim2r}. In
fact, consider a characteristic value $\sigma \sim 1$ mm for the
very near zone, and $\sigma= 10 ~\mu$m for the far zone. For the
very near zone field, the size at the observation plane in the
order of $1$ mm (basically, the unvaried value of $\sigma$). For
the far zone field one may estimate that the size in the detector
plane is the coherent angle $\lambdabar/\sigma \sim 10^{-2}$
multiplied a typical value $z \sim 10$ cm. As a result, the size
of the far-zone field at the observation plane is also about $1$
mm, hence the superposition of the two diffraction patterns, which
makes difficult to practically realize the lensless
diffractive-imaging setup.

A straightforward way to overcome this problem is to manipulate
the distribution of the scattered field. Actually, the problem is
solved when the OTR halo is suppressed. 
A square lattice of high-reflectivity metal disks with lattice
parameter $a$ can be formed on the OTR screen. The array should
cover the region of interest only (region B in Fig. \ref{halo}),
which is smaller than the halo region. In this way, the electron
bunch is practically sampled by the lattice of disks. The lattice
parameter should be small enough, that when an electron bunch
impinges on the screen one has $d \ll a \ll \sigma_r$, $\sigma_r$
being the transverse rms size of the electron bunch, and $d$ the
disk diameter. For example, an array of small disks of about
$d=1~\mu$m diameter, with separation of about $a = 5~\mu$m may be
chosen, as indicated in Fig. \ref{diffim2l}.  The (Fraunhofer)
diffraction pattern from each disk can be estimated as $\lambdabar
z/d \sim 10~$cm. Therefore, it will only be limited by the
aperture. In other words, with a resolution given by the numerical
aperture of the system, we can consider these disks behaving like
$\delta$-functions, compared to the electron-bunch scale.

The Fraunhofer diffraction pattern of the sampled electron bunch
has now a much larger dimension compared with the original one.
Note that the implementation of this technique requires no optics,
and does not impose stringent requirements on the detector.

The field in the Fraunhofer plane is given by the far-zone
expression

\begin{eqnarray}
\vec{\widetilde{E}}(z,\vec{r}) = \frac{i \omega}{2\pi c z}
\exp\left[\frac{i \omega r^2}{2 c z}\right]
\vec{F}\left(0,-\frac{\omega \vec{r}}{c z} \right)~,
\label{fraunespr}
\end{eqnarray}
where $\vec{F}(0,\vec{u})$ is the spatial Fourier transform of
$\vec{\widetilde{E}}(0,\vec{r})$ (see Eq. (\ref{vecF})). Thus,
assuming that we are sampling the field at the OTR screen at
positions $\vec{r}_k$, the sampled field in Eq. (\ref{fraunespr})
amounts to

\begin{eqnarray}
\vec{\widetilde{E}}(z,\vec{r}) \propto  \sum_k
\vec{\widetilde{E}}(0,\vec{r}_k) \exp\left[-\frac{i\omega\vec{r}_k
\cdot \vec{r}}{c z}\right]~.\label{FTsample}
\end{eqnarray}
The problem rising due to the manipulation of the OTR screen is
significant. For the lensless setup we have $I(\omega,\vec{r})$
proportional to the squared modulus of Eq. (\ref{FTsample}). In
Eq. (\ref{FTsample}) we perform a Fourier transform of the
electric field at the OTR screen with a fixed boundary, that is
the array frame. Even without sampling scatters (imagine in place
of the scatter array e.g. a rectangular screen with high
reflectivity and size of the order of the region B) one has a
complicated diffraction pattern for a single electron due, first,
to the fixed frame profile and, second, to the vectorial
properties of field. As a result, the single electron diffraction
pattern cannot be deconvolved anymore from the 2D Fourier
Transform of the charge density distribution.

In this situation we should first reconstruct the field
distribution on the OTR screen. The complexity of the problem is
now linked with the vectorial nature of the field. The squared
modulus of Eq. (\ref{FTsample}) actually consists in the sum of
two diffraction intensity patterns from two orthogonal
polarizations, and one should separately reconstruct both
polarizations on the screen, i.e. field amplitude and direction.
The problem can be equivalently stated considering a complex
representation of the field as $\widetilde{E}_x+i\widetilde{E}_y$.
The squared modulus of Eq. (\ref{FTsample}) is the diffraction
pattern from such field. Thus, our problem is to recover the
complex-valued function $\widetilde{E}_x+i\widetilde{E}_y$.

Note that $\widetilde{E}_{x,y} >0$. This is strongly related to
two facts. First, within the Ginzburg-Frank theory the electric
field from a single electron at the OTR screen is not
complex-valued. Physically this means that, in this approximation,
the OTR field is characterized by a plane wavefront on the OTR
screen. Second, the field on the OTR screen is a convolution of
the single-electron field with a positive function, the temporal
projection of the electron density distribution, i.e.
$\bar{\rho}_0(0,\vec{r})$. In this case, the reconstruction of the
complex-valued function $\widetilde{E}_x+i\widetilde{E}_y$
directly presents information about the vectorial properties of
the field, because the fact that
$\widetilde{E}_x+i\widetilde{E}_y$ is a complex-valued function is
only related to such vectorial nature.

\subsubsection{Strong support constraint based on tailoring of the OTR screen
reflectivity}

As usual in phase retrieval problems, a-priori information about
the object greatly enhances the efficiency of the iteration
algorithm.

In our case, we have a very well-defined boundary consisting in
the sharp rectangular corners\footnote{Note that the sharpness of
this corners is assumed to be diffraction limited. In the optical
wavelength range modern lithography can easily satisfy this
requirement.} of the sampling grid. The grid is placed within the
region B in Fig. \ref{halo}, meaning that all electrons are within
the sampled area, and that the sampled area is completely
illuminated, up to the boundary of the OTR screen. Thus, the edges
of the grid-frame, i.e. of our object, are guaranteed to have a
good contrast. This a-priori information is extremely useful as
concerns the ability of reconstructing the electric field at the
object plane.

It is worth to restate this concept for the case described in Eq.
(\ref{FTsample}). The solution of the 2D phase retrieval problem
is unique, while that of the 1D problem is not. Note that the
Fourier transform of the sampled object can be expressed in terms
of a polynomial. In particular, Eq. (\ref{FTsample}) is a 2D
discrete Fourier transform, i.e. a 2D polynomial. The lack of
uniqueness of the solution for the phase-retrieval problem is
equivalent to the factorability of this polynomial. Because of the
Fundamental Theorem of Algebra, the 1D phase-retrieval problem is
known to present ambiguities. In contrast to this case, since
polynomials in two variables are only rarely factorable, 2D
sampled objects usually present a unique solution to the
phase-retrieval problem.

Nevertheless, the reader should realize that the lensless imaging
method stands on less solid ground, with respect to the lens-based
setup discussed in the previous Section \ref{sub:HR}. In that
Section, we based our technique on well-known methods that can
always be applied, without any problem, in the case of
non-negative objects. In other words, proposals considered there
are based on already experimented techniques, and will surely work
without difficulties.  In the present case we outlined a novel
technique. We recognized a problem related to the fact that the
OTR halo is not small enough for its Fraunhofer diffraction
pattern to be formed in the detector plane\footnote{There is no
fundamental problem related to the far-zone plane. Difficulties
only arise on practical grounds here due to $z \sim 10~$cm.}. As a
result, the Fraunhofer diffraction pattern of the object (i.e. of
the electron bunch) includes significant interference from the
halo. We sketched a possible solution to this problem, but we did
not go deep into models and simulations.

\subsubsection{Retrieval of the electron density distribution from the electric field}

The last step in our method consists in retrieving the charge
density distribution from the knowledge of the electric field. To
be specific, since, as before, we used a bandpass filter centered
at the modulation frequency $\omega_m$, the electric field
reconstructed by the iterative algorithm is
$\vec{\widetilde{E}}(\omega_m,\vec{r})$.

The Fourier transform of the charge density distribution at
$\omega=\omega_m$ (i.e. $\bar{\rho}(\omega_m,\vec{r}) = a_f
\bar{\rho}_0(0,\vec{r})$) is related to the field by the Gauss
law, and therefore

\begin{eqnarray}
\bar{\rho}_0(0,\vec{r}) \propto
\vec{\nabla}\cdot\vec{\widetilde{E}}(\omega_m,\vec{r})~.
\label{gausslaw}
\end{eqnarray}
Outside the electron bunch, the electric field has vanishing
divergence.

The validity of Eq. (\ref{gausslaw}) is strongly related to our
analogy of the single-electron field with the field from a wire.
Such analogy is valid (see Eq. (\ref{proptoff})) for $r \ll \gamma
\lambdabar$ and, in particular, for $r \lesssim \sigma_r \ll
\gamma \lambdabar$, which is within the bunch region. Then, our
object can be expressed as a convolution between the charge
density distribution (that is, within the electrostatic analogy,
the wires distribution) and the electric field of each wire,
scaling as $1/r$. 


Finally, it should be stressed out that the grid of scatters can
be considered as a fixed reference frame. This allows to measure
with high accuracy the absolute size of the electron bunch and the
absolute position of its center of gravity as soon as the absolute
position of the grid is known.

\subsection{\label{sub:mmulti}Technique to characterize the 3D electron bunch structure
with a multi-shot measurement}

\subsubsection{Combination of real and reciprocal space imaging
spectrometers}

Up to now we used coherent OTR imaging techniques to obtain
(directly or after solution of a phase retrieval problem) a
projection of the charge density distribution, i.e. the time
integral $\int_{-\infty}^{\infty} dt \rho_0(t,\vec{r})$. Here we
will discuss how it is possible to combine diffractive imaging and
direct imaging techniques to obtain the 3D electron bunch
structure.

Very recently in \cite{BRAG} a combination of diffractive and
direct imaging to obtain the spatio-temporal structure of a
cylindrical-symmetric optical pulse was proposed and demonstrated.
In this case, the 3D problem reduces to a 2D phase-retrieval
problem. Then, a spectrometer with a dispersive element could be
used, rather than a plurality of bandpass filters which would have
been necessary in the 3D case. The Gerchberg-Saxton algorithm
\cite{GERC} could then be used to retrieve the phase along the
spatial direction, and an independent FROG measurement at a fixed
transverse position allowed phase renormalization of the phases
between different frequency slices.

Here we propose to use a similar idea for the full 3D problem. Our
setup records two 3D "cubes" of spectral data from the image both
in real space and reciprocal (i.e. spatially Fourier-transformed)
space. These 3D cubes of information include two spatial
dimensions and one spectral dimension in both domains.

The technique is not based on a single-shot, but rather on a
multi-shot measurement. The obvious requirement is, of course, the
reproducibility of the 3D electron bunch density from shot to
shot.

\begin{figure}
\begin{center}
\includegraphics*[width=140mm]{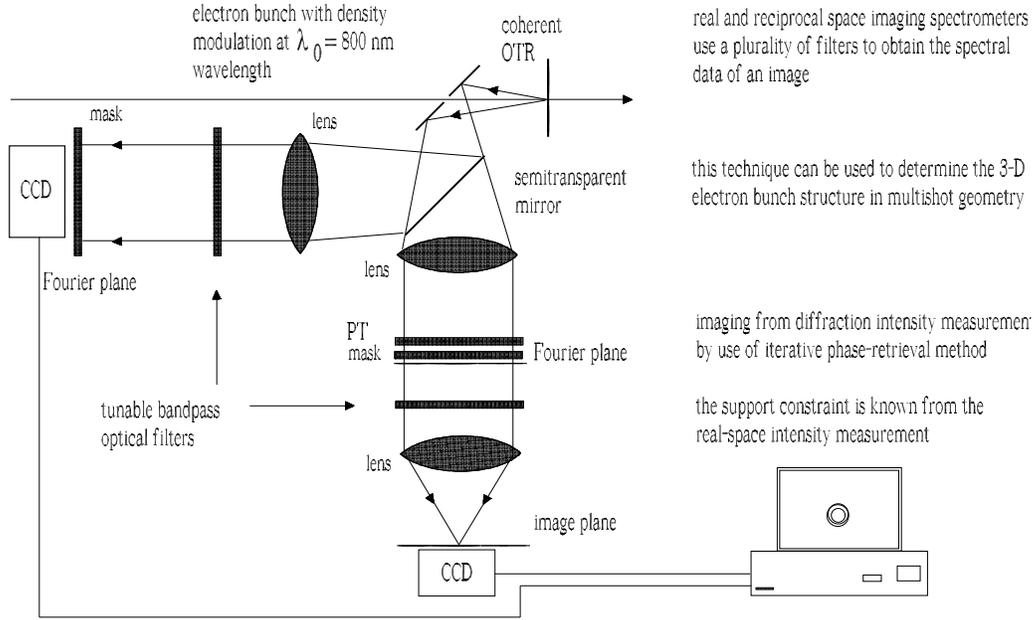}
\caption{\label{fgate} Combination of real and reciprocal space
imaging spectrometers for characterizing the 3D electron density
distribution by use of multishot measurements. }
\end{center}
\end{figure}

A possible setup for the implementation of our method is sketched
in Fig. \ref{fgate}. The OTR signal is split by a semitransparent
mirror into two parts, for the diffractive imaging, and for the
direct imaging. The only difference to the setups described above
for direct and diffractive imaging is that, now, the central
frequency of the bandpass filter can be changed with high
repetition rate, up to $10$ Hz, that is the macropulse repetition
rate of the European XFEL.

Due to the adiabatic approximation, the diffractive imaging setup
records the distribution $|\varrho_0(\Delta \omega, \vec{u})|^2$,
where $\vec{u}=\omega_m \vec{r}_f/(cf)$, and $\Delta \omega$ is,
as usual, the shift of the central frequency of the bandpass
filter from $\omega_m$.  The direct imaging setup records
$|\bar{\rho}_0(\Delta \omega, \vec{r})|^2$ instead. Thus, by
scanning through $\Delta \omega > 0$, we obtain information about
the modulus of the spatial Fourier transform of the charge density
distribution and about the modulus of the charge density
distribution. Both these distributions are recorded in the
temporal-frequency domain, at different frequencies $\Delta \omega
\geqslant 0$.


Note that, for $\Delta \omega \neq 0$, the quantity $\bar{\rho}_0$
is not real. In other words, for each value $\Delta \omega$, our
setup records information about the modulus of the temporal
Fourier transform of the electron density distribution and its
spatial Fourier transform, but misses all information about
phases. We are once more confronted with a phase retrieval
problem, where we know $|\varrho_0(\Delta \omega, \vec{u})|^2$,
and $|\bar{\rho}_0(\Delta \omega, \vec{r})|^2$, and we want to
know $\bar{\rho}_0$ (or, equivalently, $\varrho_0$). The object to
be reconstructed is now complex-valued, but we have strong
a-priori information in both real and reciprocal space.

The Gerchberg-Saxton algorithm \cite{GERC}, which was the first
practical iterative algorithm to solve the Fourier-phase-retrieval
problem, constitutes a way to achieve the recovery of the phase
from two intensity measurements in the image and in the Fourier
planes in an imaging system. Since the introduction of the
Gerchberg-Saxton algorithm, numerous variations have been studied
\cite{FINN,MIL2,LUKE} and may be possibly used instead of the
original one.

The Gerchberg-Saxton procedure gives amplitude and phase in
$\vec{r}$ (or $\vec{u}$) for each frequency. The algorithm is
applied only along the spatial dimension, so that each frequency
slice is independent of the other. In other words, there is a
random phase jump between each frequency slice. Yet, if we want to
perform an inverse temporal Fourier transform, we need to have
information about the relative phases between different frequency
slices. The setup in Fig. \ref{fgate} cannot provide this
information. It should be presented as a-priori knowledge.


Fortunately, this information is available from undulator optical
replica synthesizer (ORS) measurements \cite{ORSS}. As discussed
before, from the FROG trace we can reconstruct amplitude and phase
of the Fourier transform of the longitudinal bunch profile. As a
result, if we integrate the charge density distribution over both
transverse coordinates along a given slice, we can use the ORS
data as reference point for the phase and obtain the renormalized
phase distribution within the slice. Then, phases acquire sense,
not only relative to other points within the same slice, but also
relative to each point in other slices. We can then obtain the
full 3D electron bunch structure by performing a final temporal
Fourier transform.

\subsubsection{Diffractive imaging spectrometer. Projection algorithm.}

The technique described above takes advantage of both real and
reciprocal space imaging spectrometers to reconstruct the 3D
complex-valued Fourier transform of the electron density
distribution. The major difficulty in applying this technique is
that the 3D cubes of available spectral data are sliced at
different frequencies, thus yielding $\varrho_0(\Delta \omega,
\omega_m \vec{r}_f/(cf))$ in the reciprocal space, and
$\bar{\rho}_0(\Delta \omega, \vec{r})$ in the real space. When
$\Delta \omega \ne 0$, the object in real space, $\bar{\rho}_0$,
is complex. Once the phase problem is solved with the help of the
Gerchberg-Saxton algorithm and reconstruction is done slice by
slice, one still needs already available data to renormalize the
relative phases between different slices.

\begin{figure}
\begin{center}
\includegraphics*[width=140mm]{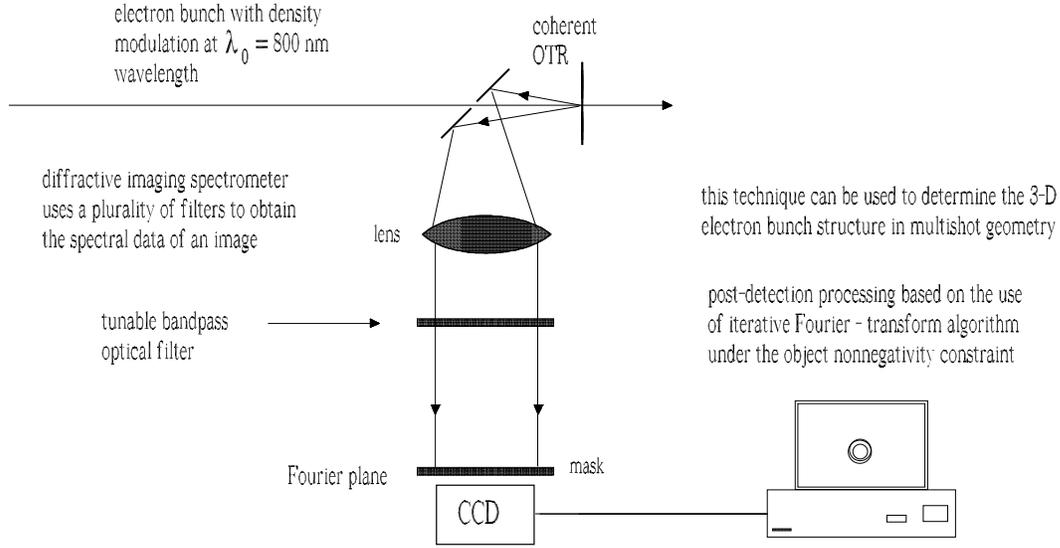}
\caption{\label{dspec} Diffractive imaging spectrometer  for
characterizing the 3D electron density distribution by use of
multishot measurements. }
\end{center}
\end{figure}
Another possible approach to obtain the full 3D electron bunch
structure for a multi-shot measurement without these drawbacks
makes use of a diffractive imaging spectrometer alone, as shown in
Fig. \ref{dspec}. Such setup allow to record the 3D cube of
spectral data $|\varrho_0(\Delta \omega, \omega_m
\vec{r}_f/(cf))|^2$, which consists of less information compared
to the technique described above.

The method takes advantage of a different way of postprocessing
the available data. For example, one can slice $|\varrho_0(\Delta
\omega, \vec{u})|^2$ at $\Delta \omega=0$ in the $(u_x,u_y)$
plane, at $u_x=0$ in the $(\Delta \omega,u_y)$ plane, or at
$u_y=0$, in the $(\Delta \omega,u_x)$ plane.

The advantage of slicing the available spectral data in this way
is clear if one remembers that the iterative reconstruction
algorithm aims to reconstruct e.g. $\bar{\rho}_0(0, x,y)$, from
the knowledge of $|\varrho_0(0, u_x,u_y)|^2$. As we have seen
before, the quantity $\bar{\rho}_0(0, x,y)$ is just the projection
of $\rho_0(t,x,y)$ along the temporal axis, i.e.

\begin{eqnarray}
\bar{\rho}_0(0, x,y) = \left \langle \rho_0(t,x,y) \right
\rangle_t = \int dt {\rho}_0(t, x,y)~,\label{slic1}
\end{eqnarray}
where we introduced $\left \langle \rho_0(t,x,y) \right \rangle_t
$ as the temporal projection of $\rho_0$. Now we use the same
relation between projections and inverse two-dimensional Fourier
transforms. In other words, the projections $\left \langle
\rho_0(t,x,y) \right \rangle_{x,y}$ of $\rho_0$  along the $x$
axis, or the $y$ axis, defined by

\begin{eqnarray}
\left \langle \rho_0(t,x,y) \right \rangle_{x} = \int d x
\rho_0(t,x,y) ~\label{slic2}
\end{eqnarray}
and

\begin{eqnarray}
\left \langle \rho_0(t,x,y) \right \rangle_{y} = \int d y
\rho_0(t,x,y)~,\label{slic3}
\end{eqnarray}
are found by solving the phase retrieval problem for
$|\varrho_0(\Delta \omega, 0,u_y)|^2$, or $|\varrho_0(\Delta
\omega, u_x ,0)|^2$. An example of how such projections should
look like, based on start-to-end simulations, is given in Fig.
\ref{BLBL}.

One concludes that the objects to be reconstructed are, in these
cases, real and non-negative. As we have seen before, this fact
drastically simplifies the use of iterative reconstruction
algorithms granting quick convergence.

We are thus capable of recovering the three orthogonal projections
of $\rho_0$ along $t$, $x$ and $y$. One may also recover other
projections as well. For example, suppose that, instead of cutting
the spectral data at $u_x=0$ (or $u_y=0$) we cut it along another
line, passing by the point $u_x=u_y=0$ but tilted of an angle
$\alpha$ with respect to the $u_x$ axis in clockwise sense. This
means that we are simply performing a rotation

\begin{eqnarray}
&&u_x' = ~~~u_x \cos\alpha+u_y \sin\alpha \cr && u_y' = -u_x
\sin\alpha+u_y \cos\alpha ~.\label{ruota} \end{eqnarray}
With respect to the new system $(\Delta \omega, u_x',u_y')$, we
are considering the cut $|\varrho_0(\Delta \omega, u_x',0)|^2$,
from which we can recover
$\left\langle\rho_0(t,x',y')\right\rangle_{y'}$, which is the
(nonnegative) projection of $\rho_0$ along the axis $y'$, where

\begin{eqnarray}
&&x' = ~~~x \cos\alpha+y \sin\alpha \cr && y' = -x \sin\alpha+y
\cos\alpha \label{ruota2}~. \end{eqnarray}
Obviously, we can choose whatever angle $\alpha$ we like in the
$(u_x,u_y)$ plane, but also in the $(\Delta \omega,u_x)$ plane, or
in the $(\Delta\omega,u_y)$ plane. As a result we can retrieve any
projection of $\rho_0(t,x,y)$ on any plane in the $(t,x,y)$ space
containing any of its axis. More in general, in our 3D cube of
data there exists a very special point where all three frequencies
$\Delta \omega$, $u_x$ and $u_y$ are zero. We may consider any 2D
plane passing through this point and reconstruct any projection in
the real space-time domain under the positivity constraint.

This information is sufficient to recover the 3D function
$\rho_0(t,x,y)$.

The method that we just proposed is a multi-shot technique
operating in the 3D Fourier domain, recording a 3D cube of
information (the intensity in the 3D Fourier domain), and
reconstructing the 3D real space-time electron density
distribution through projections, which are recovered, in their
turn, by reducing the phase-retrieval problem to the well-known 2D
phase-retrieval problem under the nonnegativity constraint of the
object. We name this technique "multi-shot FRODI"
(Frequency-Resolved Optical Diffractive Imaging).

\subsection{FRODI technique to characterize the 3D electron bunch structure
by use of a single-shot measurement}

In this Section we use the new technique Frequency-Resolved
Optical Diffractive Imaging (FRODI), which was introduced above,
for 3D imaging of an individual, ultrashort electron bunch. Using
a diffractive imaging spectrometer based on a
wavelength-dispersive element, this technique measures a 2D
spectrum of a single pulse. FRODI involves mapping the spatial
frequency onto the transverse position, so that the relevant
transverse spatial coordinate becomes the spatial-frequency axis,
e.g. $u_x$ at $u_y=0$. This can be accomplished with the help of a
slit in the Fourier plane centered on the optical axis. The
temporal frequency is the other spatial coordinate, so that a
detector can record the trace of a single pulse. In the previous
Section \ref{sub:mmulti} we saw that retrieving e.g. the
projection $(x,t)$ of the electron bunch from the trace
$(u_x,\omega)$ is equivalent to the solution of a 2D
phase-retrieval problem under non-negativity constraint. FRODI
allows measurements of the 3D structure of a single electron
bunch. This is accomplished by splitting the optical beam and
simultaneously measuring the orthogonal projections $(x,t)$,
$(y,t)$, and $(x,y)$. The respective 2D traces can be obtained on
three separate detectors.

Thus, the method of projection considered above can be
successfully applied in the single-shot case as well. The overall
idea is to acquire a set of projections of the electron bunch in
the $(\omega,u_x,u_y)$ space from a single shot, and to apply
reconstruction techniques to retrieve the electron density
distribution $\rho_0(t,x,y)$.

\begin{figure}
\begin{center}
\includegraphics*[width=140mm]{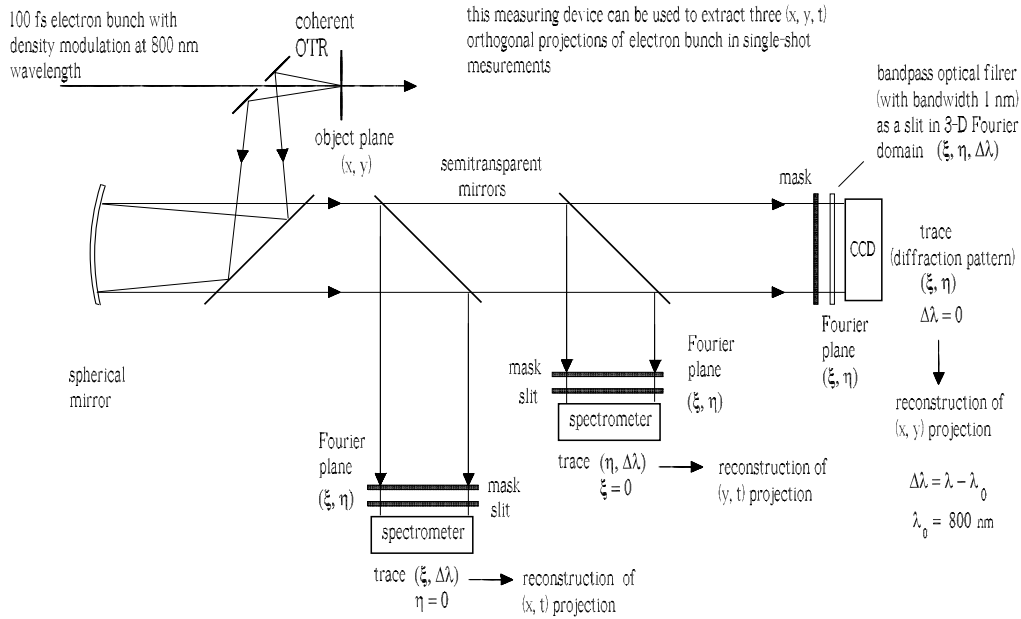}
\caption{\label{bunchfrog}  A schematic of FRODI, an apparatus
capable of measuring three orthogonal projections of the electron
density distribution for a single ultrashort electron bunch. This
will allow retrieval of the full 3D structure of the electron
bunch.}
\end{center}
\end{figure}
The main difference compared with a multi-shot measurement is that
we do not acquire the full spectral cube of information through
the multi-shot measurements, but only a few projections, those
necessary to successfully take advantage of tomographic
reconstruction methods.

A possible realization of the single-shot FRODI is shown in Fig.
\ref{bunchfrog}. The OTR radiation is redirected into three (or
more) stations, where required slices of the 3D spectral cube of
information are recorded. In the scheme in Fig. \ref{bunchfrog} we
consider the three slices at $u_x=0$, $u_y=0$ and $\Delta \omega
=0$.

The projection along the time axis is recovered from the slices at
$\Delta \omega = 0$. The setup used for this operation is the
usual diffractive imaging spectrometer in Fig. \ref{xy-proj},
where the bandpass filter is tuned at $\Delta \omega = 0$.

\begin{figure}
\begin{center}
\includegraphics*[width=120mm]{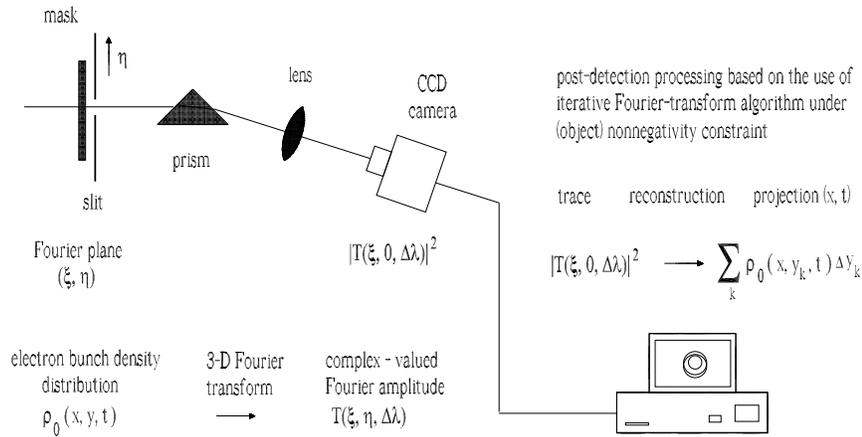}
\caption{\label{xt-proj} Beam geometry for measurements of the
(x,t) electron bunch projection. The prism-lens combination
constitutes a spectrometer.}
\end{center}
\end{figure}
\begin{figure}
\begin{center}
\includegraphics*[width=120mm]{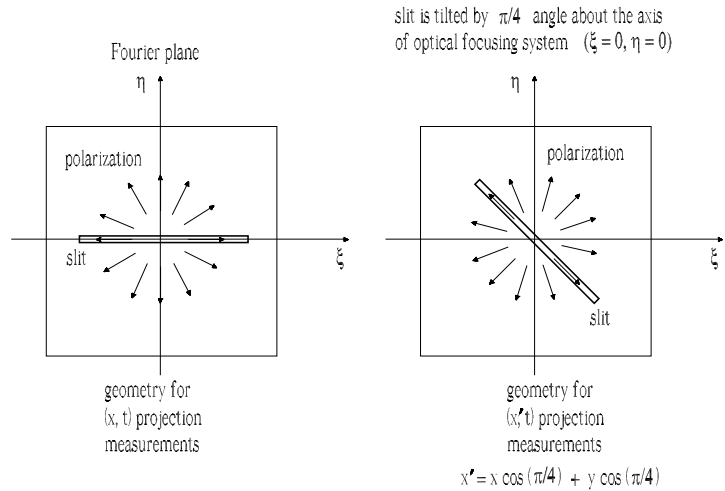}
\caption{\label{slit}  Slit geometry for measuring  different
projections of the electron bunch with a single-shot 3D imager.}
\end{center}
\end{figure}
The projections along the $x$ and $y$ axis are recovered from the
slices at $u_x=0$ and $u_y=0$. In order to obtain a
two-dimensional trace in the $(\Delta \omega,u_y)$ plane (or
$(\Delta \omega,u_x)$ plane) one needs a wavelength-dispersive
element capable of encoding $\Delta \omega$ into the spatial
transverse direction. This can be accomplished with the help of a
setup like that in Fig. \ref{xt-proj}. The usual Fourier mask is
placed in the Fourier plane, followed by a slit to select the
slice at $u_x=0$ or $u_y = 0$. In principle, one may select other
slices by tilting the slit at a given angle, as shown in Fig.
\ref{slit}. This is the equivalent of choosing a given angle
$\alpha$ in the discussion given in the previous Section about the
projection method. After the slit, the optical signal passes
through a dispersive element, e.g. a simple prism, and the 2D
trace in the $(\Delta \omega,u_y)$ or $(\Delta \omega,u_x)$ plane
is subsequently recorded by a detector.

All 2D traces are then used as an input for iterative retrieval
algorithms, yielding back the projections of the charge density
distribution $\rho_0(t,x,y)$ onto the $(x,y)$, the $(t,x)$, and
the $(t,y)$ plane. Tomographic reconstruction follows from the
knowledge of these projections.

As noted before, if more projections are needed one may extend the
scheme in Fig. \ref{bunchfrog} including slits with different
orientations. The problem of characterizing the 3D structure of an
electron bunch from a single-shot measurement can now be solved by
standard reconstruction techniques typical of tomography.

The FRODI concept proposed here is part of a new generation of
electron-beam diagnostic techniques based on coherent optical
light. It does not need any kind of synchronization, is based on
single-shot measurement, guarantees micron resolution, and
operates in real time.

\subsection{\label{sub:SN}Numerical processing of the data recorded in the Fourier plane}

Demands on the detectors for the diffractive imaging case are
qualitatively different from those for direct imaging. In the case
of diffractive imaging, a certain structure of size $\sigma$ on
the object is related to a structure of size $\lambdabar f/\sigma$
in the Fourier plane. Consider e.g. a focal distance of a few tens
of centimeters. The smallest features to be distinguished on the
object, of order of a few microns, correspond to structures of
order of a centimeter in the Fourier plane. Similarly, the largest
features, of order a few hundreds microns, correspond to
structures of order of a hundred microns in the Fourier plane.

Now, the irradiance on the detector decreases with $1/r^2$ as we
move from the center to the periphery, due to the $1/r$ behavior
of the spatial Fourier transform of the single-particle field. As
the photon density decreases, the relative importance of
shot-noise increases, i.e. the signal-to-noise ratio gets worse.
In particular, since shot-noise follows Poisson statistics, we
expect a degradation of the signal-to-noise ratio as $S/N \propto
1/r$. A way of avoiding this degradation is to bin the detector
pixels properly (or to use a \textit{ad hoc} engineered detector
with custom pixels of different shape and area). Suppose that we
bin the detector by first dividing the available area in circular
sectors and then dividing again each circular sector in many
radial slices of a given thickness. In this way, the area of each
bin would increase quadratically with the radius, thus
compensating the decrease of photon density. The number of photons
per bin can be made constant by a proper binning. As a result, the
signal-to-noise ratio linked with shot-noise would not be subject
to degradation anymore.


\section{\label{sec:holog}Measurement of the full spatio-temporal charge density of ultrashort electron bunches with Fourier transform holography}

A further natural development of our imaging techniques consists
in the introduction of a known reference wave, transforming the
previously discussed diffractive imaging setup into a setup
capable of recording Fourier holograms. Since only one Fourier
transform is required to obtain digital reconstruction, Fourier
transform holography (FTH) is far superior over all iterative
phase retrieval methods in terms of computation time.

\begin{figure}
\begin{center}
\includegraphics*[width=110mm]{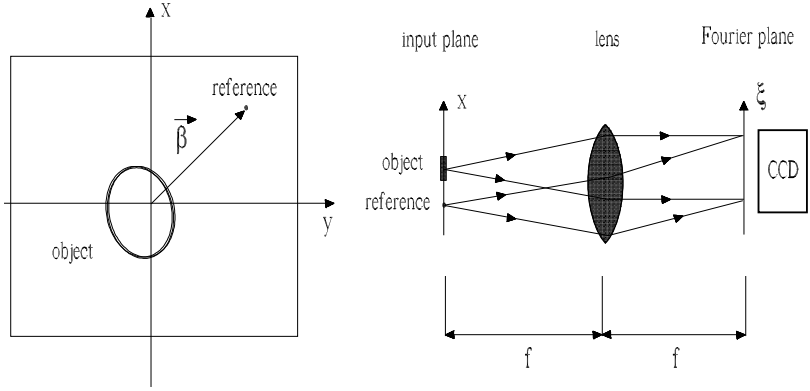}
\caption{\label{FHrecord}  Optical system used to record a Fourier
hologram. }
\end{center}
\end{figure}

A concept of a FTH setup is sketched in Fig. \ref{FHrecord} (see
\cite{STRO}). We summarize here, for future use, the working
principle of Fourier holography. Let us define the electric field
in the object plane in the space-frequency domain
${O}(\omega,\vec{r})$ for the object, and $R(\omega,\vec{r})$ for
the reference.  Later on we will see that in our case of interest
complications are introduced by the space-variant polarization of
the electric field in the object plane. However, for now we assume
that $O$ and $R$ are scalar amplitudes, i.e. amplitudes of a fixed
electric-field component.

In the Fourier plane we define $\mathcal{O}(\omega, \vec{r}_f)$
and $\mathcal{R}(\omega, \vec{r}_f)$,  where
$\mathcal{O}(\omega,\vec{r}_f) \equiv $ $\int d\vec{r}
~{O}(\omega,\vec{r})$ $  \exp[i \omega/(cf)
\vec{r}\cdot\vec{r}_f]$ is the Fourier transform of $O$, and
$\mathcal{R}$ is similarly defined. The electric field in the
Fourier plane is given by the sum
$P\cdot\mathcal{R}+P\cdot\mathcal{O}$, where $P$ is the pupil
function of the optical system, and we consider the system
isoplanatic as before. Since the pupil is considered as a simple
circular aperture of radius $a$, the pillbox function $P$ has the
property $P^2 = P$ as before. A detector will record a holographic
pattern $I(\omega,\vec{r}_f)$ given by

\begin{eqnarray}
I(\omega,\vec{r}_f) &\propto& P(\vec{r}_f)
\left|\mathcal{R}(\omega, \vec{r}_f)+\mathcal{O}(\omega,
\vec{r}_f)\right|^2 = P(\vec{r}_f)\left|\mathcal{R}(\omega,
\vec{r}_f)\right|^2 + P(\vec{r}_f)\left|\mathcal{O}(\omega,
\vec{r}_f)\right|^2 \cr && +P(\vec{r}_f)\mathcal{R}^*(\omega,
\vec{r}_f)\cdot\mathcal{O}(\omega, \vec{r}_f)\cr &&+
P(\vec{r}_f)\mathcal{R}(\omega,
\vec{r}_f)\cdot\mathcal{O}^*(\omega, \vec{r}_f)~.\label{hologram}
\end{eqnarray}
An interference pattern is thus recorded in the Fourier plane.
Such pattern is called a hologram. Inspection of terms in Eq.
(\ref{hologram}) shows that the hologram is a superposition of the
squared modulus of the Fourier transform of the object, of the
squared modulus of the Fourier transform of the reference, and of
interference terms. The squared modula carry no extra-information
compared to what can be obtained from diffractive imaging
techniques: information about the phase of the Fourier transform
of the object is not included there. Such information is included,
however, in the interference terms. Actually, it is fair to say
that Fourier holography works because the detector is a nonlinear
device, yielding the squared modulus of the input amplitude, which
includes interference terms.

The field distribution of the reference wave is known. Consider
Fig. \ref{FHrecord} and suppose, which is the case for our setup,
that the width of the reference source is of order of the optical
wavelength of interest. Then, the reference source can be
considered as a point source.  It follows that
$\mathcal{R}(\omega,\vec{r}_f) = \exp[i\omega/(cf)
\vec{\beta}\cdot \vec{r}_f]$, $\vec{\beta}$ being the vector
position of the reference. Substituting the expression for
$\mathcal{R}$ in Eq. (\ref{hologram}) and taking the inverse
Fourier transform $\mathcal{F}^{-1}$ of the holographic pattern,
one obtains

\begin{eqnarray}
\mathcal{F}^{-1}\left[I(\omega,\vec{r}_f)\right](\vec{r}) &
\propto &
\mathcal{F}^{-1}\left[P(\vec{r}_f)\left|\mathcal{R}(\omega,
\vec{r}_f)\right|^2\right](\vec{r}) +
\mathcal{F}^{-1}\left[P(\vec{r}_f)\left|\mathcal{O}(\omega,
\vec{r}_f)\right|^2\right](\vec{r}) \cr &&+\int d\vec{r'}
\mathcal{P}^*(\omega, \vec{r}+\vec{\beta}-\vec{r'})\cdot
{O}(\omega, \vec{r'}) \cr &&+\int d\vec{r'} \mathcal{P}(\omega,
\vec{r}-\vec{\beta}-\vec{r'})\cdot {O}^*(\omega, \vec{r'})
~.\label{hologram2}
\end{eqnarray}
Thus, with accuracy given by the dimension of the pupil aperture,
the inverse Fourier transform of the holographic pattern yields
straightforwardly, through the interference terms, a direct image
of the original object $O$, and a conjugate
image $O^*$. 
Note that such direct image is actually a convolution of the
object with a specific point spread function. In fact, we always
have blurring due to diffraction effects. 

The object and the reference source must be arranged with a
separation larger than twice the object width (this is called the
FTH condition). Then, by inspection of Eq. (\ref{hologram2})
follows that the object is well-separated from the second term of
Eq. (\ref{hologram2}), which amounts to the autocorrelation
function of the object.

For optical applications,  the resolution of holographic
techniques is not limited by size and quality of the reference
source. With the help of modern lithography it is not difficult to
produce an unresolved point source at optical wavelengths and let
sufficiently bright radiation through it.

\subsection{Real-time high resolution FTH imaging of electron
bunch}


\begin{figure}
\begin{center}
\includegraphics*[width=120mm]{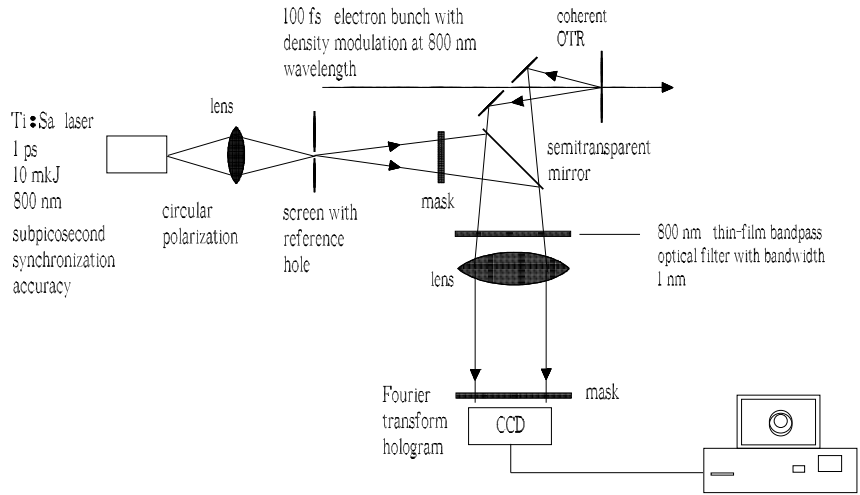}
\caption{\label{holog3} Practical setup for Fourier transform
holography using a virtual reference point source. }
\end{center}
\end{figure}
A big advantage of using FTH techniques compared to diffractive
imaging methods explored in Section \ref{sec:diffite}, is that no
iteration algorithm is needed. Thus, real-time, high-resolution
imaging of the electron bunch is granted through FTH.

A way a FTH setup may be implemented is shown in Fig.
\ref{holog3}. It can be seen to consist of the previous
diffractive imaging setup shown in Fig. \ref{recon} together with
a reference source. The reference source is constituted by a
screen with a reference pinhole illuminated by a laser pulse. The
laser pulse should be synchronized to the electron bunch with
sub-picosecond accuracy. The object wave and the reference wave
are summed  up with the help of a semitransparent mirror before
the lens.

Note that, as was discussed before in Section \ref{sec:diffite},
and for the same reasons, a narrow bandpass optical filter
centered at the modulation frequency is needed before the
detector.

\subsubsection{Laser with circular polarization as reference source}

\begin{figure}
\begin{center}
\includegraphics*[width=120mm]{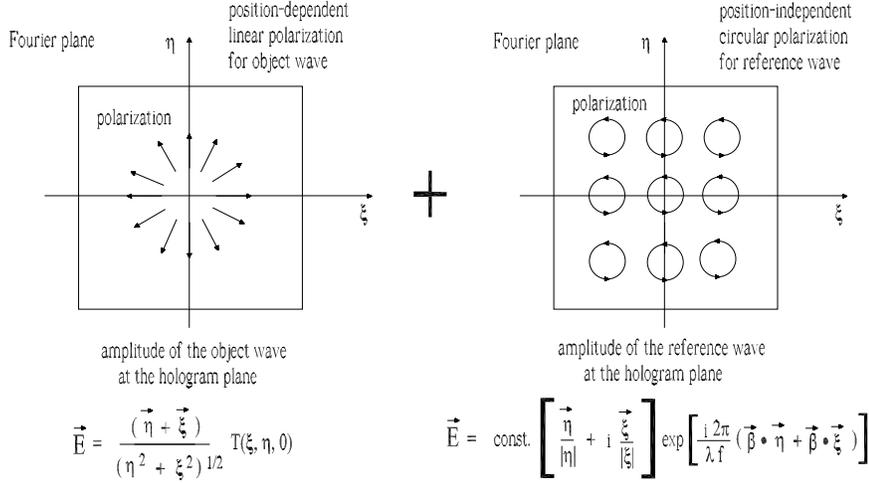}
\caption{\label{polfh} Sketch of the polarization for object and
reference wave in the hologram (Fourier) plane. }
\end{center}
\end{figure}

A specific aspect of our case study, compared to usual FTH setups,
needs further investigations. As we mentioned before, we derived
Eq. (\ref{hologram2}) under the assumption that $O$ and $R$ are
scalar objects. In our case, however, the field distribution of
the object wave in the Fourier plane is radially polarized
independently of the charge density distribution, as shown in Fig.
\ref{polfh}. Note that the object field distribution exhibits a
complicated space-variant (not radial) polarization.

Since the setup is designed to resolve the object, we assume that
the size of the Fourier pattern of the object in the detector
plane is much smaller compared to the available numerical aperture
(multiplied by $f$). Therefore, the FTH condition is still
formulated requiring that the separation between the reference
source and the object should be larger than at least two times the
characteristic size of the object.  This means that
$|\vec{\beta}|$ should be taken larger than the object
autocorrelation function in order not to superimpose the object
image with the central spot.

Space-variant polarization yields difficulties when one tries to
apply the FTH algorithm in the usual way described above. Namely,
in our case, interference terms are multiplied by a factor
depending on the position in the Fourier plane.

To show this, let us first introduce polar coordinates $\phi$ and
$r_f$, so that $\vec{r}_f = (r_f \cos{(\phi)}, r_f \sin{(\phi)})$.
Supposing that the reference wave is linearly polarized along e.g.
the $x$ direction, one has $\vec{\mathcal{R}}(\omega,\vec{r}_f) =
\exp[i\omega/(cf) \vec{\beta}\cdot \vec{r}_f] (1,0)$ and
$\vec{\mathcal{O}}(\omega,\vec{r}_f)=
\mathcal{O}(\omega,\vec{r}_f) (\cos(\phi),\sin(\phi))$.  It
follows that Eq. (\ref{hologram}) is modified to

\begin{eqnarray}
I(\omega,\vec{r}_f) &\propto&
P(\vec{r}_f)\left|\vec{\mathcal{R}}(\omega,
\vec{r}_f)+\vec{\mathcal{O}}(\omega, \vec{r}_f)\right|^2 =
P(\vec{r}_f)+ P(\vec{r}_f)\left|\mathcal{O}(\omega,
\vec{r}_f)\right|^2 \cr &&+P(\vec{r}_f)
\exp\left[-\frac{i\omega}{c f} \vec{\beta}\cdot
\vec{r}_f\right]\mathcal{O}(\omega, \vec{r}_f)\cos(\phi) \cr &&
+P(\vec{r}_f) \exp\left[~~\frac{i\omega}{cf} \vec{\beta}\cdot
\vec{r}_f\right]\mathcal{O}^*(\omega,
\vec{r}_f)\cos(\phi)~.\label{hologramb}
\end{eqnarray}
In this case, taking the inverse Fourier transform of Eq.
(\ref{hologramb}) would not give back Eq. (\ref{hologram2}), due
to the specific kind of position-dependent factor $\cos({\phi})$.
The problem is that $\cos({\phi})$ becomes zero for certain values
of $\phi$, and can be filtered out at the expenses of a reduced
resolution only.

\begin{figure}
\begin{center}
\includegraphics*[width=140mm]{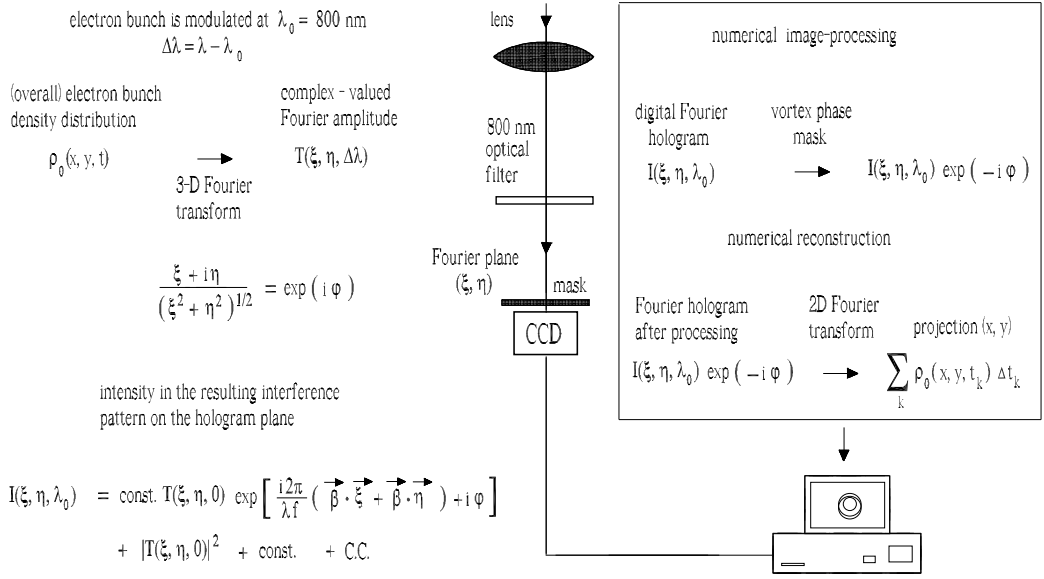}
\caption{\label{digfh}  Schematic representation of digital
Fourier hologram recording and its numerical reconstruction. }
\end{center}
\end{figure}
A possible solution consists in using a circularly polarized
reference wave, instead of a linearly polarized one, Fig.
\ref{polfh}. In this case, one has
$\vec{\mathcal{R}}(\omega,\vec{r}_f) = \exp[i\omega/(cf)
\vec{\beta}\cdot \vec{r}_f] (1,i)$. Then, Eq. (\ref{hologramb})
changes to

\begin{eqnarray}
I(\omega,\vec{r}_f) &\propto& P(\vec{r}_f) + P(\vec{r}_f)
\left|\mathcal{O}(\omega, \vec{r}_f)\right|^2 \cr
&&+P(\vec{r}_f)\exp\left[-\frac{i\omega}{c f} \vec{\beta}\cdot
\vec{r}_f\right]\mathcal{O}(\omega, \vec{r}_f)\exp[i\phi] \cr &&
+P(\vec{r}_f) \exp\left[~ \frac{i\omega}{cf} ~ \vec{\beta}\cdot
\vec{r}_f~\right]\mathcal{O}^*(\omega, \vec{r}_f)\exp[-i\phi]~.
\label{hologramc}
\end{eqnarray}
Eq. (\ref{hologramc}) includes a position-dependent factor as
well. However, such factor never becomes zero, and it can be
easily filtered out improving the resolution. A vortex phase mask
$\exp[-i\phi]$ can be applied, Fig. \ref{digfh}, for image
processing purposes. Note that one may choose whether to process
the signal digitally, or to use additional hardware for this
purpose. The same remark applies to further image processing
described in the next section as well. To fix a particular case
study we assume, for example, that the vortex mask is included
into the numerical image processing algorithm.
Applying the vortex mask one obtains

\begin{eqnarray}
I(\omega,\vec{r}_f) \exp[-i\phi] &\propto&
P(\vec{r}_f)\exp[-i\phi]+ P(\vec{r}_f)\left|\mathcal{O}(\omega,
\vec{r}_f)\right|^2\exp[-i\phi] \cr &&+
P(\vec{r}_f)\exp\left[-\frac{i\omega}{c f} \vec{\beta}\cdot
\vec{r}_f\right]\mathcal{O}(\omega, \vec{r}_f)\cr && +
P(\vec{r}_f)\exp\left[~ \frac{i\omega}{cf} ~ \vec{\beta}\cdot
\vec{r}_f~\right]\mathcal{O}^*(\omega, \vec{r}_f)\exp[-2i\phi]~.
\label{hologramc2}
\end{eqnarray}
After taking the inverse Fourier transform of the holographic
pattern in Eq. (\ref{hologramc2}), the object $O(\omega,\vec{r})$
will appear again from the third term in Eq. (\ref{hologramc2}).

Because of the extra phase $\exp[- i\phi]$, which is now present
in the first terms of Eq. (\ref{hologramc2}), the Fourier
transform will be blurred, but its size cannot be visibly changed
because the object is assumed to have much larger size compared to
the point-spread function. In fact, before multiplication by the
vortex mask, the Fourier transform of the object, $\mathcal{O}$,
is automatically limited by the pillbox function $P$, due to a
finite numerical aperture of the optical system, which is unity
for values of $r_f/f$ smaller than the numerical aperture, and
zero otherwise. This pillbox function actually acts as the Fourier
transform of the point-spread function of the optical system. Due
to the presence of the vortex phase mask, the pillbox function is
multiplied by the extra position-dependent phase-factor $\exp[-
i\phi]$, and the point-spread function of the system is now given
by the Fourier transform of the product of the pillbox function by
$\exp[- i\phi]$. Note that here we cancelled the
position-dependent phase-factor for the direct image term, thus
obtaining a non-blurred image of the object, at the expense of
blurring both the autocorrelation and the inverse image of the
object. The analysis in \cite{SACK}, where one may find plots of
the Fourier Transform of the product of a pillbox function and a
vortex, allows to conclude that this Fourier transform of the
product of pillbox function and vortex is just a  few times wider
compared to the point-spread function (which is the Fourier
transform of the pillbox only). As a result, after convolution
with e.g. the autocorrelation function, we have practically the
same size as before the convolution, and the FTH condition holds.

\subsubsection{Mask for optimum FTH imaging of electron bunch}

When we discussed diffractive imaging techniques in Section
\ref{sec:diffite} we saw that the input to the phase retrieval
algorithm is Eq. (\ref{Fplane2mody2}), which was obtained from Eq.
(\ref{Fplane2mody}) through Fourier filtering. Note that Eq.
(\ref{Fplane2mody}) is proportional to the squared modulus of the
field at the Fourier plane, Eq. (\ref{Fplane2mody0}). In our case
of interest, the Fourier transform of the object,
$\mathcal{O}(\vec{r}_f)$, coincides with Eq. (\ref{Fplane2mody0}).
The difference with respect to that case is that, now,
$\mathcal{O}(\vec{r}_f)$ does not enter as a squared modulus in
Eq. (\ref{hologramc2}), but only linearly. Therefore, at first
glance, using the same spatial Fourier filter as in the previous
Sections we cannot optimize the image resolution anymore.

However, if we properly introduce another filter after the screen
with the reference source in Fig. \ref{holog3}, we can optimize
the image resolution and still keep the previous filter in the
Fourier plane. Since the filter after the screen should work as a
Fourier filter partly countering the contribution of the filter in
the Fourier plane, it must be placed in the far-zone of the
reference wave. If we indicate with $z_m$ the distance between the
screen and the mask, and with $d$ the dimension of the reference
source, the far-zone condition is $d^2/(\lambdabar z_m) \ll 1$.
This is not a restriction. In fact $d \sim \lambda$, and the
Fraunhofer condition will always be satisfied behind a screen with
such a pinhole. Then, if we define the amplitude transmittance of
the mask in the Fourier plane $T_m(r)$, which is explicitly given
by Eq. (\ref{TMMM}) with $T_0=1$, we should introduce a second
mask with amplitude transmittance proportional to $T_m^{-1}(r)$
after the screen with the reference source. Let us name such
transmission function $T_{m,2}$. The mask will be installed at
some distance $z$ from the pinhole, in the far zone. Due to the
far-zone approximation, angular openings $\theta$ are related to
transverse sizes on the mask by $r= \theta z$. Since we are only
interested in $\theta > 1/\gamma$ we can set $T_{m,2}=0$ for
$r<z/\gamma$. Then, if the numerical aperture of the lens is given
by $NA$, one can set $T_{m,2}=(NA) T_{0,2} z/r$ for $r>z/\gamma$,
with the condition $(NA) T_{0,2} z < 1$. When this is done, given
the input field

\begin{eqnarray}
\vec{\widetilde{E}}(\vec{r}) =-\frac{2\omega e}{c^2 \gamma} \int
d\vec{r'} \bar{\rho}(\omega,\vec{r'})
\frac{\vec{r}-\vec{r'}}{\left|\vec{r}-\vec{r'}\right|}
K_1\left(\frac{\omega}{c\gamma}\left|\vec{r}-\vec{r'}\right|\right)~,
\label{inpfil}
\end{eqnarray}
one can rewrite the third term in the pattern in Eq.
(\ref{hologramc2}) with the help of Eq. (\ref{Fplane2mody}) as

\begin{eqnarray}
I(\omega,\vec{r}_f) \exp[-i\phi]&\propto&
\exp\left[-\frac{i\omega}{c f} \vec{\beta}\cdot \vec{r}_f\right]
\varrho\left(\omega, \frac{\omega \vec{r}_f}{c f}\right)
P(\vec{r}_f) ~. \label{hologramc4}
\end{eqnarray}
%

Note that, as concerns the object signal, we deconvolve the
electron bunch structure factor in the Fourier plane. One should
have a good signal-to-noise ratio, which is mandatory for success
in the implementation of our scheme. Such ratio is related to the
manipulation of the reference source, i.e. of a laser source
passing through two spatial Fourier filters. The method of binning
discussed in Section \ref{sub:SN} for increasing the
signal-to-noise ratio may be taken advantage of.

\subsection{Self-referencing FTH imaging}

A natural extension of the FTH imaging idea is to use the object
wave to measure itself. This allows one to image the electron
bunch without the need for a separate reference source.

There are several approaches to self-referencing schemes in FTH
imaging. One version of these techniques makes use of the
reference pulse which can be obtained from the object pulse itself
using spatial filtering. An alternative to the scheme proposed in
Fig. \ref{holog3} would be, in fact, to split the optical replica
pulse from the OTR screen into two, and spatially filter one copy
in order to obtain a reference pulse, i.e. a quasi-planar (being
spatially filtered) wave in the detector (Fourier) plane. This
wave\footnote{Note that this wave has linear polarization, but can
now be transformed into circular polarization (in self-referencing
schemes). Then, numerical processing including the vortex phase in
the hologram can be applied as discussed before.} interferes with
the second copy of the original pulse at some angle provided by
the FTH condition. Of course, both pulses should be spectrally
filtered with a narrow band-pass filter at frequency $\omega_m$.
Note that in this geometry the object wave, i.e. the original
optical-replica pulse, should be reflected a few times off mirrors
and beam-splitters so that both object and reference pulses can be
synchronized and combined in the detector plane. Additionally, in
our case the object wave should be attenuated of about two orders
of magnitude to increase the contrast of the hologram. This kind
of self-referencing FTH schemes is based on the use of standard
techniques and are widely discussed in literature, see e.g.
\cite{GABO}.

In this Section we consider a particular realization of
self-referencing FTH imaging, which is based on the use of the
coherent OTR halo to produce a reference wave. In this case we
need to consider a relatively large size of the reference source.
Moreover, the reference wave has linear polarization. These facts
imply the production of a lower-resolution image compared to the
high-resolution image where an optimized reference source is used.
However, due to the absence of an external reference source, the
holographic setup turns out to be the same as for diffractive
imaging. Then, the low-resolution image can be used in combination
with previously discussed diffractive imaging techniques.

In this way, an implementation of FTH schemes only consists in the
addition of some modification in the OTR screen to an already
existing (diffractive imaging) setup, Fig. \ref{recon}. This is an
obvious advantage, which opens up e.g. the possibility of
simultaneously performing diffractive imaging and FTH. For
example, the original holographic pattern can be processed in
parallel with Eq. (\ref{hologramc4}). Then, standard
phase-retrieval techniques can be applied to the second term in
Eq. (\ref{hologram2}). The advantage of the combination of FTH
with iterative phase retrieval methods in X-ray applications has
been already demonstrated experimentally in \cite{STAD}.

\subsubsection{\label{sub:low}Use of the coherent OTR halo to produce the reference wave}

\begin{figure}
\begin{center}
\includegraphics*[width=130mm]{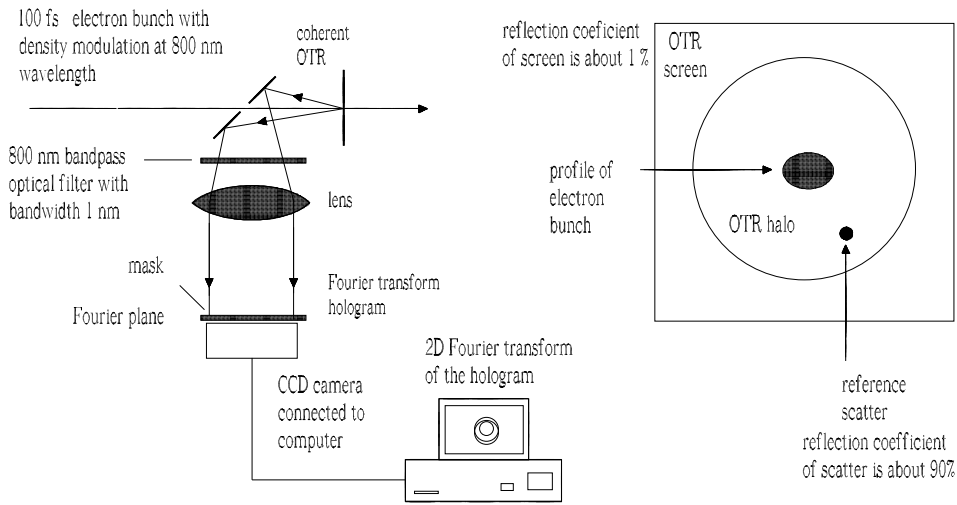}
\caption{\label{holog} Setup for Fourier holography, where the OTR
halo outside of the electron bunch profile is used to create a
reference wave.}
\end{center}
\end{figure}
A straightforward idea to implement an FTH setup for
low-resolution FTH imaging without the help of an external laser
is to take advantage of the large OTR halo outside the electron
bunch profile on the OTR screen. A small disk with high
reflectivity coefficient may be formed on the OTR screen, like in
Fig. \ref{holog}, and we also assume that the reflectivity of the
entire OTR screen is made lower.

The reliability of the image obtained by holography is an
important advantage of FTH over diffractive imaging. No iterative
process is involved in the numerical reconstruction of the object
image from the hologram. There are no convergence and uniqueness
problems as in the iterative phase retrieval approach.

However, in all FTH techniques, the spatial resolution limit
achievable is defined by the size of the reference source. The
size of the reference source has a strong effect on the reference
photon flux too. The smaller the source, the smaller the number of
photons in the reference signal, the smaller the signal-to-noise
ratio. Thus, there is a limit to how small the size of the source
can be. Of course, in case an external laser is used to define a
reference, one can think of focusing the laser on the reference
pinhole, or to increase the power of the laser, up to some extent.
In the setup in Fig. \ref{holog}, there is no such possibility.
Hence we need to consider a relatively large size of the reference
source, Fig. \ref{holog}, which implies, in its turn, a
lower-resolution image. Such low-resolution image can be used, as
explained before, as a strong support for diffractive imaging
techniques, so that the iterative algorithm can run in real-time.

However, it should be noted that since the OTR halo is used to
illuminate the reference source, one does not deal anymore with a
circularly polarized reference wave. The electric field in the OTR
halo region is radially polarized. Therefore, the polarization of
the reference wave is linear. Its direction coincides with the
direction $\vec{\beta}$ of the vector identifying the position of
the reference scatter with respect to the center of the OTR
screen.

In this case one has $\vec{\mathcal{R}}(\omega,\vec{r}_f) =
\exp[i\omega/(cf) \vec{\beta}\cdot \vec{r}_f] \vec{\beta}/\beta$
and $\vec{\mathcal{O}}(\omega,\vec{r}_f)=
\mathcal{O}(\omega,\vec{r}_f) \vec{r}_f/r_f$. It follows that Eq.
(\ref{hologramb}) is modified to

\begin{eqnarray}
I(\omega,\vec{r}_f) &\propto&
P(\vec{r}_f)\left|\vec{\mathcal{R}}(\omega,
\vec{r}_f)+\vec{\mathcal{O}}(\omega, \vec{r}_f)\right|^2 =
P(\vec{r}_f) + P(\vec{r}_f)\left|\mathcal{O}(\omega,
\vec{r}_f)\right|^2 \cr &&+P(\vec{r}_f)
\exp\left[-\frac{i\omega}{c f} \vec{\beta}\cdot
\vec{r}_f\right]\mathcal{O}(\omega,
\vec{r}_f)~\frac{\vec{\beta}\cdot{\vec{r}_f}}{\beta r_f}\cr &&
+P(\vec{r}_f) \exp\left[~~\frac{i\omega}{cf} \vec{\beta}\cdot
\vec{r}_f\right]\mathcal{O}^*(\omega,
\vec{r}_f)~\frac{\vec{\beta}\cdot{\vec{r}_f}}{\beta
r_f}~.\label{hologrambsimp}
\end{eqnarray}
Since we want to use diffractive imaging as the main technique
here, and the FTH only provides a better support for the iterative
algorithm, we may want to use a mask with amplitude transmittance
$T_m(r)$ given in Eq. (\ref{TMMM}) in the Fourier plane. However,
in the case under consideration we cannot use an additional mask
for the reference source. As a result, we have a new situation
concerning FTH imaging. After the Fourier mask, the third term in
Eq. (\ref{hologrambsimp}) reads

\begin{eqnarray}
I(\omega,\vec{r}_f) &\propto&  \exp\left[-\frac{i\omega}{c f}
\vec{\beta}\cdot \vec{r}_f\right] \varrho\left(\omega,
\frac{\omega \vec{r}_f}{c f}\right)
P(\vec{r}_f)~\frac{\vec{\beta}\cdot{\vec{r}_f}}{\beta}~.
\label{hologramc4simp}
\end{eqnarray}
%

The factor ${\vec{\beta}\cdot{\vec{r}_f}}/{\beta}$ appearing in
the interference terms now yields a difference compared to the FTH
imaging technique described above.

Taking the inverse Fourier transform of Eq.
(\ref{hologramc4simp}), we see that such interference term will
not yield the function $\bar{\rho}$ but, rather, the directional
derivative of $\bar{\rho}$ along the polarization direction of the
reference signal, given by the unit vector $\vec{n}_{\beta}$,
shifted of $\vec{\beta}$ and properly scaled. This fact can be
easily understood. First, we remember that, given the function
$f(\vec{r}_f)$, its directional derivative along the direction
$\vec{n}_{\beta}$ is given by

\begin{eqnarray}
\frac{d f}{d\vec{n}_{\beta}} \equiv \vec{n}_{\beta}\cdot
\vec{\nabla} f(\vec{r}_f)~. \label{dirder}
\end{eqnarray}
From Eq. (\ref{dirder}) and from the properties of Fourier
transforms follows that

\begin{eqnarray}
\mathcal{F}^{-1} \left[\vec{n}_{\beta} \cdot
\vec{r}_f~f(\vec{r}_f)\right](\vec{r}) = \frac{i c
f}{\omega}\vec{n}_{\beta}\cdot
\left\{\vec{\nabla}\mathcal{F}^{-1}\left[f\right](\vec{r}_f)\right\}~.
\label{tfprop}
\end{eqnarray}
Thus, our method foresees the use of a known directional
derivative of $\bar{\rho}_0$ at $\Delta \omega = 0$ as a support
for diffractive image. The iterative algorithm will act on the
autocorrelation term in the holographic pattern with a strong
support, together with the positivity condition for
$\bar{\rho}_0(0,\vec{r})$. This should be sufficient to obtain
real-time convergence of the iterative algorithm.

Note that the reference scatter has a high reflection coefficient.
In order to provide a reflectivity contrast we should actually
introduce an attenuator for the OTR pulse. When we additionally
introduce a mask we might end up with an insufficient photon flux
in the detector plane. However, starting with diffractive imaging
setups, we have a choice between two possibilities of imaging
processing. We can use a transparency with amplitude transmittance
as we discussed before, or we can use numerical processing. In the
case of Diffractive imaging and FTH with external reference source
we have such choice as well. Since we already suppressed the OTR
screen reflectivity, the numerical image processing option is to
be preferred. In this case, the numerical processing can be
separately optimized for the FTH image and for the DI, which
operates with the autocorrelation of the object.

\subsubsection{\label{sub:mrf}Multiple reference FTH}

A natural generalization of ideas and observations in Section
\ref{sub:low} consists in increasing the amount of a-priori
information used in the iterative algorithm. In reference
\cite{SCHL}, a multiple reference source FTH was used in the soft
x-ray range to simultaneously record five images of the sample. It
was demonstrated that averaging these images yields an improvement
in the image quality due to enhancement of the signal-to-noise
ratio.

\begin{figure}
\begin{center}
\includegraphics*[width=130mm]{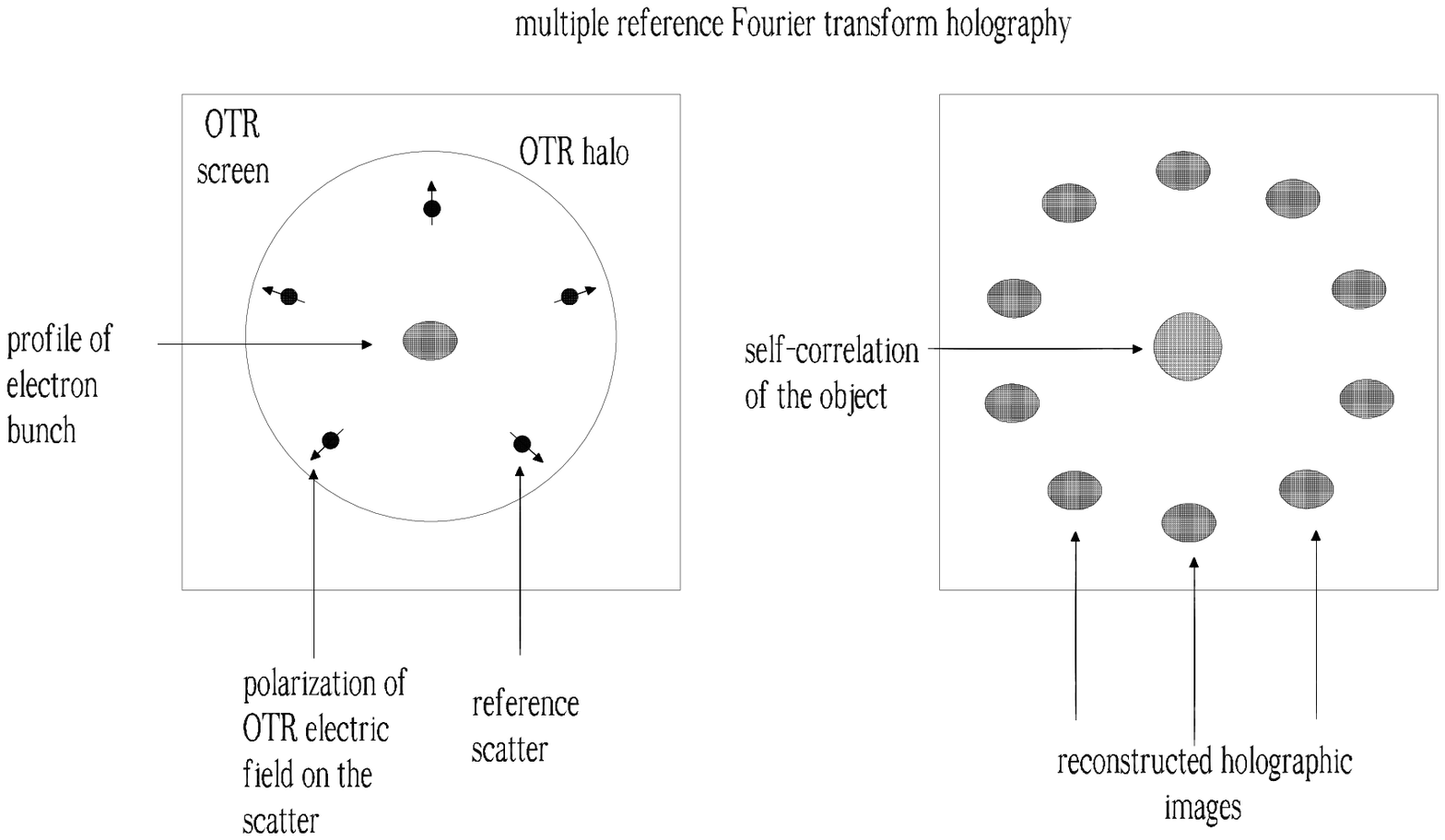}
\caption{\label{holog4} A straightforward extension to the Fourier
transform holography technique is to introduce multiple reference
sources. This method can be used to increase the amount of
a-priori information used in iterative algorithm. The picture
shows arrangement of the reference scatters on the vertices of a
regular pentagon \cite{SCHL}. The arrows show directions of the
polarization of the OTR field at the scatters. Two redundant
images of the bunch appear for each reference source. Five
independent images of the bunch can be recovered from the hologram
and then used as support constraint for high resolution
diffractive imaging.}
\end{center}
\end{figure}
Here we still consider the setup in Fig. \ref{holog}, but with
several reference sources. The situation is illustrated in Fig.
\ref{holog4} where, for the sake of illustration, five references
are considered. From each reference source one can reconstruct a
directional derivative of the object. The a-priori knowledge of
the (different) polarization directions of the field at the
reference points on the OTR screen can be used to retrieve the
image from the hologram. One reconstructs different directional
derivatives. Therefore, the bunch profile can be reconstructed.

Note that the reconstructed Fourier transform contains the
autocorrelation of all reflecting structures. Two redundant images
of the object appear for each reference scatter. One image is the
cross-correlation of the reference with the object, while the
other, its complex conjugate, is located radially opposite to the
origin. The autocorrelation  contains five independent images of
the object. The center is occupied by the autocorrelation of the
object and of each reference scatter. Images should tile, and not
overlap on the autocorrelation. This is realized in the
arrangement of the reference scatters on the vertices of a regular
pentagon.

It should also be noted that the previously discussed schemes
using the OTR halo as a reference source are particular cases of
self-referencing FTH measurements. In this case, the
low-resolution image would serve as an input for the iterative
phase retrieval algorithm, giving to the numerical algorithm a
starting point closer to the real solution. In such combined
approach, the larger features of the bunch are first imaged
unambiguously by holography. By subsequent oversampling phasing,
one can try to further improve the image resolution up to the
diffraction limit. A combined Fourier holography-oversampling
approach has the advantage that high-resolution and real-time
reconstruction can be reached simultaneously.

\subsection{Technique to characterize the 3D electron bunch structure
by use of multi-shot measurements}

\subsubsection{Frequency-gated Fourier transform holography}

In Section \ref{sec:diffite} we proposed a technique to
characterize the 3D electron density distribution taking advantage
of a diffraction-imaging spectrometer, in multi-shot geometry.
Here we propose another technique for multishot measurement of 3D
charge density distribution based on frequency-gated Fourier
Holography.

\begin{figure}
\begin{center}
\includegraphics*[width=140mm]{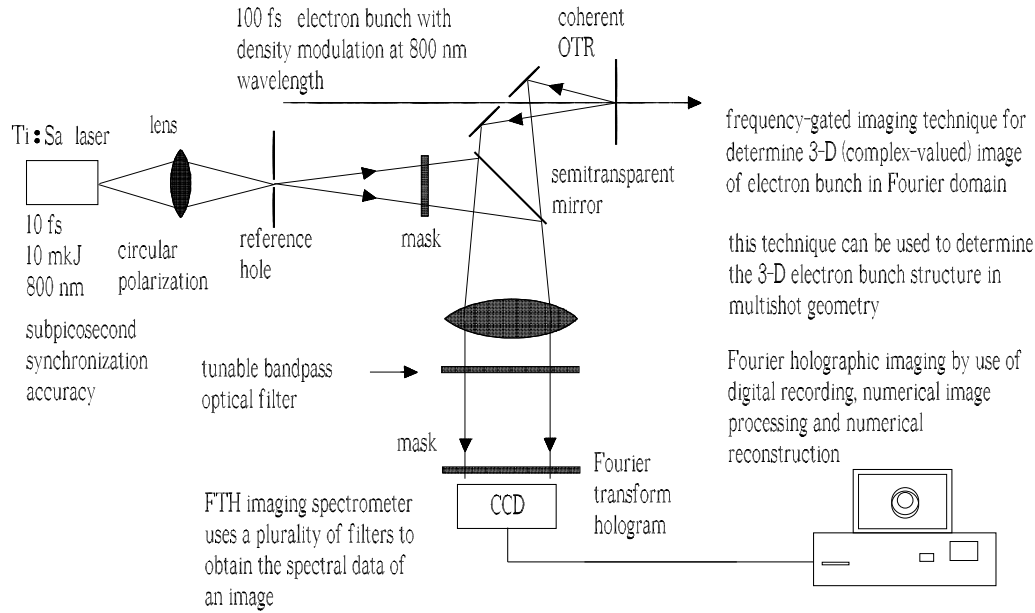}
\caption{\label{3dmulti}  Multishot measurement of the 3D
structure of electron bunches by use of frequency-gated Fourier
transform holography.}
\end{center}
\end{figure}
A possible realization of the method is schematically shown in
Fig. \ref{3dmulti}. The setup is very similar to that proposed in
Fig. \ref{holog3} for Fourier Holography imaging, except for two
particulars. First, instead of a relatively long ($1$ ps) laser
pulse we should use a $10$ fs laser pulse. Second, instead of a
fixed bandpass optical filter we consider an interchangeable
bandpass filter with the same $1$ nm bandwidth, whose central
frequency can be rapidly changed, with a repetition rate similar
to that of the electron-bunch trains.

The duration of a pulse $\Delta T$ is related to the spectral
range $\Delta \lambda$ (in terms of wavelengths) by $\Delta T \sim
\lambda^2/(c \Delta \lambda)$. The laser pulse is chosen short
enough so that the reference pulse has a spectral range (about
$100$ nm) of the order of that associated with the minimal time
scale which we are planning to resolve (about $10$ fs). Note that
we do not have synchronization problem because when the combined
electromagnetic disturbance from reference and object passes
through the usual bandpass filter, the filter stretches it
yielding a wave packet about $1$ ps long (for a $1$ nm-bandwidth
filter). As a result for overlapping reference and object pulses
on the detector we only need sub-ps synchronization. If we now
tune the bandpass filter to other frequencies, we can perform
sliced measurements of $\bar{\rho}_0(\Delta \omega, \vec{r})$
within the predicted wide spectrum of the optical replica pulse
due to varying transverse size along the electron bunch.


We should reconstruct the 3D cube of data given by Fourier
transforming $\bar{\rho}_0(\Delta \omega, \vec{r})$ with respect
to $\Delta \omega$. Note that, for $\Delta \omega \neq 0$, the
quantity $\bar{\rho}_0$ is not real, so we should address the
problem that we can only reconstruct the relative phases in each
frequency slice. Yet, if we want perform inverse temporal Fourier
transform we need to have information about the relative phases
between slices. Our setup cannot provide this information. It
should be presented as a-priori knowledge. The relative phases
between slices can be measured using an undulator Optical replica
Synthesizer (ORS) \cite{ORSS}. In other words, we assume that we
are performing an extra FROG measurement of the peak-current
profile to correctly set the relative phase of each
frequency-slice \footnote{It should be stressed that the scheme in
Fig. \ref{3dmulti} is a purely linear optical scheme and does not
require intense pulses.  We assume that we perform an additional
FROG measurement over the undulator radiation in the ORS setup
\cite{ORSS}, to define the relative phase of each Fourier
component. The undulator radiator in the ORS setup yields energies
of about a few tens microjoule per pulse, which is much more than
the energy from the coherent OTR pulse, and more than sufficient
for FROG measurements.}.

We can then perform numerically, and with negligible computation
time, the Fourier Transform of $\bar{\rho}_0(\Delta \omega,
\vec{r})$ with respect to $\Delta
\omega$ and have a high resolution image. 

We may note here that the possibility of reconstructing complex
objects like $\bar{\rho}_0(\Delta \omega, \vec{r})$  is relatively
recent and specific of digital recording only. We have a
possibility to reconstruct taking the Fourier transform of a
complex object. Because the reconstructed wave field is a complex
function, both the intensity as well as the phase can be
calculated. This is in contrast to the case of optical hologram
reconstruction, in which only intensity is made visible. During
many decades, only optical reconstruction of holograms was used,
and only the intensity distribution of the object was recorded.

The present technique may be very attractive in practice. Again,
problems with iterative technique are avoided, and there are no
difficulties related to synchronization for 3D characterization.
The only limitation of this method is that it relies on a
multishot geometry.

\subsubsection{\label{sub:stF}Spatio-temporal Fourier transform
holography}

In the following we discuss an extension of the FTH technique from
the space into the space-time or, rather, into the space-frequency
domain. In time-domain holography, the reference pulse is a short
pulse. During holographic recording, the complex amplitude of the
spectral component of each signal pulse is stored in a series of
fringes, which arise as a result of interference with the
corresponding spectral component from the reference pulse. This
spectral holography technique was first proposed in
\cite{MAZ1,MAZ2}, and experimentally demonstrated on the
femtosecond time-scale in \cite{WEIN}. A close analogy exists
between FTH of monochromatic waves and spectral holography of
time-varying signals. This analogy results from the similarity
between the equation describing spatial diffraction and temporal
dispersion. Spectral holography is also named by others "spectral
interferometry" \cite{GEIN}. It should be mentioned that
interferometric stability is not required in our case, since this
method is based on single-shot interference-pattern measurements.
Spectral interferometry may be coupled with spatial interferometry
to preserve the spatial-phase information \cite{TANB}.

In \cite{TREO}, an extension of spatial holography and spectral
interferometry into spatio-temporal holography was proposed and
experimentally demonstrated, allowing for the reconstruction of
the full spatio-temporal electric field from a single ultrashort
laser pulse. Such technique constitutes the state of the art as
regards the diagnostics of ultrashort optical pulses.

Authors of \cite{TREO} deal with the problem of measuring an
arbitrary pulse. As a result, they only use self-referencing
schemes. Therefore, they must take advantage of a-priori known
information about the reference pulse in the temporal direction,
meaning that they need FROG measurement of the reference pulse
after spatial filtering. Since the technique is self-referencing,
this requires a signal pulse with large energy. We deal, however,
with a simpler situation. In our case, the optical replica pulse
has a duration of $100\div 200$ fs, so that a laser with about
$10$ fs-long pulses, which is available, can be used as an
external reference source. Therefore, there are no strict
requirements on the energy of the signal pulse.

\begin{figure}
\begin{center}
\includegraphics*[width=140mm]{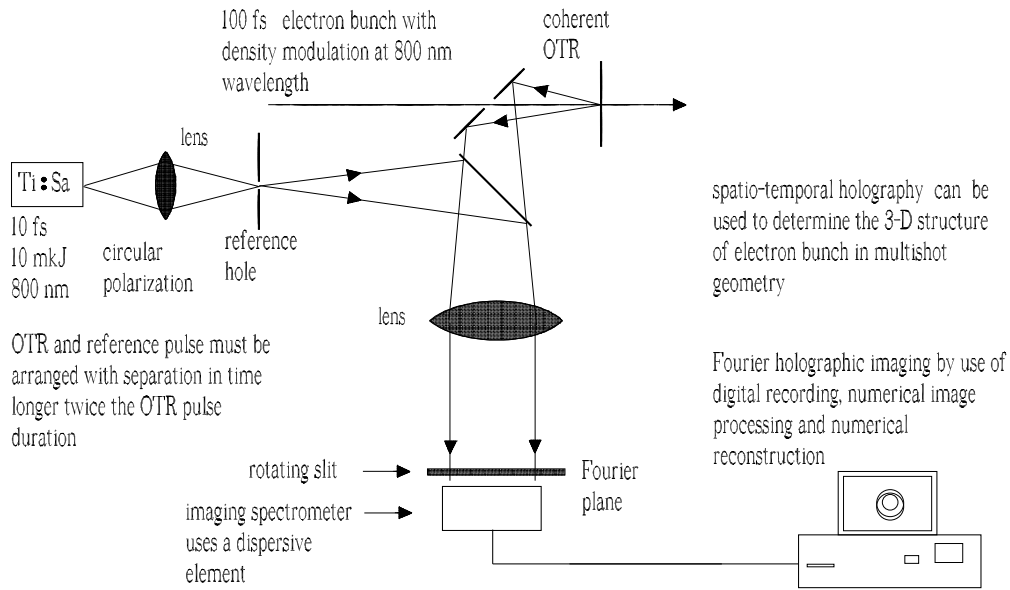}
\caption{\label{stft}  Multishot measurement of the 3D structure
of electron bunches by use of spatio-temporal Fourier transform
holography.}
\end{center}
\end{figure}
A possibility to extend FTH techniques from the space to the
frequency-space domain is illustrated in Fig. \ref{stft}. We
propose to use an imaging spectrometer with a dispersive element
installed in the Fourier plane of the coherent imaging system. The
slit of the spectrometer is centered on the optical axis and can
be rotated around the optical axis. In this way, we can record a
two dimensional slice of data with spatial frequency on one axis
and spectral wavelength information on the other.

The reference source can be considered as a point source in the
space-domain and as "temporal point source", i.e. a "short"
optical pulse in the time domain. Object and reference source must
be arranged in space with a separation twice larger than the
object width (this is the usual FTH condition) and must be
arranged with a separation in time domain twice longer than the
object pulse duration (this is the temporal equivalent of the FTH
condition).

Problems with the polarization of the object are solved in the
same way as for the frequency-gated FTH imaging i.e. by using a
circularly polarized reference pulse. The position of the slit and
its relation with the polarization of the OTR radiation pulse was
already discussed when considering the FRODI setup (see Fig.
\ref{slit}). The polarization of the OTR radiation pulse will be
selected by the slit orientation, and will always be linear
independently of the orientation of the slit, because it is
positioned in the spatial Fourier plane and it is centered around
the optical axis. As a result, with a circularly polarized
reference source we will always have the same optimal interference
for any orientation of the slit (aside for a constant, unimportant
phase in the hologram).

This technique has the same advantages of holographic techniques.
The only disadvantage is constituted by the need of additional
hardware (reference laser source), which is common to all
frequency-gated FTH techniques. However, no precise
synchronization is needed: sub-ps accuracy will suffice. In fact,
the only important point is that temporal separation will obey the
temporal equivalent of the FTH condition, also accounting for
jitter. For the holographic technique to work, it does not matter
if the temporal separation fluctuates from shot to shot.

Similarly to FRODI, this multishot geometry can be rearranged in a
single shot geometry. Several detection stations can be arranged
to simultaneously measure few projections of the charge density
distribution. Moreover, as for FRODI too, there is no need for
a-priori information from FROG measurements. The advantage with
respect to FRODI is that no ITA is needed, which will
substantially speed up the reconstruction process.


\subsubsection{\label{sub:srefff}Self-referencing measurements in
spatio-temporal FTH}

In the previous Section we discussed spatio-temporal FTH. This
approach requires a well-characterized reference pulse. In this
Section we restrict our attention on self-referencing
spatio-temporal FTH, in which the Optical Replica pulse is used to
measure itself. This allows one to reconstruct the 3D structure of
the electron bunch without the need for a reference laser source.

In particular, in this Section we discuss a method based on 2D
shearing interferometry\footnote{There is no important physical
difference between interferometry and holography.} \cite{DORR} in
the spatio-temporal frequency domain $(\omega, u_x)$. Shearing
interferometry is a well-known technique for the measurement of
wavefronts in the spatial domain. It operates by having the
wavefront with phase function $\phi(x,y)$ interfering with a
test-replica of itself shifted (laterally sheared) along the $x$
axis by a distance $\Delta x$. Thus, at a given spatial point
$(x,y)$ on the detector, the phase difference $\phi(x+\Delta
x,y)-\phi(x,y)$ can be extracted from the measured interferogram.
If $\Delta x$ is sufficiently small, such phase difference is
proportional to the gradient $\partial \phi/(\partial x)$. The
unknown spatial phase $\phi(x,y)$ can then be obtained (up to an
additional constant) by integration.

There is a well-known one-dimensional version of this technique to
characterize ultrashort laser pulses, which is called Spectral
Phase Interferometry of the Direct Electric-field Reconstruction
(SPIDER) \cite{MONM}. The SPIDER technique uses the principle of
spectral shearing interferometry to measure the spectral phase.
Typically, SPIDER is based on the interference of two time-delayed
replicas of the input pulse, which are also frequency-shifted by
$\Delta \omega$ with the help of non-linear optical methods. The
spectral gradient $\partial \phi/(\partial \omega)$ of the phase
$\phi(\omega)$ is then recovered from the one-dimensional
interferogram. Further integration yields back the phase
$\phi(\omega)$. A one-dimensional shearing-interferometry setup in
the temporal-frequency domain is more complicated than its spatial
counterpart, because it requires non-linear optics to generate the
spectral shear. However, the SPIDER technique can be extended to
allow for the reconstruction of the spatio-temporal phase from a
measurement of the spatial gradient of the phase \cite{DORR}.

\begin{figure}
\begin{center}
\includegraphics*[width=120mm]{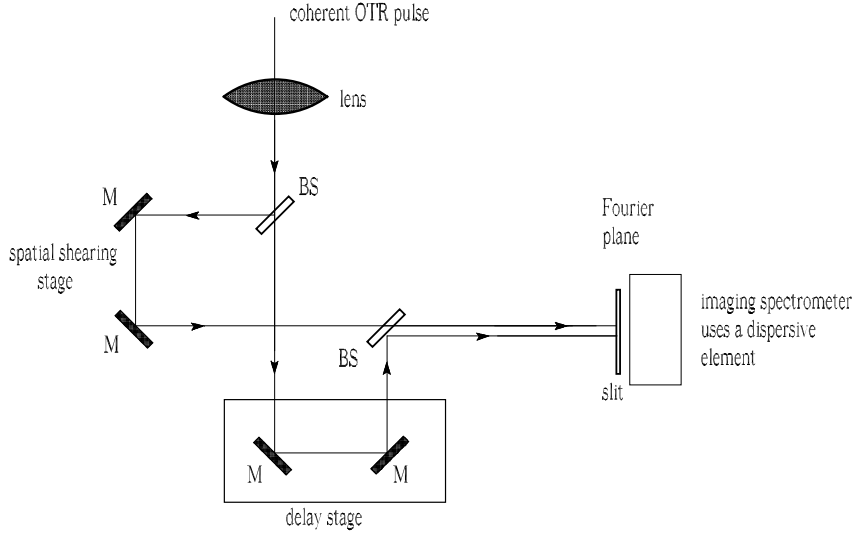}
\caption{\label{spide} Self-referencing spatio-temporal FTH
concept: BS, beam splitter; M, mirror. A Michelson interferometer
provides independent control of the spatial shear and delay
between two copies of the input coherent OTR pulse.}
\end{center}
\end{figure}
In our case of interest, it is possible to completely characterize
a slice of electric field $\widetilde{E}(\omega,\vec{u})$ as a
function of $\omega$ and e.g. $u_x$ for a given $u_y=0$, and
subsequently scan the angle of the slice in the $(u_x,u_y)$ plane,
corresponding to the $(x_f, y_f)$ plane, i.e. the Fourier plane.
We refer to the scheme in Fig. \ref{spide}. The phase
$\phi(\omega, u_x, 0)$ can be measured with a spatial shearing
interferometer, which generates two copies of the input field, one
of which is spatially sheared by $\Delta u_x$ with respect to the
other, and with relative spectral phase $\omega \tau$. The field
at a given plane is imaged through the two arms of a Michelson
interferometer onto the slit of the imaging spectrometer, which is
equipped with a dispersive element, so that the imaged intensity
is given by

\begin{eqnarray}
I(\omega, u_x) \propto
\left|\widetilde{E}(\omega,{u}_x,0)+\widetilde{E}(\omega,{u}_x+\Delta
u_x,0)\exp[i\omega \tau]\right|^2~. \label{Iomux}
\end{eqnarray}
If the fringe period due to the reference phase $\omega\tau$ is
sufficiently small, the interferometric component
$\widetilde{E}^*(\omega,{u}_x+\Delta
u_x,0)\widetilde{E}(\omega,{u}_x,0)\exp[i\omega \tau]$ can be
extracted with the help of Fourier processing techniques. This can
be obtained by using a temporal delay $\tau$ larger than a few
times the duration of the Optical Replica pulse. The argument of
the interferometric component is

\begin{eqnarray}
\phi(\omega,{u}_x+\Delta u_x,0)-\phi(\omega, u_x,0) = \Delta u_x
\frac{\partial\phi}{\partial u_x} (\omega,{u}_x,0)+\omega \tau~.
\label{phiphi}
\end{eqnarray}
Through the settings of the Michelson interferometer, one can
control the shear $\Delta u_x$ and the delay $\tau$. After
subtraction of the reference phase from the right hand side of Eq.
(\ref{phiphi}), one obtains the frequency-resolved
spatial-gradient of the phase $\partial \phi/(\partial u_x)
(\omega, u_x,0)$, which can be then integrated to give the phase
$\phi(\omega, u_x,0)+\phi_0(\omega)$, where $\phi_0(\omega)$ is an
arbitrary  function of the frequency only. This function can be
eliminated by renormalizing the phase at each $\omega$ for the
fixed spatial frequency $u_x=0$ and $u_y=0$ to the phase obtained
from the FROG measurement of the peak-current profile in the ORS
undulator setup. In fact,

\begin{eqnarray}
\phi_\mathrm{FROG} (\omega) = \phi(\omega,0,0)+\phi_0(\omega)~.
\label{phfrog}
\end{eqnarray}
The technique that we just described shares the same advantage of
the spatio-temporal FTH approach discussed in Section
\ref{sub:stF} and, additionally, it does not need a reference
laser source. The only disadvantage is constituted by the need for
a-priori information from FROG measurements.

\subsubsection{\label{sub:mxp}Time-gated Fourier-Transform Holography}

Up to now we introduced 3D FTH techniques, which operate in the
Fourier domain. These techniques require no special
synchronization\footnote{Sub-picosecond synchronization is still
required in our proposed scheme, but it may be relaxed using a
longer laser pulse, which poses more stringent requirements for
the average power of the laser beam.}. This is one of their
advantages. Here we will introduce time-resolved FTH techniques
operating in real space-time domain, which are based on the fact
that holograms record information about the object only under
simultaneous illumination by a coherent reference wave
\cite{ABRA}.

A laser capable of producing single pulses much shorter than the
electron bunch is needed for time-resolved imaging. Reference
lasers with $10$-fs pulse duration are available. One of the main
technical problems for time-resolved holography is the
synchronization between reference and optical pulses. Here we
propose a way of getting around this obstacle by shifting the
attention from the problem of synchronization to the problem of
measurement of the slice peak-current.

\begin{figure}
\begin{center}
\includegraphics*[width=110mm]{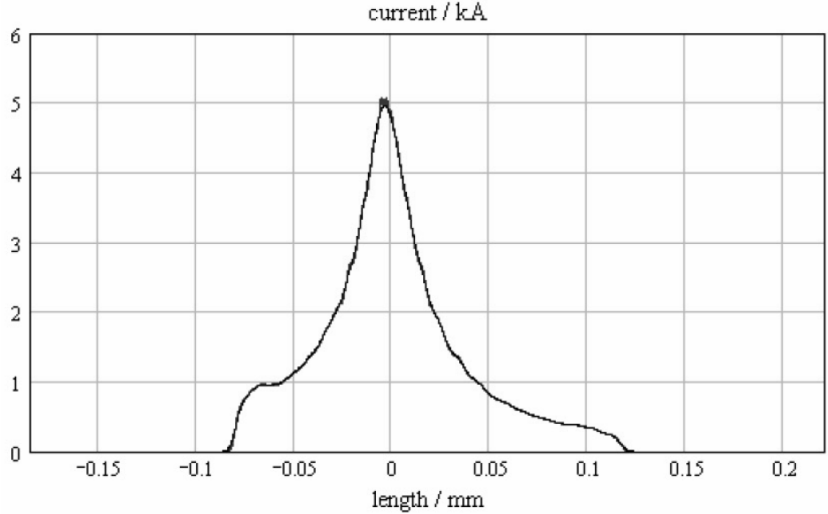}
\caption{\label{currXFEL} Peak-current profile at the entrance of
the SASE undulator (after XFEL TDR). }
\end{center}
\end{figure}

Consider Fig. \ref{currXFEL}, which shows the foreseen
longitudinal bunch profile (cited from \cite{XFEL}). Due to the
strong non-linear dependence of the FEL process on the
peak-current, only the central part is suitable for lasing.

\begin{figure}
\begin{center}
\includegraphics*[width=140mm]{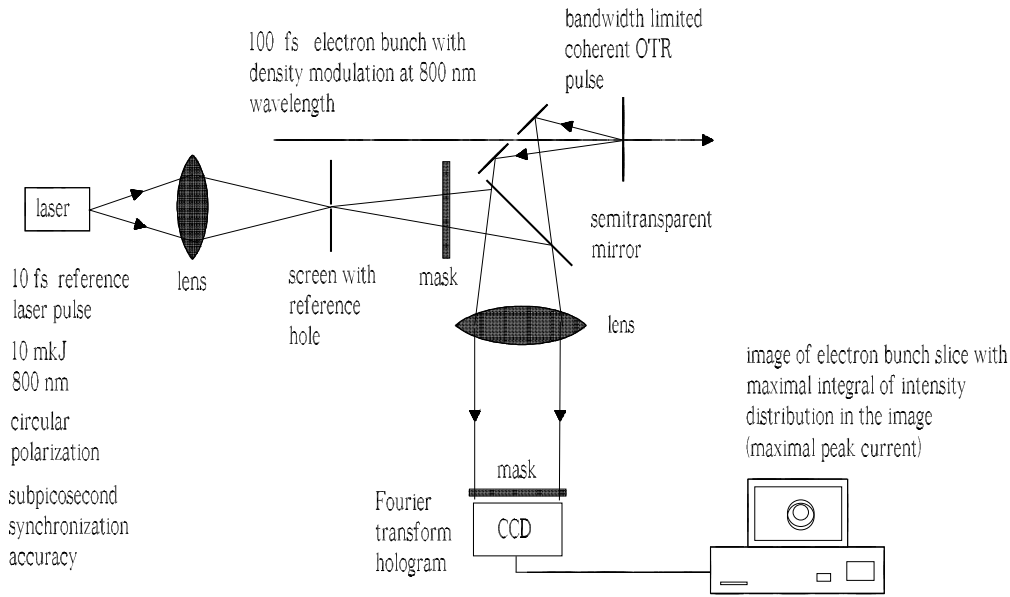}
\caption{\label{sliceim2}  Time-gated FTH setup for imaging the
slice of electron bunch with maximal peak-current. Since the
integral of the image of the electron bunch density distribution
is known, the slice with maximal peak-current can be uniquely
determined even when time jitter is present. }
\end{center}
\end{figure}
\begin{figure}
\begin{center}
\includegraphics*[width=130mm]{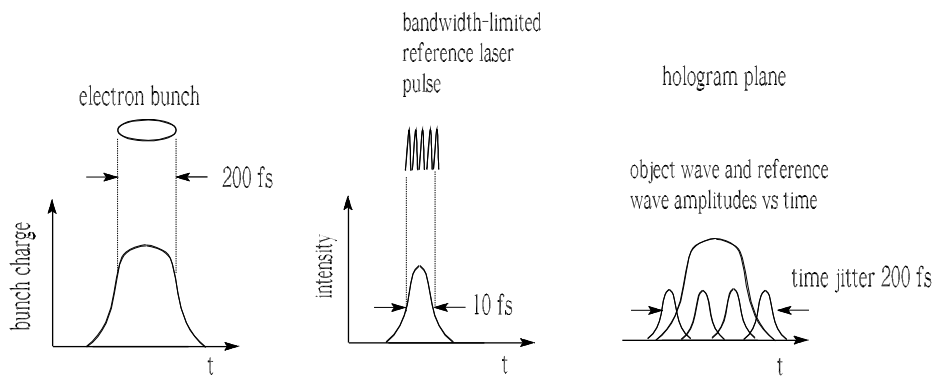}
\caption{\label{slicimfin} Sketch of principle for time-gated
FTH-imaging of the bunch slice with maximal peak-current.  }
\end{center}
\end{figure}
It is therefore of special interest to develop techniques capable
of imaging the slice with maximal peak-current, with a typical
width of about $3~\mu$m (or $10~$fs). The setup of the proposed
time-gated FTH technique is shown in Fig. \ref{sliceim2}.

The method is based, as before, on the Fourier Transform
holography. The idea is to use a very short reference laser pulse
of the order of $10$ fs, i.e. the slice longitudinal dimension, in
order to obtain contributions to the interference terms in the
hologram from a single slice\footnote{Note that the actual
thickness of the slice can be arbitrary, from $3 \mu$m up to the
limit for projected emittance, depending on the reference pulse
length.} within the electron bunch.

In contrast to the frequency-gated FTH, no spectral filter is
used\footnote{The Fourier masks are present instead. However,
their presence (in all schemes starting with those in Section
\ref{sec:diffite}) can be replaced by numerical image processing
before reconstruction.}, so that the reference and the optical
replica are not easily synchronized in the hologram plane. In
fact, both optical replica and reference laser pulse are subject
to time jitter in the order of $100\div 200$ fs, and
sub-picosecond synchronization does not guarantee that the
reference pulse overlaps with the maximum of the optical replica
pulse in the hologram plane. In order to solve this problem one
needs to discriminate the slice with maximal peak-current within
the electron bunch from the rest of the beam.

The duration $\tau_\mathrm{ref} \simeq 10$ fs of the reference
pulse is relatively long due to the applicability of the adiabatic
approximation, as $c \tau_\mathrm{ref}/\lambdabar \sim 30$ at our
selected optical wavelength, and it is relatively short compared
to the electron bunch duration, as $\tau_p/\tau_\mathrm{ref} \sim
20$.

\begin{figure}
\begin{center}
\includegraphics*[width=140mm]{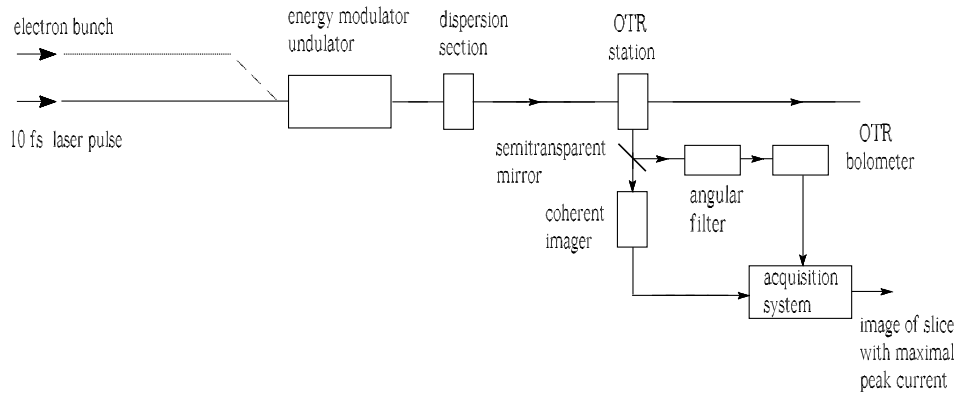}
\caption{\label{slirec}  Setup for recording image of bunch slice
with maximal peak-current using ultrashort seed laser pulse for
optical replica synthesizer. }
\end{center}
\end{figure}
A method for imaging the bunch slice with maximal peak-current is
illustrated in Fig. \ref{slicimfin}. The hologram records
information about the object only when it is illuminated with the
coherent reference wave. One of the main technical problems for
time resolved holography techniques is the relative
synchronization of the reference laser pulse and the optical
replica pulse from the OTR screen. In fact, both optical replica
and reference laser pulses are subject to time jitter. As
mentioned above, we shift the attention from the problem of
synchronization to the problem of measurement of the peak-current
of the slice. If a relative peak-current is known for each slice,
a sorting of results according to peak-current measurements
uniquely gives the slice with maximal peak-current without
knowledge of the delay between reference and object pulses.

Another technique closely related to the time-gated method for
imaging of the electron bunch with maximal peak-current is shown
in Fig. \ref{slirec}. The idea is to use a very short (order of
$10$ fs) reference laser pulse in order to obtain the contribution
from a single slice within the electron bunch (see \cite{ANG2,
ANGE}). As before, the arrival times between the laser pulse and
electron bunches jitter from shot to shot. In the case of Fig.
\ref{slirec}, we have the additional possibility to control the
slice peak-current using a bolometer, i.e. measuring the total
charge in the modulated slice on a shot-to-shot basis.
Subsequently, the acquisition system will discriminate the slice
with maximal peak-current. In this case, angular filtering  of
coherent OTR is needed before the bolometer in order to select the
halo part, whose energy-per-pulse does not depend on the
transverse size of the electron bunch.

\begin{figure}
\begin{center}
\includegraphics*[width=120mm]{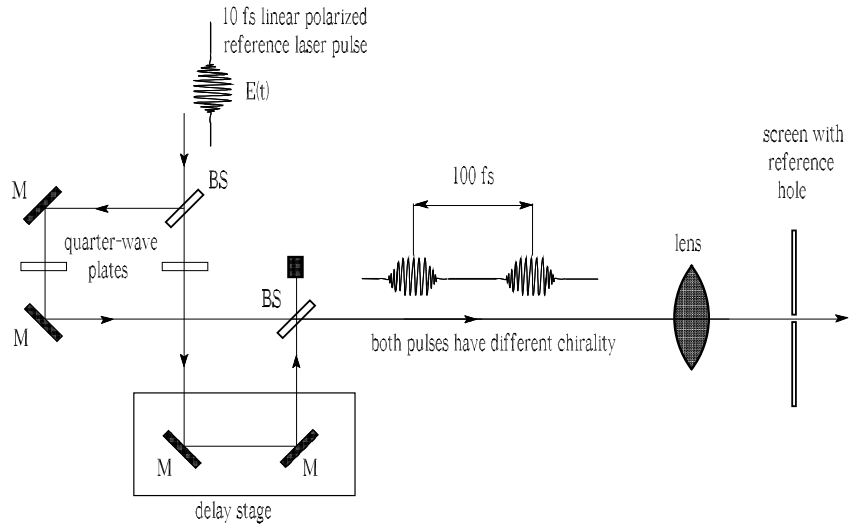}
\caption{\label{chira1} Simple method for solution of the
double-valued (peak-current) function problem in multi-shot 3D
electron bunch FTH imaging. The reference laser pulse is split
into two parts, where one is delayed with respect to the other of
about half electron bunch duration. Both pulses are passed through
different quarter wave plates and as a result have different
chirality. The pulses are recombined before being sent onto the
focusing optics (see Fig. \ref{sliceim2}).  }
\end{center}
\end{figure}
The time-gated FTH technique for imaging the electron-bunch slice
with maximal peak-current can be extended to a scheme to provide
the 3D image of the electron bunch in a multi-shot geometry. The
idea exploits the jitter between the reference (laser) pulse and
the electron bunch, i.e. the optical replica pulse. In fact, due
to the presence of jitter, not only the slice with maximal
peak-current will be recorded. The main problem is to
discriminate, shot by shot, the longitudinal position of the slice
in order to reconstruct the 3D image of the electron bunch. One
can measure the peak-current of the slice as discussed before,
with reference to Fig. \ref{sliceim2}, but there is always an
ambiguity on whether a slice with a given peak-current is located
before or after the slice with maximal peak-current. In other
words, with the exception of the slice with maximal peak-current,
we have a difficulty with a double-valued function (peak-current
as a function of position) for other peak-current values.

\begin{figure}
\begin{center}
\includegraphics*[width=120mm]{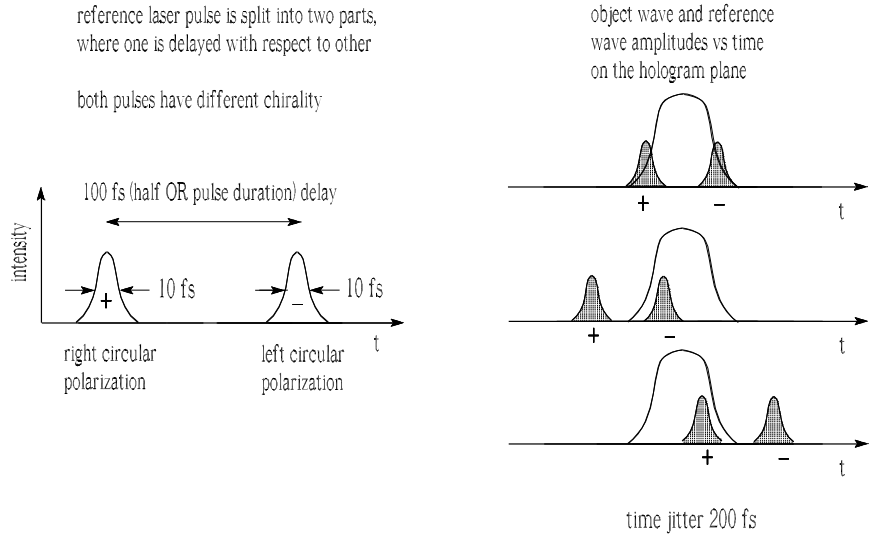}
\caption{\label{chira2} Schematic representation of 3D multi-shot
measurements in a time-resolved FTH setup. Left: time diagram of
the reference wave. Right: temporal profiles as they appear in the
hologram plane. The new method attempts to get around the jitter
obstacle, by measuring the slice peak-current. If the relative
peak-current is known for each slice, sorting the results
according to the vortex sign at the central (low spatial
frequency) region of the hologram uniquely gives the position of
the slice within the electron bunch.}
\end{center}
\end{figure}

Here we propose a solution to this problem with the help of a
double reference pulse, which may be produced as shown in Fig.
\ref{chira1}. A $10$-fs linearly polarized laser pulse is first
split into two identical pulses. The polarization of these pulses
is changed from linear to circular with the help of two different
quarter-wave plates, emerging with different chirality. Finally,
one of the two pulses passes through a delay stage of about half
electron bunch duration or more before being sent to the focusing
optics of the FTH setup (see Fig. \ref{sliceim2}), so that one
obtains two $10$-fs reference pulses with opposite chirality
delayed of about $100$ fs as shown in Fig. \ref{chira2}), left.

We should now be able to distinguish among the three situations
shown on the right of Fig. \ref{chira2}. When this is
accomplished, the 3D reconstruction can be successfully performed.
Due to the particular delay choice, there are no other beam
geometries to be studied. The idea is to distinguish among the
three situations on the right of Fig. \ref{chira2} by studying the
low-frequency part of the hologram, related to the halo regions C
and D in Fig. \ref{halo}. In these regions, the hologram shape
will not depend on the transverse electron density distribution.
Use of numerical  post-processing deconvolution (instead of using
two masks as in Fig. \ref{sliceim2}) will serve our purpose here.

\begin{figure}
\begin{center}
\includegraphics*[width=130mm]{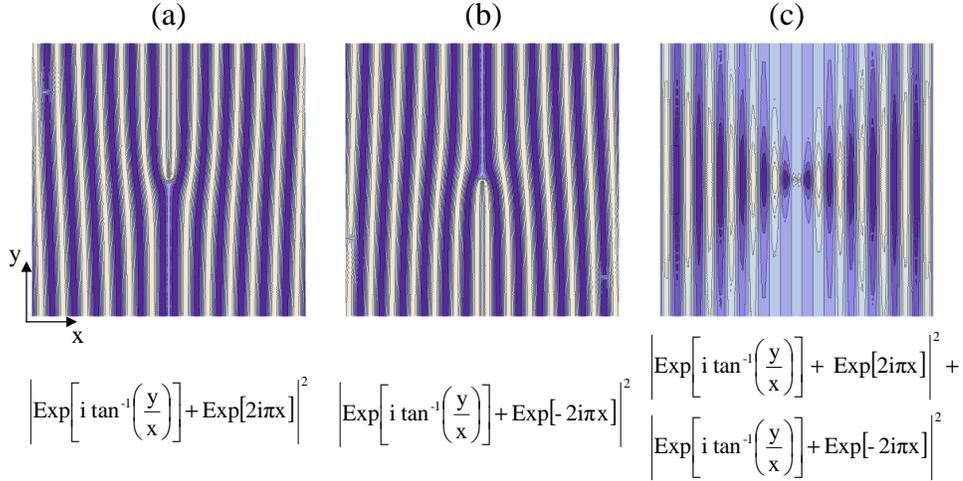}
\caption{\label{forks} Holographic patterns corresponding to (a)
chirality $+1$ of the reference, (b) chirality $-1$ of the
reference and (c) sum of the previous two cases.}
\end{center}
\end{figure}
The shape of such kind of holograms "is characterized by nearly
parallel" fringes, "except for a forking pattern in the vicinity
of the core" \cite{SAKS}, as shown in the example of Fig.
\ref{forks}. For chirality $+1$ (bottom case on the right of Fig.
\ref{chira2}, corresponding to case $(a)$ in Fig. \ref{forks}) the
forking pattern will be directed upwards, for chirality $-1$
(upper case on the right of Fig. \ref{chira2}, corresponding to
case $(b)$ in Fig. \ref{forks}), downwards. When we have
simultaneous overlapping of the two reference pulses with the
Optical replica pulse (center case on the right of Fig.
\ref{chira2}, corresponding to case $(c)$ in Fig. \ref{forks}) we
have a combination of two forking patterns one upwards and the
other downwards.

It remains to be discussed whether the separation of the
holographic fringes is enough to guarantee distinguishable
patterns. The separation between the fringes depends on the
transverse offset of the reference source relative to the object.
When the offset of the reference source increases up to about $3
\sigma$, $\sigma$ being the rms transverse electron bunch
dimension, the fringes will be separated of a distance
proportional to $1/3 \sigma$, which is much larger compared to
$1/(\gamma \lambdabar)$. As a result, for the first fringe we will
obtain a fork which is well distinguishable, meaning that it does
not overlap with the bunch structure and it is not influenced by
the structure of the single electron pattern.

We can speculate that computerized analysis will easily
distinguish among these three kinds of events, solving the problem
of identifying the position of the slice within the bunch. There
are many advantages related to the application of this technique.
In fact, it avoids complications like the need for ITA, it does
not require a-priori information about the electron bunch
structure (i.e. can be used  to measure ultra-short bunches) and
it does not require synchronization.

\subsection{HOTRI technique to provide a full 3D image of individual, arbitrary electron
bunches}

In this Section we introduce a novel technique to characterize the
3D structure of the electron bunch. This technique combines
multiple-reference FTH (Section \ref{sub:mrf}) and time-gated FTH
(Section \ref{sub:mxp}) methods. The fact that multiple-reference
FTH may be naturally used for time-resolved experiments was first
discussed in a concept presented in \cite{SCH2}.

\begin{figure}
\begin{center}
\includegraphics*[width=140mm]{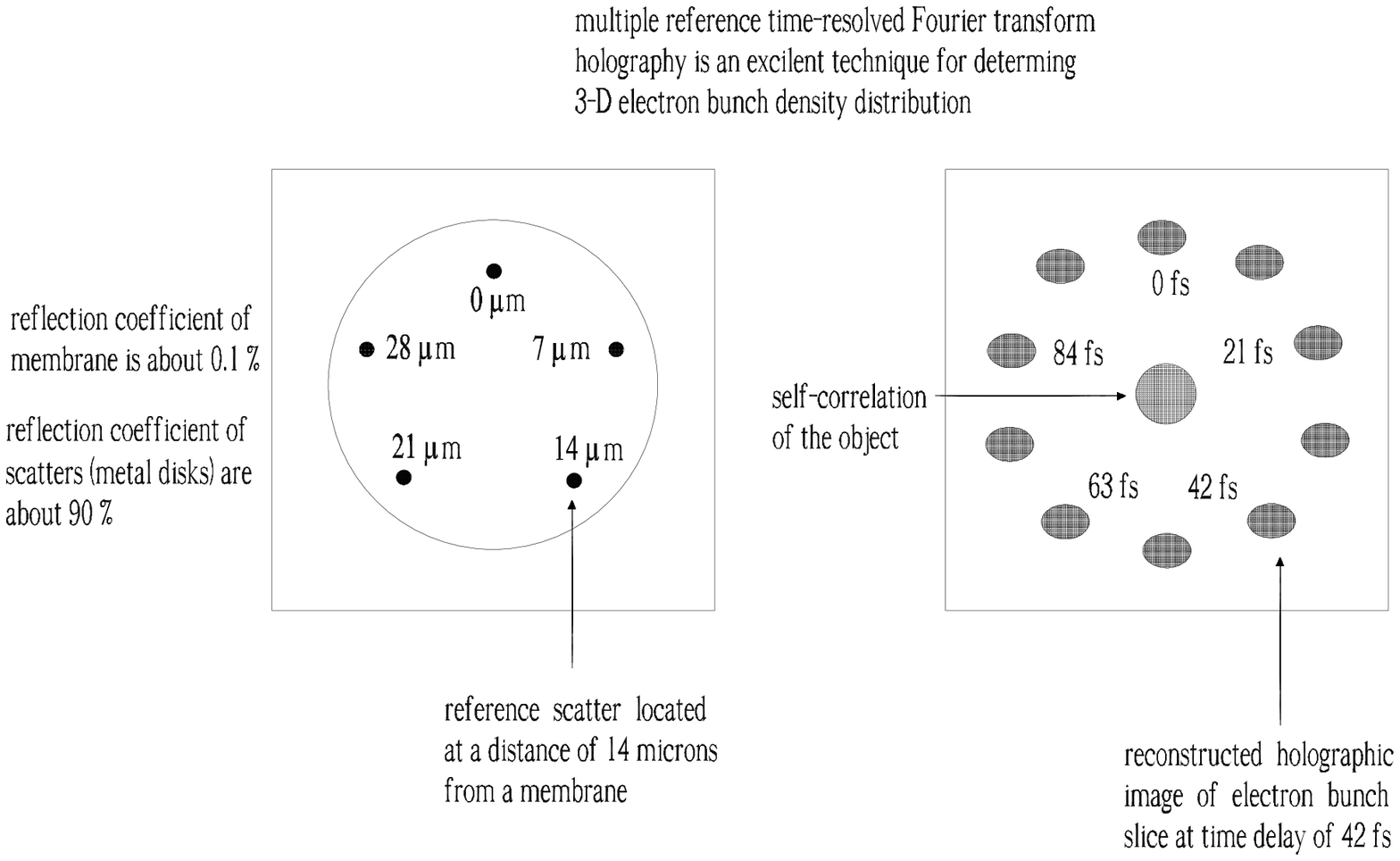}
\caption{\label{movie2} Membrane for time-resolved FTH with
reference scatters (left) and reconstructed holographic images of
electron bunch slices (right). }
\end{center}
\end{figure}

Multiple-reference FTH can allow us to simultaneously record
images of different slices of the electron bunch, thus providing a
3D image of a single electron bunch. Our idea is to vary the
time-delay of the reference sources. Each reference source is a
small disk with high reflectivity, and the distance from the
reference scatter to the membrane is varied in order to create a
time-delay between reference wave and OTR object wave, Fig.
\ref{movie2}, left. As a result, once the hologram is processed,
different reconstructed object images will correspond to different
slices with given time-delays along the electron bunch, Fig.
\ref{movie2}, right.

\begin{figure}
\begin{center}
\includegraphics*[width=140mm]{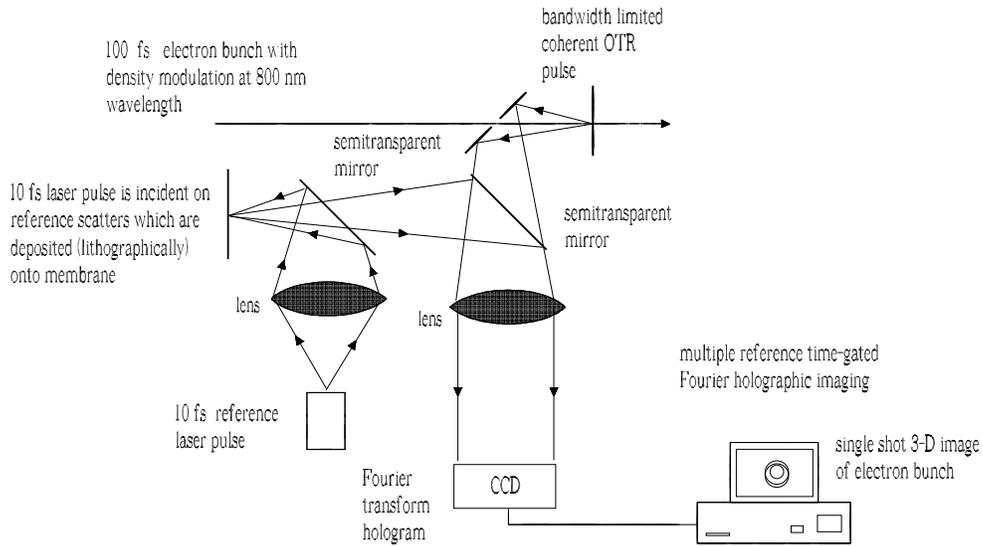}
\caption{\label{movie1}  3D electron bunch structure monitor by
use of the HOTRI technique. }
\end{center}
\end{figure}

From the measurement using the setup shown in Fig. \ref{movie1},
we can make a movie (five frames with $10$ fs exposure-time and
$50$ THz frame-rate) of the electron bunch propagating through the
OTR screen, showing the electron density distribution against
space and time. We named this novel technique Holography Optical
Time Resolved Imaging (HOTRI).

\section{\label{sec:chirp}Sensitivity to the energy chirp of the
electron bunch}

As we mentioned above, some coherent imaging techniques take
advantage of the fact that the test pulse is bandwidth-limited. In
this Section we will discuss what conditions should be met for the
optical replica pulse to be considered bandwidth-limited, and what
is the impact of a phase-chirp on the different techniques
proposed in this work.

In Section \ref{sec:setup} we discussed the impact of
self-interactions on the optical replica. We demonstrated that, in
practical cases of interest, self-interactions effects and
energy-spread influence can be neglected, and the amplitude of the
bunching can be considered uniform along the bunch. This is
sufficient to produce an optical replica of the electron bunch,
that is a radiation pulse whose electric field amplitude is a
replica of the charge density distribution of the electron bunch.
The optical replica pulse can be used to characterize the
longitudinal bunch profile of the electron bunch \cite{ORSS} with
the FROG technique.

\begin{figure}
\begin{center}
\includegraphics*[width=110mm]{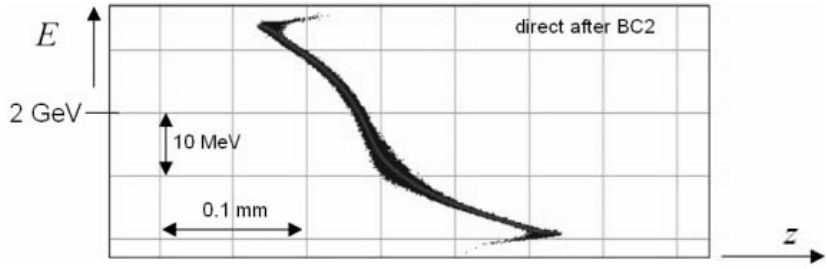}
\caption{\label{phspl}   The foreseen longitudinal phase-space
distribution of electrons after the magnetic compressor BC2 at the
European XFEL, extracted from \cite{MART}.}
\end{center}
\end{figure}

It should be remarked, however, that the bunch compression
procedure introduces an energy chirp in the electron beam after
the magnetic chicane BC2 in the case of the European XFEL, Fig.
\ref{compe}. This chirp introduces, as it will be discussed below,
a distortion into the phase of the bunching which is not important
for the ORS setup as concerns peak-current measurements, but is
important when dealing with electron bunch imaging. A
comprehensive analysis of the evolution of the longitudinal phase
space after compression, including nearly all important physical
effects (space charge interactions, Coherent Synchrotron Radiation
(CSR), shape variation, resistive walls) is given in \cite{MART}.
The longitudinal phase-space of the electron bunch just after the
magnetic compressor BC2, taken from \cite{MART} is reproduced in
Fig. \ref{phspl}.

The energy chirp shown in Fig. \ref{phspl} is due to a combination
of different effects. In fact, energy chirp is intrinsically
introduced in the framework of a single-particle dynamics during
the bunch compression procedure, but it is modified by the
presence of other wakefields. Moreover, as noted in \cite{MART},
the bunch formation system is composed by several chicanes, which
complicates furthermore the situation. A qualitatively similar
conclusion can be found in simulations for LCLS (see the study on longitudinal phase-space in \cite{LCLS}).

The measurement of the longitudinal bunch profile by FROG is
insensitive to the presence of the energy chirp and, better, as
shown in \cite{SLCH}, the ORS can be used to measure the energy
chirp. In fact, a FROG device can measure the phase of the optical
replica, and the phase is strongly related to the energy chirp by


\begin{eqnarray}
\delta \phi = \frac{R_{56}}{\lambdabar_m} \frac{\delta
\gamma}{\gamma}~. \label{chirphi}
\end{eqnarray}
However, the phase chirp of the optical replica pulse is, in its
turn, related to whether or not the optical replica pulse is
bandwidth limited.

For a bandwidth-limited pulse of duration $\tau_p$ and bandwidth
$\Delta \omega$ one has $\Delta \omega \tau_p \sim 1$. This
implies that $\Delta \omega/\omega \sim (\omega \tau_p)^{-1}$.
Also, for a bandwidth-limited pulse, one should have a constant
phase i.e., from a practical viewpoint, a phase-shift smaller than
a radian along the pulse. A phase-chirp linear in time corresponds
to a shift in the carrier frequency. One can account for such a
chirp by introducing an effective carrier frequency. Thus, we will
neglect the linear phase-chirp and account for the non-linear
phase-chirp only. We may restate the condition above, requiring
that the non-linear chirp should be sufficiently small, and the
phase distortion across the pulse should be smaller than a radian.

Here we consider the low energy case of $2$ GeV after the second
bunch compressor, where the longitudinal phase-space is depicted
in Fig. \ref{phspl}. The value of the momentum compaction factor
is that of the ORS dispersion section, i.e. $R_{56} \simeq 50
~\mu$m. With the help of Eq. (\ref{chirphi}), one can see that a
phase shift $\delta \phi$ of a radian corresponds to an energy
shift of about $\Delta\gamma \sim 5$ MeV along the pulse.
Considering the European XFEL case, one should extract from the
large linear chirp Fig. \ref{phspl}, the non-linear energy chirp
component.  A quick estimation shows that the non-linear component
of the chirp is in the order of $2\div 3$ MeV per bunch length,
yielding a phase shift smaller than a radian. There is some room
to optimize the situation by increasing the energy modulation
introduced by the seed laser, increasing the energy in the seed
laser pulse and subsequently decreasing the $R_{56}$ parameter in
Eq. (\ref{chirphi}) of a factor $2\div 3$.

It should be noted that the large chirp shown in Fig. \ref{phspl}
is introduced \textit{ad hoc} at the European XFEL (and at LCLS
too) in order to diminish CSR effects in the chicane. Additionally
this also allows to compensate for the linear component of the
energy chirp introduced by wakefields after the bunch formation
system, which has opposite slope. In several situations of
interest, the non-linear component of the chirp can be smaller at
the end of the bunch formation system.

Having said this, it is interesting to discuss which techniques
presented in this paper are influenced by the energy chirp of the
electron beam, and which are not.

First, let us consider methods that are not influenced. As we have
seen in Section \ref{sec:ft}, direct (coherent) imaging without
spectral filter allows one to record the integral of the squared
modulus of the optical replica pulse, i.e. a quadratic projection.
In this case, any phase deviation along the optical replica pulse
is irrelevant. In fact, energy chirp only introduces a phase
perturbation which depends on time and not on transverse
coordinates, at least within a single-particle dynamics
approximation in the bunch-formation system.

Also techniques based on the a-priori use of information from
independent FROG measurements work, without modification, with and
without chirp.  This is the case when we combine real and
reciprocal space spectrometers, Fig. \ref{fgate}, and when we
consider frequency-gated FTH, Fig. \ref{3dmulti}. Other techniques
which are completely insensitive to energy chirp are
spatio-temporal FTH, Fig. \ref{stft}  and HOTRI, Fig. \ref{movie1}
and, more in general, all time-gated FTH techniques. This follows
from the short duration of the reference pulse in all these cases,
which is $10$ fs long and has a bandwidth of about $100$ nm. In
fact, the non-linear energy chirp necessary to significantly
perturb the operation of time-resolved techniques based on such
kind of reference pulses can be estimated to be about $100$ MeV,
which is far from the design parameters of the European XFEL and
LCLS. As a result, direct (coherent) imaging without spectral
filter, techniques based on independent FROG measurements, the
spatio-temporal FTH method and all time-gated FTH techniques are
insensitive to energy chirp.

Second, there are techniques that are influenced by the presence
of energy chirp, but that may be easily modified to account for
such presence. These are schemes operating in the frequency domain
in the 3D imaging mode, i.e. the multishot measurement method
based on the use of the projection algorithm, Fig. \ref{dspec},
and FRODI, Fig. \ref{bunchfrog}. In order to account for the
effects of the energy chirp, in these cases we can make use of the
knowledge about the phase deviation along the optical replica
pulse as measured by FROG. This requires modification of the
technique, because it implies an extra FROG measurement. Note that
the required information is automatically available if we perform
peak-current profile measurements.

Once the phase along the optical replica pulse is known, we can
use it e.g. in the reconstruction of 2D traces for FRODI.
Consider, for the sake of illustration, the $(x,t)$ projection. If
we assume that the optical pulse is bandwidth-limited, its phase
is constant, and $\rho_0$ and $a_f$ are real and positive (see Eq.
(\ref{duerho0})), i.e. one can talk about the slowly varying real
and positive envelope of the pulse\footnote{Note that
$\bar{\rho}_0(\Delta \omega,\vec{r})$ is still a complex-valued
function of $\Delta \omega$. Also note that in this case,
$\bar{\rho}_0(0,\vec{r}) = a_f \int \rho_0(t,\vec{r}) dt$. In
other words $\bar{\rho}_0$ at $\Delta \omega=0$ coincides with the
temporal projection of $\rho_0$.}. If, instead, the optical pulse
is not bandwidth limited, one has $a_f = a_0 \exp[i\phi(t)]$ with
$a_0$ constant and $\phi(t)$ a non-zero phase due to energy chirp
within the electron bunch.

Therefore, when the energy chirp is negligible, the envelope of
the optical replica pulse is an exact replica of $\rho_0(t,x,y)$,
i.e. of the electron density distribution, which is real and
nonnegative. As seen before, this is enough to guarantee unique
reconstruction of the object. The fact that the optical replica
pulse envelope is real and nonnegative actually amounts to
complete information about the phase of the pulse envelope, which
is just constant. When non-negligible energy chirp is present, the
pulse envelope is no more real, but the phase along the $t$ axis
is nevertheless fully determined by the FROG measurements, and
this is enough to guarantee reconstruction. This allows to
reconstruct any arbitrary projection $(x',t)$, where $x'$ is
rotated of an arbitrary angle with respect to $x$ in the $(x,y)$
plane, but not the $(x,y)$ trace. Anyway, if a sufficient number
of $(x',t)$ traces are known, this information is enough for
reconstruction. Note that the 3D optical replica pulse in the
space-time domain can now be reconstructed as a cube of
complex-valued data. The amplitude is proportional to
$\rho_0(t,x,y)$.


Finally, third, all other techniques presented in this work,
including direct (coherent) imaging with spectral filter, Fig.
\ref{schemecoh}, linear projection diffraction imaging, Fig.
\ref{recon}, and FTH imaging of linear projection, Fig.
\ref{holog3}, can operate only with a bandwidth-limited optical
replica pulse. Note that when we extend these techniques to 3D
reconstruction methods we can easily overcome the energy-chirp
problem. However, when we deal with single $(x,y)$ projection
measurement only, like in Fig. \ref{schemecoh}, Fig. \ref{recon}
and Fig. \ref{holog3}, this is not possible anymore and one should
take care that the energy chirp is actually negligible.

\section{\label{sec:conc}Conclusions}

The study presented in this work opens up a novel field in
high-energy, ultrashort electron beam diagnostic methods, based on
the use of coherent Optical Transition Radiation (OTR) for 3D
imaging of an ultrashort electron bunch by means of the Optical
Replica Synthesizer (ORS) \cite{ORSS}.

The ORS technique was proposed as an attempt to solve a very
challenging problem, namely the longitudinal electron bunch
diagnostics, by producing a 1D optical replica radiation pulse,
thus reducing the electron bunch characterization problem to the
1D optical pulse characterization problem. The ORS performs
longitudinal diagnostics of ultrarelativistic, ultrashort electron
bunches in two steps. First, an optical replica of the electron
bunch is prepared.  Second, 1D characterization of the electron
bunch is performed by a femtosecond oscilloscope. For this purpose
the ORS setup makes use of a commercially available
Frequency-Resolved Optical Gating (FROG) device. The availability
of methods and instrumentation for measuring ultrashort optical
pulses constitutes the main advantage of using an optical replica
compared to other diagnostics schemes.

In the case of 3D characterization of an electron bunch we
demonstrated that the same approach can be realized. In other
words, the 3D imaging of an electron bunch with spatio-temporal
coupling can be reduced to 3D characterization of an optical pulse
with spatio-temporal coupling. However, the 3D problem has a
different status compared to its 1D counterpart. In fact, the 3D
problem was formulated only recently, and no commercial devices
are yet available. This does not mean that the schemes discussed
in this work constitute a challenge as concerns their technical
realization. The 3D problem has a number of possible solutions,
and this explains why some of the techniques proposed here (FRODI,
HOTRI) are completely original. Note that our formulation of the
3D problem is less general than the one addressed by the optical
community. In our case of interest, the optical pulse is not
arbitrary, but being the optical replica of an electron bunch, has
a duration of about $100$ fs and has specific constraints like the
bandwidth limit, or in more general cases, a phase which is a
function of time only. This opens up the possibility to use some
approaches, as Fourier Transform Holography (FTH) with reference
source and its variations, which cannot be applied for an
arbitrary pulse.

We proposed to use coherent OTR as the source of radiation for the
optical replica pulse. In the case of coherent OTR we can take
advantage of the large number of photons and of the coherent
properties of the radiation pulse. We developed schemes based on
direct coherent OTR imaging, diffractive imaging, FTH, and a
combination of these techniques. In particular, novel techniques
like FRODI and HOTRI allow for the 3D characterization of
ultrashort electron bunches by use of single and multishot
measurements.

We proposed to exploit the highly-developed software algorithms
for Diffractive Imaging and Fourier Transform Holography.
Reconstruction of 3D images in real space-time domain can be based
on methods like the Gerchberg-Saxton algorithm, the Fineup
algorithm or their generalizations. Moreover, similarly to the ORS
setup, we proposed to make use of the large selection of
commercially available instrumentation for optical pulse
diagnostics. Performing diagnostics in the range of visible optics
will greatly simplify a choice of optical elements or their
manufacturing. Also, ultrashort, up to $10$ fs, optical laser
sources with any state of polarization are available as a
reference sources. The same approach, to use already
highly-developed devices, is proposed when considering 3D-imaging
methods in the 3D Fourier domain. For this kind of applications
there is a need for optical systems capable of forming
high-resolution images at different wavelengths. Imaging
spectrometers are instruments capable of providing this
functionality. There are two types of imaging spectrometers. One
type is based on the use of a plurality of bandpass filters to
measure a 3D data cube in the 3D Fourier domain for different
frequencies. A second type is based on the use of a dispersive
element, and it is particularly suitable for single shot 3D
measurements.

Due to the users' needs, future XFELs will operate with shorter
and shorter electron bunches. Our proposed diagnostic techniques
have the potential for extensions from the $100$-fs time scale,
which we discussed here, to the $10$-fs time scale of electron
bunch duration by straightforward rescaling to shorter
wavelengths. Exploitation of shorter wavelengths is possible
because our setups are based on linear optical elements only. The
shortest possible wavelength compatible with glass optics is about
$200$ nm, which corresponds to the fourth harmonic of a Ti:Sa
laser. Using particular geometries, FROG devices can operate in
this range as well \cite{FROG}.

We regard the concepts presented here, based on the combination of
ORS with coherent OTR imager setups, as the start of a novel
direction in diagnostic methods. Therefore, the present work
should not be considered as comprehensive of all foreseeable
applications. On the contrary, we hope that our work will
stimulate interest and open the door to many new possibilities for
ultra-fast electron bunch diagnostics.


\section{\label{sec:graz}Acknowledgements}

We thank our colleagues Dirk Noelle, Harald Redlin and William
Schlotter for useful discussions, Massimo Altarelli, Reinhard
Brinkmann and Edgar Weckert for their interest in this work.


\begin{thebibliography}{99}



\bibitem{LCLS} J. Arthur et al. Linac Coherent Light Source
(LCLS). Conceptual Design Report, SLAC-R593, Stanford (2002) (See
also http://www-ssrl.slac.stanford.edu/lcls/cdr).


\bibitem{SCSS} Tanaka, T. \& Shintake, T. (Eds.): SCSS X-FEL Conceptual Design
Report. Riken Harima Institute, Hyogo, Japan, 2005 (see also
http://www-xfel.spring8.or.jp).

\bibitem{XFEL} M. Altarelli et al. (Eds.), XFEL: The European X-Ray Free-Electron
Laser. Technical Design Report, DESY 2006-097, DESY, Hamburg
(2006) (See also http://xfel.desy.de).

\bibitem{KOND}
A.M.~Kondratenko and E.L.~Saldin, Part. Accelerators 10 (1980)
207.

\bibitem{DERB}
Ya.S.~Derbenev, A.M.~Kondratenko and E.L.~Saldin, Nucl. Instrum.
and Methods 193 (1982) 415.

\bibitem{BONI} R. Bonifacio, C. Pellegrini and L. Narducci, Opt.
Commun. 50 (1984) 373.

\bibitem{PELL}
J.B.~Murphy and C.~Pellegrini, Nucl. Instrum. and Methods A 237
(1985) 159.

\bibitem{ORSS}  E. Saldin, E. Schneidmiller and M.
Yurkov,  Nucl. Instrum. and Methods 539, 3 (2005) 499.

\bibitem{FROG} R. Trebino, "Frequency-resolved optical gating: the measurement of ultrashort laser pulses",
Boston, Kluwer Academic Publishers (2002).


\bibitem{ANG2} A. Angelova et al., Phys. Rev. ST Accel. Beams 11
(2008) 070702.

\bibitem{ANGE} A. Angelova et al., TUPC114, Proceedings of EPAC
2008, Genova, Italy (2008) 1332.


\bibitem{TIME} G. Geloni, E. Saldin, E. Schneidmiller and M.
Yurkov, Opt. Commun. 281 (2008) 3762.

\bibitem{STAB} G. Geloni, E. Saldin, E. Schneidmiller and M.
Yurkov, Phys. Rev. ST AB 11 (2008) 120701.

\bibitem{BRAG} F. Bragheri et al., Opt. Lett. 33, 24 (2008) 2952.

\bibitem{GERC} R.W. Gerchberg and W. Saxton, Optik 35 (1972) 237.

\bibitem{ABRA} N. Abramson, Opt. Lett. 3, 4 (1978) 121.

\bibitem{CZON} P.L. Csonka, Part. Accel. 8 (1978) 225.

\bibitem{OURX} G. Geloni, E. Saldin, E. Schneidmiller and M. Yurkov, Nucl.
Instrum. and Meth. in Phys. Res. A 583 (2007) 228.


\bibitem{LOO1} H. Loos et al., Observation of Coherent Optical Transition Radiation in the
LCLS Linac, SLAC-PUB-13395 (2008).

\bibitem{LASH} E. Saldin, E. Schneidmiller and M. Yurkov, Nucl.
Instrum. and Methods in Phys. Res. A 528 (2004) 355.

\bibitem{GINZ} V.L. Ginzburg and I. M. Frank, Soviet Phys. JETP
16 (1946) 15.

\bibitem{WIDE} H. Wiedemann, Synchrotron Radiation, Springer Berlin,
2003.

\bibitem{OUREDGE} G. Geloni, E. Saldin, E. Schneidmiller and M.
Yurkov, "Theory of Edge Radiation" DESY 08-118 (2008),
http://arxiv.org/abs/0808.1846, submitted to Nucl. Instrum. and
Methods in Phys. Res. A.

\bibitem{BEAM} European XFEL Start-to-End Simulations,
http://www.desy.de/xfel-beam/s2e/xfel$\_$v4.html (2006).

\bibitem{ARTF} G. Geloni, E. Saldin, E. Schneidmiller and M.
Yurkov, Optics Communications 276, 1 (2007) 167.

\bibitem{JACK} J. Jackson, Classical Electrodynamics, 3rd ed., Wiley, New York,
1999

\bibitem{ROSE}  J. Rosenzweig, G. Travish, A. Tremaine Nucl. Instr. and
Methods A 365 (1995) 255.

\bibitem{TREM} Tremaine, et al,  Phys. Rev. Letters 81 (1998)
5816.

\bibitem{TRE2}  A. Tremaine, et al,  Nucl. Instrum. and
Methods in Phys. Res. A 429 (1999)  209.

\bibitem{LUMP}  A. H. Lumpkin, et al., Phys. Rev. Lett. 86 (2001)
79.

\bibitem{LUM2} A. H. Lumpkin, et al.,  Phys. Rev. Lett. 88 (2002)
23801.

\bibitem{LUM3} A. H. Lumpkin, et al., Nucl. Instrum. and
Methods in Phys. Res. A 507 (2003) 200.

\bibitem{LUM4}  A. H. Lumpkin, et al,  Nucl. Instrum. and
Methods in Phys. Res. A 528 (2004) 179.

\bibitem{BENG} B. Yang, A design report for the optical
transition radiation imager for the LCLS undulator, LCLS-TN-05-21
(2005).

\bibitem{GOOD} J. W. Goodman, Introduction to Fourier Optics, Mc Graw-Hill Book
Company (1968).

\bibitem{DUMO} P. Dumontet, Opt. Acta 2 (1955) 53.

\bibitem{TICH} D.A. Tichenor and J.W. Goodman, J. Opt. Soc. Am.
62 (2008) 293.

\bibitem{MILL} J.P. Mills and B.J. Thompson, J. Opt. Soc. Am. A, 3, 5,
694 (1986).

\bibitem{BRAI} E. Brainis, C. Muldoon, L. Brandt and A. Kuhn, Optics Communications 282 (2009)
465.

\bibitem{CAST} M. Castellano and V.A. Verzilov, Spatial Rosolution
in Optical Transition Radiation (OTR) beam diagnostics,
LNF-98/017(P) (1998).

\bibitem{LEBE} V.A. Lebedev, Nucl. Instrum. and Meth. in Phys. Res.
A 372 (1996) 344.

\bibitem{BAKA} R. Bakarat, J. Opt. Soc. Am. 52 (1962) 276.

\bibitem{POLA} G. Machavariani et al., Optics Communications 281 (2008)
732.

\bibitem{POL2} P.B. Phua, High power radially polarized light generated from photonic crystal segmented
half-wave-plate, http://arxiv.org/abs/0710.4979v1 (2007)

\bibitem{BOO1} S. G. Lipson, H. Lipson  and D. S.
Tannhauser, Optical Physics, Cambridge University press (1995).

\bibitem{BORN} M. Born and E. Wolf, "Principles of Optics", 5th ed.,
Pergamon, Oxford (1975).

\bibitem{FINN} J. R. Fineup, Appl. Optics, 21 (1982) 2758.

\bibitem{MIL2}  R. P. Millane, J. Opt. Soc. Am. A 7 (1990) 394.

\bibitem{FIN9} J. R. Fineup, Appl. Optics 32 (1993) 1737.

\bibitem{FIN2} J. R. Fineup and A. M. Kowalczyk, Opt. Soc. Am. A 7
(1990) 450.

\bibitem{LUKE} P.R. Luke, J.V. Buske and R.G. Lyon, SIAM Review,
44, 2  (2002) 169.

\bibitem{STRO} G. W. Stroke and D. Falconer, Phys. Lett. 13
(1964) 306.

\bibitem{SACK} Z.S. Sacks et al. J. Opt. Soc. Am. B, 15, 8
(1998) 2226.

\bibitem{STAD} L-M. Stadler et al.,  Phys. Rev. Lett. 100
(2008) 245503.

\bibitem{SCHL} W. F. Schlotter et. al., Appl. Phys. Lett. 89
(2006) 163112.

\bibitem{GABO} P. Gabolde and R. Trebino Optics Express 12,
19 (2004) 4223.

\bibitem{MAZ1} Y. T. Mazurenko, Appl. Phys. B 50, (1990) 101.

\bibitem{MAZ2} Y. T. Mazurenko, Opt. Eng. 31, (1992) 739.

\bibitem{WEIN} A. M. Weiner and D.E. Leaird, OPt. Lett. 19, (1994)
123.

\bibitem{GEIN} J. P. Geindre et al., Opt. Lett. 19 (1994) 1997.

\bibitem{TANB} T. Tanabe et al. J. Opt Soc. Am. B 19 (2002) 2795.

\bibitem{TREO} P. Gabolde and R. Trebino Optics Express 14,
23 (2006) 11460.

\bibitem{DORR} C. Dorrer, E. Kosik and I. Walmsley, Appl. Phys. B
74 (2002) S209.

\bibitem{MONM} A. Monmayrant, S. Gorza, P. Wasylczyk and I.
Walmsley, Beyond the fringe: SPIDER - the anatomy of ultrashort
laser pulses", Photonik international 2007, (2007) 2.

\bibitem{SAKS} Z. S. Sacks, D. Rozas, and G. A. Swartzlander Jr., J. Opt. Soc. Am. B, 15, 8
(1998) 2226.

\bibitem{SCH2} W. F. Schlotter et. al., Opt. lett. 32 (2007)
21.

\bibitem{MART} M. Dohlus, Modelling of space charge and CSR effects in bunch
compression systems, in Proceedings of EPAC 2006, Edinburgh,
Scotland, WEYFI01, (2006) 1897.


\bibitem{SLCH} Y. Ding, Z. Huang and P. Emma, Integration of the optical replica ultrashort electron bunch
diagnostics with the current-enhanced SASE in the LCLS, SLAC-PUB
12672 (2007).

%





\end{thebibliography}
\end{document}